\documentclass [11pt]{article}

\usepackage{setspace}
\usepackage{amssymb}
\usepackage{amsmath}
\usepackage{amsfonts}
\usepackage{amssymb}
\usepackage{setspace}
\usepackage{amsthm}

\usepackage{graphicx}
\usepackage{url}
\usepackage{color}
\usepackage[dvipsnames]{xcolor}
\definecolor{cuteBlue}{rgb}{0.258, 0.387, 0.574}
\definecolor{cuteGreen}{rgb}{0, 0.3, 0}
\usepackage{cancel}
\usepackage{comment}
\usepackage[framemethod=TikZ]{mdframed}
\usepackage{enumitem}
\usepackage{wasysym}
\usepackage{listings}
\usepackage{float}
\usepackage{booktabs}
\usepackage{fixltx2e}
\usepackage{threeparttable}
\usepackage{titling}
\usepackage{zref-base}
\usepackage{makecell}
\usepackage{array}
\usepackage{hhline}
\usepackage{titlesec}
\usepackage{gensymb}

\usepackage[
  labelfont=bf,
  font=small,
  aboveskip=30pt,
  belowskip=10pt,
  labelsep=period,
  justification=raggedright,
  singlelinecheck=off
]{caption}
\usepackage{subcaption}
\usepackage{longtable}

\usepackage[colorlinks=true, urlcolor=cuteBlue, citecolor=cuteGreen, linkcolor=black]{hyperref}

\usepackage[normalem]{ulem}

\usepackage[utf8]{inputenc}

\usepackage[T1]{fontenc}

\usepackage{textcomp}

\usepackage{authblk}

\reversemarginpar
\allowdisplaybreaks

\date{\vspace{-5em}}  

\titlespacing\section{0pt}{12pt plus 4pt minus 2pt}{-2pt plus 2pt minus 2pt}
\titlespacing\subsection{0pt}{12pt plus 4pt minus 2pt}{-2pt plus 2pt minus 2pt}
\titlespacing\subsubsection{0pt}{12pt plus 4pt minus 2pt}{-2pt plus 2pt minus 2pt}

\setlist{itemsep=0pt, topsep=0pt}

\setitemize{listparindent=\parindent}
\setenumerate{listparindent=\parindent}

\newcolumntype{M}[1]{>{\centering\arraybackslash}m{#1}}
\newcolumntype{N}{@{}m{0pt}@{}}

\usepackage{orcidlink}

	\usepackage[
		backend=bibtex,
		style=nature, 
		sorting=none,				
		url=false, 					
		doi=false, 					
		isbn=false, 				
		eprint=false, 			
		maxbibnames=9, 			
		firstinits=true,
	]{biblatex}

	\usepackage[
		aboveskip=30pt,
		belowskip=10pt,
		labelfont=bf,
		labelsep=period,
		justification=raggedright,
		singlelinecheck=off
	]{caption}

        \usepackage{gensymb}
	\usepackage{textgreek}

	\usepackage{titling}

	\usepackage{adjustbox}

	\usepackage{ifthen}
	\newboolean{maintext}
	\newboolean{sitext}

	\usepackage{xparse}

	\newcommand{\beginsupplement}{
					\setcounter{section}{0} 
			\renewcommand{\thesection}{S\arabic{section}}%
			\setcounter{table}{0} 
			\renewcommand{\thetable}{S\arabic{table}}%
			\setcounter{figure}{0} 
			\renewcommand{\thefigure}{S\arabic{figure}}%
			\setcounter{equation}{0} 
			\renewcommand{\theequation}{S\arabic{equation}}%
		}

	\title{The Environment-Dependent Regulatory Landscape of the {\it E. coli} Genome }
	
	\author[1]{Tom R\"oschinger\,\orcidlink{0000-0002-4900-3216}}
	\author[2]{Heun Jin Lee\,\orcidlink{0009-0002-4615-9271}}
	\author[1]{Rosalind Wenshan Pan\,\orcidlink{0009-0005-0778-3794}}
	\author[1]{Grace Solini\,\orcidlink{0009-0009-9523-6393}}
	\author[1]{Kian Faizi\,\orcidlink{0000-0003-1306-0320}}
	\author[3]{Baiyi Quan}
	\author[1, 3]{Tsui‑Fen Chou}
	\author[4]{Madhav Mani\,\orcidlink{0000-0002-5812-4167}}
	\author[5, 6]{Stephen Quake}
	\author[1, 7, +]{Rob Phillips\,\orcidlink{0000-0003-3082-2809}}

	\affil[1]{Division of Biology and Biological Engineering, California Institute
	of Technology, Pasadena, CA 91125, USA}
	\affil[2]{Department of Applied Physics, California Institute of Technology, Pasadena,
	CA 91125, USA}
	\affil[3]{Proteome Exploration Laboratory, Beckman Institute, California Institute of Technology, Pasadena, CA 91125, USA}
	\affil[4]{Department of Engineering Sciences and Applied Mathematics, Northwestern University, Evanston, IL 60208, USA}
	\affil[5]{Department of Bioengineering, Stanford University, Stanford, CA 94305, USA}
	\affil[6]{Department of Applied Physics, Stanford University, Stanford, CA 94305, USA}
	\affil[7]{Department of Physics, California Institute of Technology, Pasadena,
	CA 91125, USA}
	\affil[+]{Correspondence: tom@caltech.edu, phillips@pboc.caltech.edu}

\begin{document}


	\addtocontents{toc}{\protect\setcounter{tocdepth}{-1}}
		\begin{refsegment}
			\defbibfilter{notother}{not segment=\therefsegment}
			\setboolean{maintext}{true}
			\ifthenelse{\boolean{maintext}}{
			\maketitle 

\section{Abstract}

All cells respond to changes in both their internal milieu and the environment around them through the regulation of their genes.  Despite decades of effort, there remain huge gaps in our knowledge of both the function of many genes (the so-called y-ome) and how they  adapt to changing environments via regulation.  Here we describe a joint experimental and theoretical dissection of the regulation of a broad array of over 100 biologically interesting genes in {\it E. coli} across 39 diverse environments, permitting us to discover the binding sites  and  transcription factors that mediate regulatory control.  Using a combination of mutagenesis, massively parallel reporter assays, mass spectrometry and tools from information theory and statistical physics, we  go from complete ignorance of a promoter's environment-dependent regulatory architecture to predictive models of its behavior.  As a proof of principle of the biological insights to be gained from such a study,  we chose a combination of genes from the y-ome,  toxin-antitoxin pairs, and genes hypothesized to be part of regulatory modules; in all cases, we discovered a host of new insights into their underlying  regulatory landscape and resulting biological function.

\section{Introduction} 
\noindent The discovery in the early 1960s that there are genes whose job it is to control other genes and how that control is exercised through environmental influences was heralded as ``the second secret of life,''~\cite{Judson1996,ullmann2011memoriam} vastly expanding the original conception of the gene itself.  It has now been more than sixty years since Jacob and Monod ushered in their repressor-operator model and the allied discovery of allosteric regulation~\cite{jacob1961genetic,Echols2001}.  And yet, although more than $10^{17}$ bases have been deposited in the Sequence Read Archive (SRA) database~\cite{sragrowth}, we are still extremely far from understanding how all of the genes of any organism are regulated, or even  what the functions of all of those genes might be. The regulatory landscape of the genome requires building a bridge between its base pairs, the molecules that bind to them, and its environmental context. This is the language we wish to learn how to speak, read, and write.   However, despite a prodigious effort in the case of \textit{Escherichia coli}~\cite{Ishihama2017,belliveau2018systematic,Larsen2019,ireland2020deciphering, GaoPalsson2021,Freddolino2022,trouillon2023genomic,Lally2025}, one of biology's best studied model organisms, for roughly  60\% of its genes, databases lack any description of their regulatory architectures~\cite{Keseler2011,Santos-Zavaleta2019}. 
Note that we use the words ``regulatory architecture'' to imply that the binding site positions are known with base pair resolution and which transcription factors bind those binding sites is explicitly known, meaning that simple cartoons like those shown in the upper right of Figure~\ref{fig1}(A) can be assembled. Further, for roughly 35\% of its genes we lack sufficient evidence to  report their function.  These genes of unknown function have been christened the  y-ome~\cite{ghatak2019ome,moore2024revisiting} since many of them  have names that begin with the letter y, dating back to the first complete annotation of the \textit{E. coli} genome \cite{blattner1997complete}.  There is yet another crucial challenge to understanding the regulation-based physiology and evolution of these organisms, namely, our vast ignorance of the ways in which genes are coupled to environmental stimuli.  A beautiful demonstration of the
often hidden influence of environment on phenotype  was carried out more than a decade ago~\cite{Nichols2011}.  Just as the  y-ome is a powerful and concise nomenclature for our ignorance of the functional properties of the proteome, the ``allosterome'' refers to our ignorance of the ways in which environmental signals couple to those very same proteins, changing their  activity~\cite{Lindsley2006}. Said differently, different proteins ``care'' about different environmental perturbations.   Making progress on all of these fronts is essential to a modern, genome-based understanding of the physiology and evolution of all organisms.

The goal of our work is to systematically address these questions by providing a simultaneous promoter-by-promoter and high-throughput quantitative dissection of a variety of biologically interesting genes, as well as genes whose function or context have not yet been discovered. We place particular emphasis on the all-important question of how  genes are regulated in response to a myriad of different environmental conditions.  
In contrast to the painstaking and hugely successful gene-by-gene dissections of classical molecular biology (for several excellent examples from a huge literature, see~\cite{Weickert1993,Muller-Hill1996,Schleif1993Book,Schleif2000,Ptashne2002,Browning2004,Dodd2005,Oppenheim2005}), our aim is to determine promoter function and regulation for many genes,  in the presence of a broad canvas of distinct environments, all in a single experiment. By using a combination of mutagenesis, massively parallel reporter assays (MPRAs), mass spectrometry and statistical physics, we can go from complete ignorance of a promoter's regulatory architecture to predictive models based upon thermodynamic or kinetic models of gene expression (Figure~\ref{fig1}(A)). Our work builds on and is inspired by  many brilliant studies using MPRAs~\cite{kinney2010using,Melnikov2012,Kosuri2013,Urtecho2019,Regev2020,Lagator2022,Barkai2024}.
In particular, the present work is founded upon the MPRA studies known as Sort-Seq~\cite{kinney2010using, Melnikov2012, belliveau2018systematic} and Reg-Seq~\cite{ireland2020deciphering},  which share the philosophy of using mutated promoters and gene expression measurements in conjunction with information theory to generate high-throughput hypotheses for binding site locations.  In conjunction with these binding site hypotheses, both experimental and computational approaches are then used to determine which transcription factors bind them.  For reasons we will explain in detail later in the paper, we note that the problem of figuring out which transcription factors bind to which binding sites is very challenging.  Here, we developed and exploited a next generation version of Reg-Seq, including a streamlined protocol and genome integrated reporters, to study the regulatory architecture of more than 100 promoters in 39 different growth conditions, while paving the way to study the entire regulatory genome of an organism in one experiment. A detailed discussion of related methods and literature can be found in the Supplementary Information~\ref{sec:SI_literature}.

\begin{figure}
    \centering
    \includegraphics[height=\dimexpr\textheight-124pt\relax]{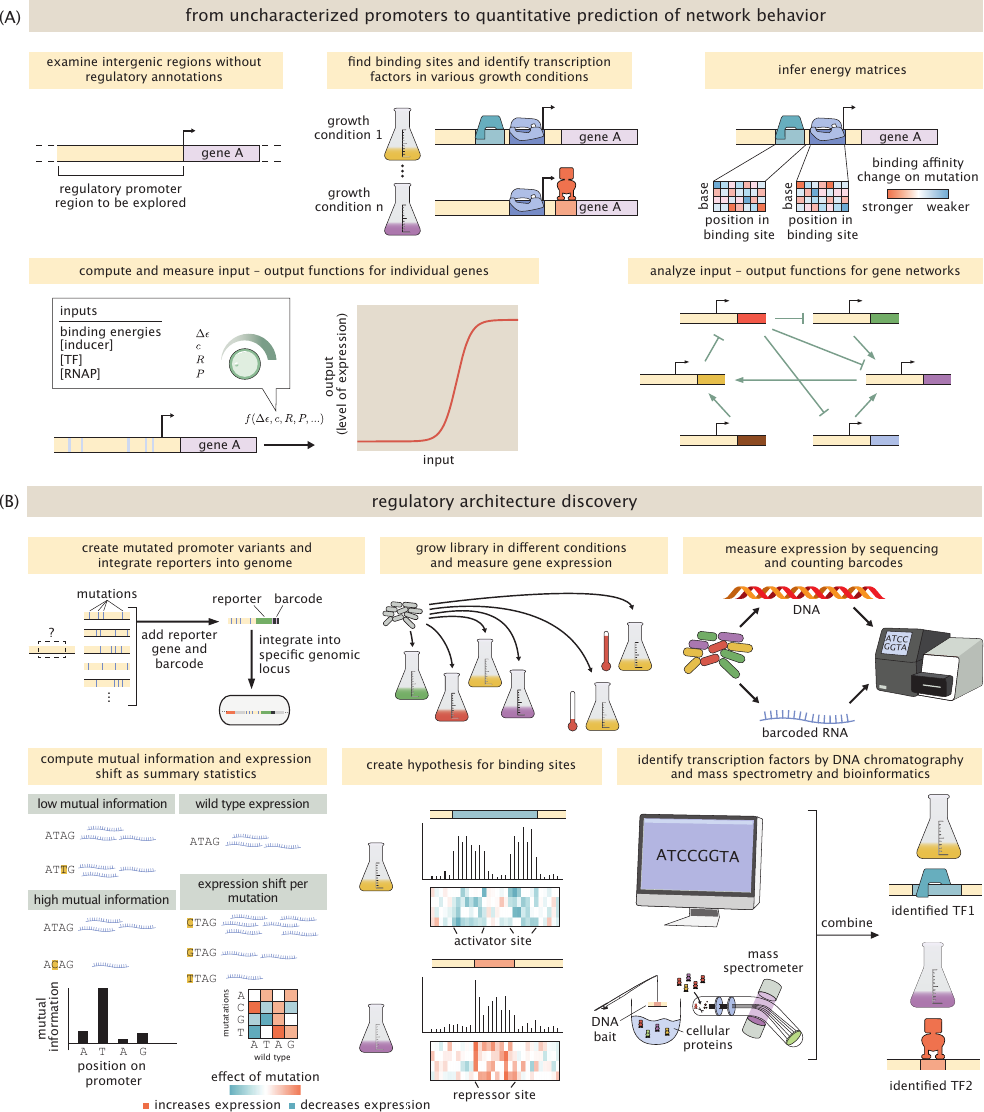}
    \caption{\textbf{Discovering regulatory architectures.}
    (A) High level strategy for complete characterization of the regulatory architecture of a previously uncharacterized gene.  The procedure starts with genes of unknown regulatory architecture and aims for a quantitative analysis of the input-output properties of the regulatory network.  (B) Experimental procedure for characterizing a previously uncharacterized gene. The six elements shown here provide a schematic of the steps needed to go from an uncharacterized promoter to one in which we have a well-defined, environmentally-dependent model of the binding sites, the allied transcription factors, and the energy matrix (shown under the information footprints) describing their binding interaction.}
    \label{fig1}
\end{figure}

Our ultimate objective is to carry out a systematic analysis for every gene in {\it E. coli} (and later other organisms such as {\it P. aeruginosa}) under a broad array of environmental conditions. 
To achieve such a condition-dependent dissection of the regulatory landscape, we adopt the protocol shown in Figure~\ref{fig1}(B).   Briefly, the procedure entails first using massively parallel reporter assays in conjunction with information theory and simple probabilistic models to generate hypotheses for the locations of binding sites.  These binding sites are then used as a fishing hook for DNA chromatography and  mass spectrometry to identify bound transcription factors (see lower right of Figure~\ref{fig1}(B)) or used for computational motif scanning~\cite{Gupta2007} in which we compare our putative binding sites to databases of known binding sites.  In addition to identifying which transcription factors bind  these putative binding sites, the goal is to ultimately infer a nucleotide-by-nucleotide binding energy matrix (for a deep analysis of this approach see~\cite{Stormo2013}) as we have done in the past~\cite{kinney2010using,ireland2020deciphering, belliveau2018systematic, barnes2019mapping}, which predicts a promoter's input-output function based on statistical mechanics~\cite{barnes2019mapping, garcia2011quantitative, brewster2012tuning, razo2018tuning, chure2019predictive}. 

As a proof of principle of the biological insights to be garnered from a study that carries out high-throughput and quantitative promoter discovery across a large diversity of environmental conditions,  we chose a suite of genes of wide biological significance. In particular, we focus on 104 promoters that struck us as particularly exciting.
We explored these promoters in 39 growth conditions, where we perturb the environment in varying ways (see upper middle of Figure~\ref{fig1}(B)), ranging from providing cells with a variety of carbon sources to the presence of antibiotics to growth in anaerobic conditions. 16 of the promoters were chosen as so called ``gold standards.'' These genes have well-defined regulatory architectures and have been studied in detail in previous experiments~\cite{ireland2020deciphering, belliveau2018systematic}, many as famous case studies. 
Including these gold-standard genes in our repertoire is important because  it allows us to compare the approaches presented in this work to previous studies of these promoters and verify the results, as well as to discover refinements or contradictions with the results present in databases of regulatory knowledge. 

Of course, despite the great interest in recapitulating decades of molecular biology experiments on gene regulation in a high-throughput fashion, our approach shows its real power when brought to bear on previously unexplored promoters.  To that end, 18 genes were chosen that display sensitivity to their environmental contexts  as discovered in a  seminal mass spectrometry proteome study in \textit{E. coli}~\cite{schmidt2016quantitative}. In those experiments, cells were grown under 22 different growth conditions and the copy numbers of roughly half the $\approx 4500$ genes were measured and summarized in a fascinating giant spreadsheet.  The 18 genes here were chosen because their copy numbers were found to be highly variable across conditions in that mass spectrometry census.  This high variation in copy number suggests that in certain environmental conditions,  these genes are under strong regulatory control. We performed a systematic analysis of this data which makes it abundantly clear how different conditions can yield very different protein copy numbers~\cite{Belliveau2021}.  For example, carbon transporters for sugars other than glucose reveal large changes in copy number only in the presence of the carbon source they specifically transport. Of the 18 high-variance genes we considered from that mass spectrometry study, 9 had no function annotated at the time of this study, meaning that not only were they interesting candidates from a regulatory perspective, but they were also part of the y-ome~\cite{ghatak2019ome}.  
 
 We expanded our analysis of  y-ome genes by choosing  another  13 such genes lacking any functional description from EcoCyc \cite{moore2024revisiting}. We hypothesize that these genes offer an interesting opportunity for discovering new regulatory networks and gene functions,  since the identification of transcription factor binding sites offers clues as to  which pathways these genes are involved in, and acts as a starting point to discover their function.  As another proof-of-principle biological category, we chose 18 genes that are part of toxin/anti-toxin systems. Expression of toxin genes can have drastic effects on cellular physiology and can be triggered by various stresses \cite{yamaguchi2011regulation}, requiring tightly controlled regulation.  Since one of our primary biological emphases here is on the mapping between environmental conditions and regulatory architecture, toxin-antitoxin systems provide a ready-made case study in regulatory response to environmental dynamics.  Our next set of biologically interesting case studies focused on the recent introduction of the so-called iModulons in the work of Lamoureux et al.~\cite{lamoureux2021precise,lamoureux2023multi}.  They dubbed groups of genes that are controlled by the same transcription factor iModulons. We chose two newly identified groups, responding to the putative transcription factors YmfT and YgeV, respectively. Including these sets of genes is an interesting opportunity to investigate how genes respond to environments as a collective. Additionally, both iModulons contain even more y-ome genes. We rounded out our list of case studies by choosing  6 genes that are part of gene regulatory networks with feedforward loop motifs, which present a compelling starting point for discovering how pertubations in the environment are transmitted through gene networks~\cite{Milo2002,shen2002network,mangan2003structure,mangan2003coherent}. The entire list of genes that serve as the basis of our study can be found in Table \ref{tab:genes_chosen}.

We note that much beautiful earlier work has provided deep genome-wide insights into how an organism responds to specific environments.  We summarize some of these results in Section~\ref{sec:SI_literature}.  Here we switch the narrative to a perspective that focuses on a smaller set of genes than the whole genome, but with the objective of rigorously characterizing the way in which the environment alters the regulatory landscape. To that end, the remainder of the paper is organized to illustrate how we carried out these case studies. In the next section, we provide a wide-ranging exploration of the principal results of our study. There we present both general conclusions as well as specific biological descriptions of particularly interesting genes.  The Supplemental Information goes much further by providing both the summary data and a description of each of the more than 100 genes that were the object of our study.

\vfill\eject
{\fontsize{5pt}{8pt}\selectfont
\begin{longtable}{|l|l|l l |l|l|l l |l|l|} 
\hline
group & gene & & & group & gene & & & group & gene\\
\hline
gold Standard & & & & antibiotic/toxin & & & & incoherent feed forward, type 4 &\\
 & rspA  &  & & & tisB&  & & & ihfA (X) \\
& araA &  & & & blr&  & & & ompR (Y)  \\
& araB &  & & & gyrA&  & & & (z) ompF (Z) \\
& znuB &  & & &ghoT&  & & & \\  
& znuC &  & & & emrA&  & & &\\  
& xylA &  & & & emrB&  & & &\\  
& xylF &  & & & yagB&  & & &\\ 
& dicC &  & & & yjjJ&  & & &\\  
& relE &  & & & prlF&  & & &\\  
& relB &  & & & yhaV&  & & &\\  
& ftsK &  & & & acrB&  & & &\\ 
& lacI &  & & & acrZ&  & & &\\ 
& marR &  & & & ldrD&  & & &\\  
& dgoR &  & & & rdlD&  & & &\\ 
& dicA &  & & & tabA&  & & &\\  
& araC &  & & & ratA&  & & &\\
&  &  & & & dinQ&  & & &   \\
&  &  & & & tolC&  & & &   \\
\hline
Schmidt et el. & & & & Schmidt et el. uncharacterized & & & &  incoherent feed forward, type 1 &\\
& hdeA & & & &cusF & & & & crp (X)  \\
& aceA  & & & & yjbJ & & & & galS (Y)\\
& ecnB  & & & & elaB & & & & mglB (Z$_1$)\\
& mcbA  & & & & yncE & & & & galE (Z$_2$\\
& tnaA  & & & & yqjD & & & &\\
& mglB  & & & & ybaY & & & &\\
& gatA  & & & & ygiW & & & &\\
& lpp   & & & & zapB & & & &\\
& tmaR  & & & & ybeD & & & &\\
\hline
YmfT imodulon  & & & & YgeV imodulon  & & & & uncharacterized protein (y-ome) & \\ 
& fur    & & & & ybiY  & & & &   yacC              \\
& sulA   & & & & rcsB  & & & &   yacH              \\
& intE   & & & & xdhA  & & & &   yadG              \\
& xisE   & & & & xdhB  & & & &   yadI               \\
& ymfH   & & & & xdhC  & & & &   yadE               \\
& ymfJ   & & & &  ygeW & & & &   yadM               \\
& ymfT   & & & & ygeX  & & & &   yadN               \\
& ymfL   & & & & ygeY  & & & &   yadS               \\
& ymfM   & & & & hyuA  & & & &   ykgR               \\
& ymfN   & & & & ygfK  & & & &   yahC               \\
& beeE   & & & & ssnA  & & & &   yahL               \\
& jayE   & & & & ygfM  & & & &   yahM               \\
& ymfQ   & & & & xdhD  & & & &   yqaE               \\
& stfE   & & & & ygfT  & & & &                  \\
& icdC   & & & & uacT  & & & &                  \\
& recN   & & & & cpxR  & & & &                  \\
\hline

\caption{\textbf{Genes subjected to Reg-Seq study.} A total of 104 genes were chosen for this study that can be classified into different groups as seen in the nine categories. The gold standard genes were chosen because they have been well characterized in the past~\cite{belliveau2018systematic, ireland2020deciphering}. A group of genes were chosen because of their role in antibiotic resistance or toxin/anti-toxin systems. Two feed-forward loops were chosen. Two groups were chosen from the dataset by Schmidt et al., 2016 \cite{schmidt2016quantitative} by identifying genes with strongly varying expression across their 22 different growth conditions, indicating transcriptional regulation that senses these conditions. The y-ome genes were selected from the  EcoCyc database~\cite{karp2023ecocyc} by screening for  genes with minimal functional annotation. Finally, two groups of genes (YmfT and YgeV imodulons) were chosen from the PRECISE2.0 dataset \cite{lamoureux2021precise}, as they form iModulons.}
\label{tab:genes_chosen}
\end{longtable}
}
\section{Results}
\label{Section:Results}

Before describing the results of our investigation of the 104 genes described above, we briefly recount some of the key elements of the full experimental and computational approach used to achieve them.  The main elements of our work are schematized in Figure~\ref{fig1}(B).  For those interested in the precise details, the Methods and Supplementary Information attempt to transparently provide enough explanation for others to repeat the work described here for themselves.  There are several main experimental and computational progressions needed to carry out environmentally-dependent regulatory discovery  that are the main substance of this section, and we describe them now in turn.

As noted in the introduction, massively parallel reporter assays are an excellent way to discern regulatory behavior with base-pair specificity, since there is a direct link between measured expression and nucleotide identity in the promoter region.  
We use a modified Reg-Seq~\cite{ireland2020deciphering} protocol with the long-term goal of increasing the scale of the method from hundreds of genes at a time to thousands. Here, we give a summary of these modifications; a detailed breakdown and protocols can be found in the Materials and Methods in Section~\ref{methods}. For each gene included in this study, we found its promoters and computationally generated a collection of mutated variants and ordered the sequences as a synthesized oligonucleotide pool. These oligonucleotides were ligated to sequencing barcodes and cloned into a plasmid vector. For the 104 genes in this study, we found 119 promoters, leading to 178,619 promoter variants that were ordered and more than 95\% (170,167) were recovered during mapping. Across all promoter variants, 5,316,504 unique barcodes were identified, with a median of 28 barcodes per variant. Because we previously observed that expression from plasmids depends significantly on plasmid copy number, possibly leading to non-physiological expression levels~\cite{brewster2014transcription}, we used genome-integrated versions of our libraries.  Following genome integration, 168,952 promoter variants and 2,232,542 barcodes were found, with a median of 13 barcodes per promoter variant. Having multiple unique barcodes per variant is essential for handling possible biases that could be introduced by different sequences during gene expression or library preparation. The entire library was then grown in 39 unique growth conditions, which are described in detail in Materials and Methods section~\ref{sec:media_and_growth}. Once the cultures reached the desired state, they were harvested and prepared for DNA and RNA sequencing to count abundances of barcodes. 

\begin{figure}[h]
    \centering
    \includegraphics[width=6.2truein]{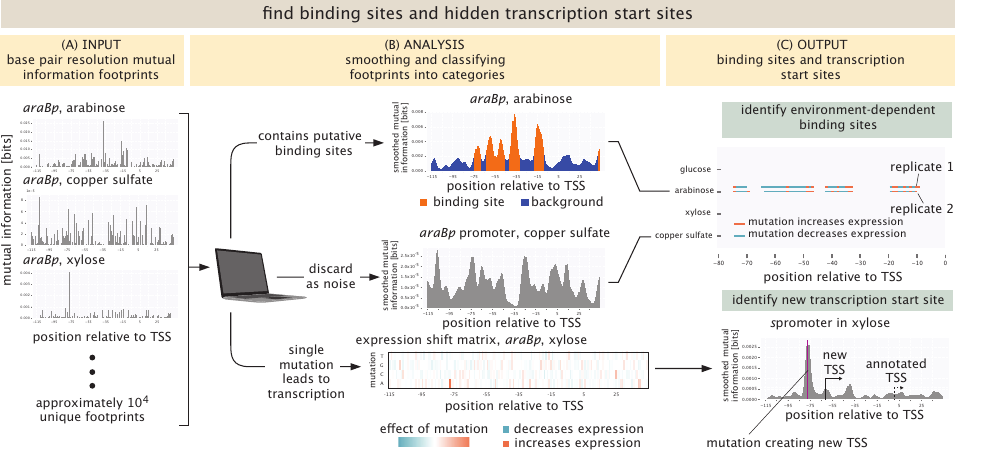}
    \caption{\textbf{Schematic of the steps taken  for analyzing the data needed for regulatory architecture discovery.}  (A) One of the key summary statistics that emerges from our experiments is the information footprint. There is an information footprint for every promoter in every growth condition. Here the top three panels show the information footprint for the {\it araBp} promoter in three growth conditions. (B)  Information footprints and expression shift matrices provide clues about binding sites and transcription start sites. (C) As shown in the top panel, one of the outcomes of this analysis is binding site hypotheses for a given promoter under each set of growth conditions.  As shown in the bottom panel, a second outcome of this analysis  is the discovery of new transcription start sites that are mutationally ``close'' to the original promoter sequence.
    }
    \label{figure_method}
\end{figure}

The outcome of these measurements is vast quantities of data which reports both on the abundance of the different cells harboring the mutant libraries as well as the abundance of RNAs within those cells which reports on the level of gene expression.  For that data to yield promoter discovery, we have to next analyze it in such a way that we have a bridge between sequence identify and gene expression, providing hypotheses for the presence of binding sites and new transcription start sites.  As shown in Figure~\ref{figure_method}, our primary analytic tools are summary statistics that provide base pair by base pair insight into how sequence controls expression.  Using barcode counts, we compute information footprints by calculating the mutual information between the identity of a base (mutated vs. wild-type) and the level of expression measured~\cite{kinney2010using,belliveau2018systematic}. Additionally, expression shifts are calculated to determine if a mutation increases or decreases expression. See Supplementary Information section~\ref{sec:theory} for a more detailed description of this important statistical step. A summary of the entire approach, both measurements and analysis, is shown in brief in Figure~\ref{fig1}(B) and described in detail in Materials and Methods and Supplementary Information.

\subsection{Summary statistics and hypothesis generation for binding sites}

At the most fundamental level, the data from our experiments is sequence data.  DNA sequences reveal the number of cells harboring a given promoter mutant;  mRNA sequences reveal the level of gene expression for that given promoter mutant.  As shown in Figure~\ref{figure_method}, one of the ways in which we visualize that data is through summary statistics that provide a nucleotide-by-nucleotide rendering of the importance of a given base pair for the level of gene expression. In particular, as seen in Figure~\ref{figure_method}(A), and described in detail in Section~\ref{sec:theory}, the mutual information allows us to compute a so-called information footprint that shows how much the expression changes as a result of mutating a given base pair~\cite{kinney2010using,Pan2024}.  A second summary statistic, also described in Section~\ref{sec:theory} and  shown in the bottom panel of Figure~\ref{figure_method}(B), is the expression shift matrix that illustrates whether expression goes up or down if the wild type base is substituted by any of the three others.  As a means of understanding in detail how these summary statistics work in deciphering regulatory response, in an earlier work, we constructed a ``theory of the experiment'' in which we used known energy matrices and statistical mechanical models to generate synthetic Reg-Seq datasets which could then be subjected to all of the same summary statistics as in our actual experimental data~\cite{Pan2024}. The upper panel of  Figure~\ref{figure_method}(C) provides a third way of summarizing our findings for each promoter.   Here, we have a list of the entirety of environmental conditions that the promoter was tested in  and an allied map of where putative binding sites or new transcription start sites were found.   For those interested in seeing these summary statistics in play for all of our promoters in all  environmental conditions, please refer to 
\url{http://rpdata.caltech.edu/data/interactive_footprints.html} and  \url{http://rpdata.caltech.edu/data/all_data.pdf}. 
Note that the Supplemental Information includes an exhaustive  examination of every single gene considered in our experiments.  Figure~\ref{fig:SI_legend} provides a more general key to how we represent the various data and summary statistics in this work.  The idea of the key is to explain the icons we use in all of our schematics as well as the way that we summarize the data.

The simplest way to generate binding site hypotheses from these summary statistics is by eye.  Though we have done this in the past for lower throughput versions of these experiments~\cite{belliveau2018systematic,ireland2020deciphering}, the current scale of the data makes it  clear that automated approaches have become a necessity.  Part of the suite of tools we used for binding site hypothesis formulation was to perform multiple replicates of the same experiment to gain insights into the correlations between experiments and to quantify the noise.
At least two replicates were performed for each growth condition.  In some conditions we decided to perform a third replicate if the initial two showed little correlation. In total, 90 different experiments (39 conditions in duplicate with a few done in triplicate) were performed, resulting in more than 10,000 promoter-growth condition pairs that have to be processed. In general, we find there are three distinct classes of results. First, there are promoter sequences that contain an active transcription start site (TSS) and potentially multiple binding sites for sigma factors and regulators. Then, there are cases where an otherwise inactive promoter sequence gets activated at an alternative transcription start site by a single mutation. Finally, there are promoter sequences that do not contain an active transcription start site for any of the mutations we performed. An overview of these different types of results can be found in Figure~\ref{figure_method}(B). 

To automatically classify the data into these categories, we compute the coefficient of variation of the mutual information footprint across all positions in the promoter. Next, the footprints are smoothed using a Gaussian kernel, and the coefficient of variation is computed again. By evaluating how the coefficient of variation changes upon smoothing, we can distinguish between footprints with single positions of high mutual information and footprints with putative binding sites.  A detailed description of this operation can be found in the Supplementary Information in Section~\ref{sec:ident}.

For promoters containing putative binding sites, the locations of binding sites are identified using a two-state Hidden Markov Model (HMM).   Binding sites are identified as groups of positions with high mutual information and noisy regions are identified as groups of positions with low mutual information. HMMs are commonly used in biological applications~\cite{yoon2009hidden} and are particularly useful here to study the transition between the binding sites and the noisy regions.
For promoters where we find a potential alternative transcription start site resulting from only a single mutation, we use the very helpful model of LaFleur et al.~\cite{lafleur2022automated}, which predicts transcription rates by $\sigma_{70}$  for any query sequence.  We have found this tool to provide a powerful interpretive tool for Reg-Seq data. Both models are discussed in detail in the Supplementary Information Section~\ref{sec:ident}.

\subsection{Identifying Transcription Factors}

Identification of hypothetical binding sites is only the first step in the regulatory architecture discovery process; we also have to identify which proteins bind to those sites. As with other aspects of these questions, much beautiful work has shown how to perform such identification~\cite{ishihama2016transcription,GaoPalsson2021,trouillon2023genomic,Lally2025}.  This is a particularly challenging step  since we have found that even with a number of complementary methods, successfully assigning transcription factor identity to every putative new binding site (or even previously established binding sites) is difficult across all promoters and all conditions.

As seen in Figure~\ref{TF_ident}, we have adopted several key strategies.  First, we have undertaken a broad collection of experiments in which we use the putative binding sites as ``fishing hooks'' for TF pulldown from cell lysate as shown in the lower right panel of Figure~\ref{fig1}(B) in conjunction with mass spectrometry. The promoter for the gene {\it dicC} is active in every condition we tested, as shown in the left panel of Figure~\ref{TF_ident}(A). We found a putative, unknown repressor site, as shown in the middle panel of the figure which is therefore a very interesting candidate for transcription factor identification.  To that end,   the upper right panel of Figure~\ref{TF_ident}(A) shows how a DNA chromatography and mass spectrometry measurement for the {\it dicCp} promoter fishes out an entire collection of peptides, where in this case we find an enrichment of the two transcription factors ArcA and YgbI. For the data shown here,  cells are used that were induced with 2,2-dipyridyl. A complementary computational approach is to compare the putative binding site to previously identified sequences in databases such as RegulonDB and EcoCyc, which is described in detail in the supplementary information~\ref{sec:theory_comp}.  Here too (lower right panel of Figure~\ref{TF_ident}(A)), we find that the previously identified ArcA binding site sequence is very similar to the sequence we identified from the information footprint analysis described in the previous section.   Figure~\ref{TF_ident}(B) repeats the strategy shown in Figure~\ref{TF_ident}(A) but now for the {\it mhpRp2} promoter. This promoter has a binding site for the activator CRP annotated between the positions -38 to -59~\cite{manso2011escherichia}. As shown in the left panel in Figure~\ref{TF_ident}(B), we find this binding site only in stationary phase. As in the case of the {\it dicCp} promoter, mass spectrometry yields an enrichment in ArcA as well as GalS, where lysates from cells grown to stationary phase were used.  The computational approach similarly identifies ArcA to bind around the -10 region of the promoter (here shifted to the -20 position), and identifies CRP to bind to its known activator site.

\begin{figure}
    \centering
    \includegraphics[width=6.8truein]{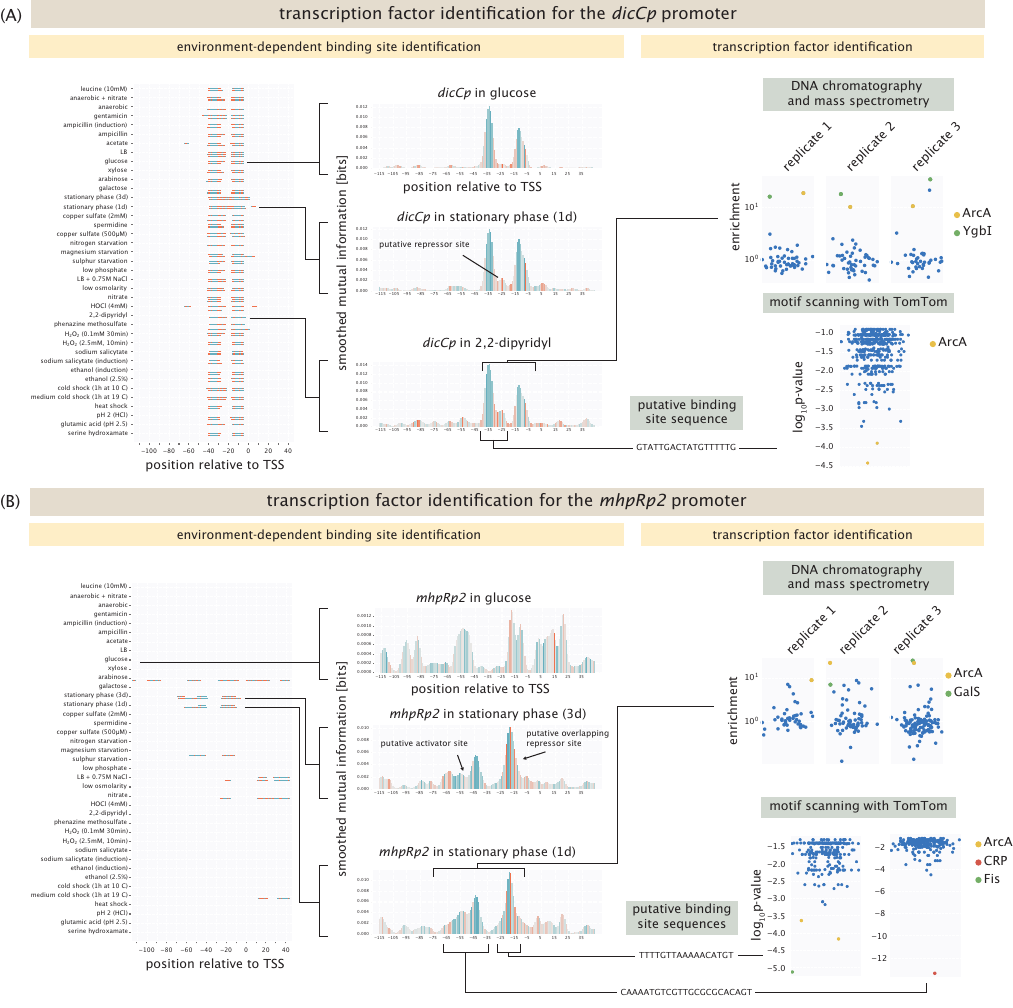}
    \caption{\textbf{Identifying transcription factors for putative binding sites.} (A) Putative binding sites identified in the information footprints for the {\it dicCp} promoter are used in mass spectrometry experiments. ArcA and YgbI are found to be highly enriched for this promoter. ArcA is also identified in a computational search using TomTom.  (B) Putative binding sites identified in the information footprints for the {\it mhpRp2} promoter are used in mass spectrometry experiments and in computational motif scanning which leads to the identification of several candidate transcription factors.
    }
    \label{TF_ident}
\end{figure}

Though we have successfully used the multiple complementary approaches described above and shown in Figure~\ref{TF_ident}, this system does not suffice to identify transcription factors for all promoters under all conditions.   One possible reason for this in the context of the mass spectrometry results is that the act of lysing cells to fish out transcription factors can have the effect of compromising the environmental conditions that led to transcription factor binding in the first place. On the other hand, computational approaches are limited to the scope of the binding sites that are reported in the databases. Unfortunately, in many cases, a potential transcription factor binding partner may not even have any consensus sequence reported in the literature.

\subsection{Dissecting the Regulatory Architecture of Gold Standard Genes}

Our exploration of the environment-dependent regulatory genome is predicated on examining a broad variety of different genes under a wide range  of different environments as summarized in Table~\ref{tab:genes_chosen}.  Our starting point is sixteen distinct ``gold standard'' genes that allow us to examine case studies for which much is already known. For example, we can examine the environmental dependence of regulation explicitly, as shown in Figure~\ref{araB_example}(A) for  well-known arabinose genes. The regulation of the \textit{araC} and \textit{araBAD} operons has been studied extensively~\cite{reeder1993arac,hahn1984upstream,lobell1991arac,lee1992repression}. The promoters for the operons are on opposite strands of the DNA, and mainly regulated by AraC. If there is no arabinose present, AraC binds two distant binding sites, leading to repression of the promoter through DNA looping. But, when arabinose is present, AraC instead binds  in proximity of the transcription start site and activates transcription from the promoter. This interaction is very specific;  hence, we only expect to detect the activating architecture in the arabinose growth condition. This is indeed the case,   as is shown by the peaks in the information footprints  in the middle panel of Figure~\ref{araB_example}(B). Mutations within these peaks lead to a strong decrease in expression as is shown by the color of the peaks. As shown in the right panel of Figure~\ref{araB_example}(B), these binding sites do not show up when cells are grown with other carbon sources. Indeed, as is shown in Figure~\ref{fig:SI_araB_BS}, the binding sites are found only when arabinose is used as a carbon source. We do not recover the architecture associated with repression primarily due to the limited size of the promoter region that we study:  the most important binding site that leads to looping is \textit{araO2}, which is 275~bp upstream of the transcription start site and therefore outside of the region that was part of our mutant library.  Of course, this is a weakness in the method which can be easily remedied with a larger library at a higher financial cost.     Additionally, the promoter is very weak when not activated, even in the absence of loop formation, which could explain why we do not find the transcription start site in any other condition other than growth on arabinose. We do find a new transcription start site when one specific base in the promoter is mutated, which is discussed in Section~\ref{Section:TSSOldNew}. 

\begin{figure}
    \centering
    \includegraphics[width=6.2truein]{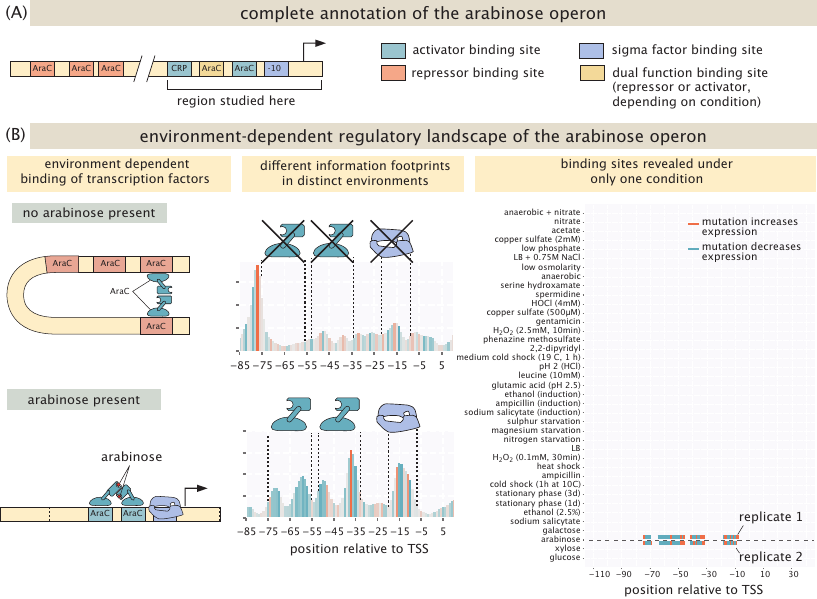}
    \caption{{\bf Environmental dependence of the regulatory architecture in the arabinose operon.} The study of the arabinose operon over the last 60 years has resulted in a complete characterization of its regulatory landscape~\cite{Schleif2000}.  Depending upon conditions, the  binding sites are occupied in very different ways.  As seen in the lower panel,  the Reg-Seq approach adopted here results in markedly different information footprints in different environments. The panel on the right shows that only in the presence of arabinose are binding sites revealed.}
    \label{araB_example}
\end{figure}

Figure~\ref{fig3} highlights several other important examples where changing environmental conditions lead to a wholesale change in
the binding sites revealed in a Reg-Seq measurement.  
For example, one important transcription factor involved in antimicrobial resistance in \textit{E. coli} is MprA~\cite{lomovskaya1995emrr,rodionov2001comparative}. As shown in Figure~\ref{fig3}(A), MprA is a transcriptional repressor that inhibits the expression of its own transcriptional unit, which contains two genes (\textit{emrA} and \textit{emrB}) encoding components of multidrug resistance efflux pumps and are important for \textit{E. coli}'s intrinsic resistance to a variety of compounds. One of the known inducers of MprA is salicylate \cite{lomovskaya1995emrr}, which binds MprA and leads to a conformational change of the repressor and consequent loss of repression. When cells are grown in glucose, we find the known repressor site for MprA as reported in databases, demonstrated by the red peak in \textit{mprA}'s information footprint. When 10 mM of sodium salicylate is added to the media for 1 h, there is a loss of repression as expected, as is shown by \textit{mprA}'s information footprint in Figure~\ref{fig3}(A).  The information footprint thus serves as a direct window onto the changes in regulatory architecture in different environments.

Another example of very specific regulation is the activation of \textit{cusC}, which encodes a component of a copper efflux system. \textit{cusC} is activated by CusR in the presence of copper through phosphorylation of CusR by CusS \cite{outten2001independent}. We find expression at the \textit{cusC} promoter when 2~mM copper sulfate is added, as shown by \textit{cusR}'s information footprint in Figure~\ref{fig3}(A). There is also activation in one of the replicates for both growth at pH 2.5 with 1 mM glutamic acid and growth with 500 \textmu M copper sulfate, as shown in Figure~\ref{fig:SI_cusC}. As shown in Figure~\ref{fig:SI_cusC_mass-spec}, there was no enrichment in mass spectrometry experiments when  the lysate used was from cells grown in copper sulfate, which may indicate that CusR does not stay in its phosphorylated state after cell lysis, and hence does not bind to the binding site \textit{in vitro}. 

The third example shown in Figure~\ref{fig3}(A) is the CpxR-CpxA two-component system consisting of the cytoplasmic transcriptional regulator CpxR and the histidine kinase CpxA, which respond to stress in the inner membrane and regulate the expression of a number of genes~\cite{raivio2014everything,cho2023envelope}. CpxR activates its own expression in the phosphorylated state~\cite{choudhary2020elucidation}. Here, we examine the promoter for \textit{cpxR} which includes the CpxR binding site.  We find the activator site when gentamicin is added to growing cells, as shown by \textit{cpxR}'s information footprint in Figure~\ref{fig3}(A).   This result is in line with the previously identified interaction of ArcA and CpxA upon treatment with gentamicin~\cite{cudic2017role}.  This changes the phosporylation activity of CpxA and hence, the regulatory activity of CpxR. As shown in Figure~\ref{fig:SI_cpxR_mass-spec}, we also found enrichment for CpxR in mass spectrometry experiments where cells are induced with gentamicin before harvest.  Additionally, we find the activator binding site in one of two replicates for both low copper sulfate concentration (500 \textmu M) and acidic shock (pH 2.5, 1 mM glutamic acid). We find a very similar result for a second promoter containing a CpxR activator site in our library, \textit{yqaE}. As is the case for \textit{cpxR}, the activator site is recovered in gentamicin conditions, and the same replicates for low copper sulfate concentration (500 \textmu M) and acidic shock (pH 2.5, 1 mM glutamic acid), see Figure~\ref{fig:SI_yqaE_BS}. Once again, the information footprint provides a direct measure of the environmental dependence of the inferred regulatory architecture.

An even more interesting case study in environmental dependence of regulatory architecture from our list of gold standard genes is provided by LexA.  The transcriptional repressor LexA is involved in the cellular stress response to DNA damage, also known as the SOS response~\cite{d1985sos}. LexA has 42 annotated binding sites in EcoCyc, and it has been observed that LexA and many of its regulatory targets are upregulated when treated with 2.5~mM hydrogen peroxide (H$_2$O$_2$) for 10 minutes \cite{roth2022transcriptomic}. The reason for this upregulation is that in the presence of hydrogen peroxide, the coprotease RecA activates self-cleavage of LexA, causing its dissocation from its binding sites and expression of its target genes~\cite{little1991mechanism,giese2008reca}.  Hence, its information footprint should disappear under this condition.  5 of the promoters studied in our experiments have annotated binding sites for LexA, namely, \textit{ftsK}, \textit{tisB}, \textit{dinQ},  \textit{sulA} and \textit{recN}. In each of these cases, we find the repressor sites, as shown for \textit{sulA}, \textit{recN} and \textit{tisB} in Figure~\ref{fig3}(B) (\textit{ftsK} and \textit{dinQ} are shown in  supplementary figures~\ref{fig:SI_ftsK} and \ref{fig:SI_dinQ}, respectively) when grown in minimal media with glucose. When hydrogen peroxide is added, the repressor binding site completely disappears. Interestingly, when gentamicin is added, binding seems to be reduced but not abolished, as can be seen by the reduced peaks in  information footprints, see Figures~\ref{fig:SI_ftsK}-\ref{fig:SI_recN}. We performed DNA chromatography and mass spectrometry experiments for the LexA binding site in the \textit{tisB} promoter, using both lysates from cells grown in stationary phase, and from cells that were induced with hydrogen peroxide. We see clear enrichment for LexA, showing specific binding to the binding site, in the stationary phase lysate. However, when the lysate from the induced cells is used, we do not find enrichment for LexA.

The results from these gold standard genes, and others like them, show that the strict environmental dependence of the regulatory landscape is revealed by our Reg-Seq experiments and the summary statistics used to envision the data.  Given these results, we now turn to the use
of our approach for discovering previously unknown regulatory architectures and the ways in which regulation influences the physiology and adaptation of {\it E. coli} to different environments.

\begin{figure}
    \centering
    \includegraphics[width=6.2truein]{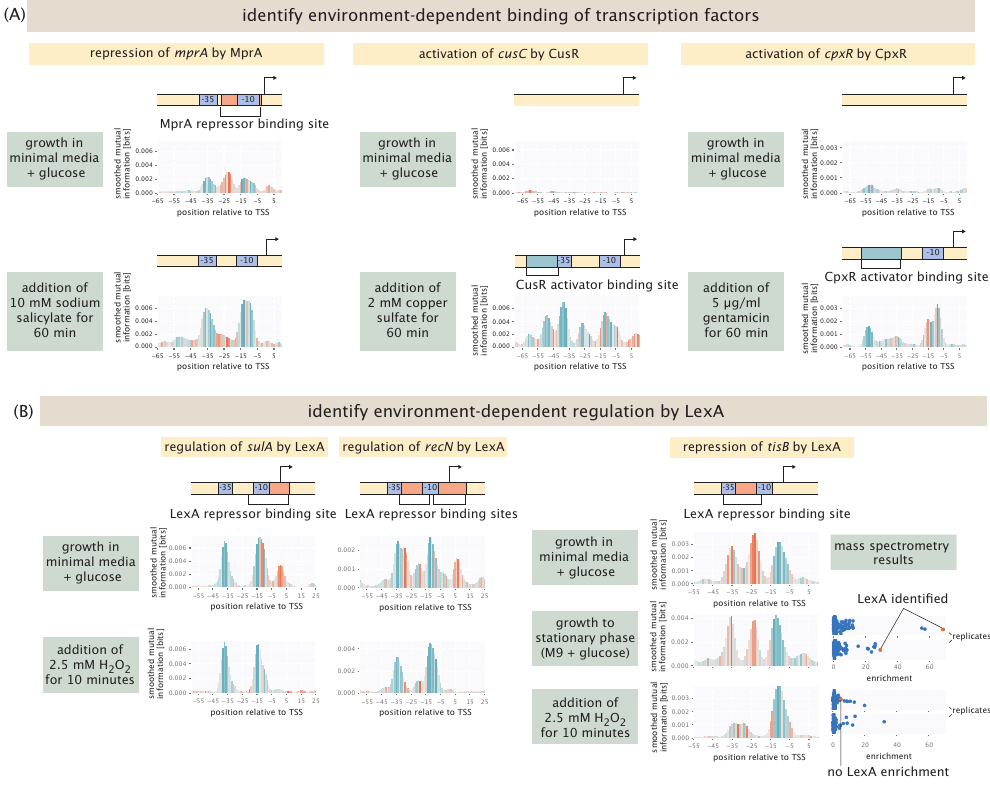}
    \caption{\textbf{Environmental dependence of the regulation of ``gold standard'' genes.} (A) Depending upon the environmental conditions, the constellation of binding sites for a given promoter as revealed by the information footprints will be different.  Three examples of genes whose regulatory landscape changes upon changing the environmental conditions.  (B) Genes regulated by LexA and changes in their regulatory landscape in different environments. Each information footprint reveals a different environment-dependent regulatory landscape.  The two conditions on the far right show how mass spectrometry was used to identify the LexA protein as the binding partner of the sites revealed in the information footprint. }
    \label{fig3}
\end{figure}

\subsection{How does \textbf{\textit{E. coli}} Sense its Environment?}
\begin{figure}
    \centering
    \includegraphics[width=5.7truein]{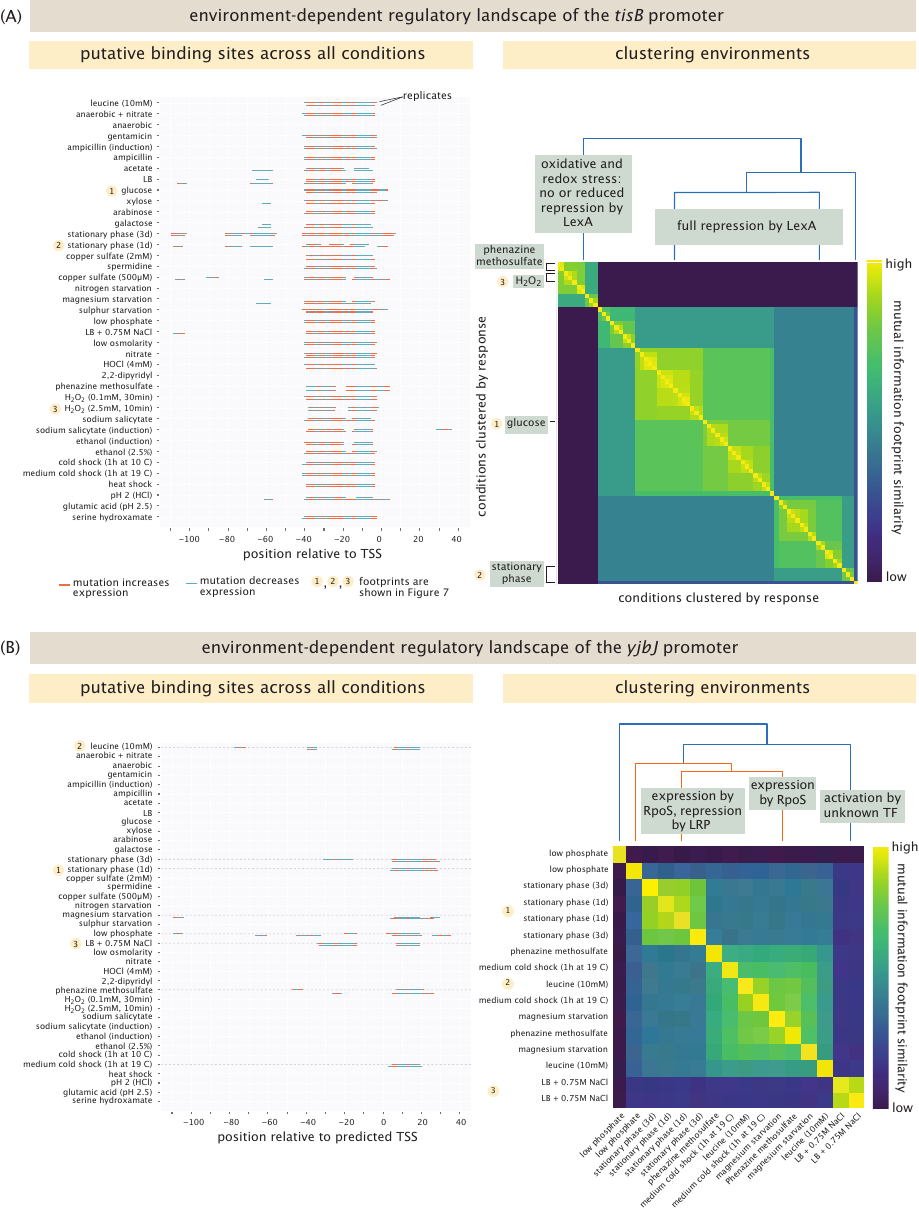}
    \caption{\textbf{Transcriptional regulation of \textbf{\textit{tisB}} and \textbf{\textit{yjbJ}} across all growth conditions.}  (A) Putative binding sites (left) and environments perceived as being similar (right) for the {\it tisB} promoter.  Note that in nearly all conditions, there are putative binding sites identified.  (B) Putative binding sites (left) and environments perceived as being similar (right) for the {\it yjbJ} promoter. Both examples attempt to capture how many ``senses''  a given promoter has. }
    \label{condition_clustering_examples1}
\end{figure}

By virtue of exploring the regulatory landscape under a diverse set of environmental conditions, we can ask whether seemingly distinct environments are ``perceived'' by a particular gene in the same way.  For example, as noted above, regulation of \textit{tisB} by LexA is affected by hydrogen peroxide. But repression is also relieved by phenazine methosulfate, a response that was previously unknown to the best of our knowledge. The cellular response to phenazines is mainly regulated by the SoxS-SoxR system \cite{amabile1991molecular}, which we can see by the activation of \textit{arcZ} by SoxS in the presence of phenazine methosulfate, as shown in Figure~\ref{fig:SI_acrZ}. The fact that LexA responds not just to hydrogen peroxide, but also to phenazine methosulfate and gentamicin indicates a similarity of response under distinct environments.  We were curious whether it might be possible to classify environments for which a given gene engender nearly the same response. Intuitively, we expect that environmental conditions that yield the same putative binding sites will have similar information footprints.  As we already saw for \textit{araB} in Figure~\ref{araB_example}, it is possible for one set of binding sites to appear only under one unique condition. 
Figure~\ref{condition_clustering_examples1} examines two 
examples in which we compare responses under different growth conditions.   We determined which conditions lead to similar responses by performing hierarchical clustering for each promoter across all the conditions in which we found binding sites.  The results show clusters with similar information footprints and hence, similar regulatory patterns. 

For example, the left panel of Figure~\ref{condition_clustering_examples1}(A) shows the putative binding sites for the {\it tisB} promoter in each  condition, while  the right panel of Figure~\ref{condition_clustering_examples1}(A) shows the
result of the clustering analysis.  The left panel is a way of summarizing all the putative binding site regions in all the different conditions at the same time.  As shown in Figure~\ref{araB_example}, we note that unlike in the arabinose operon, here we find that for positions between 0 and -40, for nearly all conditions 
there is a strong signature in the information footprints indicating that those bases are important under all those conditions.  However, under stationary phase, for example, only bases between $\approx$ -60 and -80 are relevant.  The clustering analysis shown in the right panel summarizes these distinctions.  Note that in the analysis, we have removed the itemized labels of the different conditions, but an enlarged version of the figure is shown in Figure~\ref{SI-fig:full_tsB_cluster} of the SI for those wishing to see which environments cluster together.

Figure~\ref{condition_clustering_examples1}(B) shows a y-ome gene for which the clustering analysis provides an interesting result.   The example shown here is \textit{yjbJ}, which belongs to the group of genes we  identified from the Schmidt et al. mass spectrometry dataset~\cite{schmidt2016quantitative}. As seen in the left panel, \textit{yjbJ} shows binding sites in only 8 out of the 39 conditions tested. Interestingly, of these 8 conditions, it appears there are three different stereotyped regulatory responses, as is displayed in the identified clusters.

\begin{figure}
    \centering
    \includegraphics[width=6.2truein]{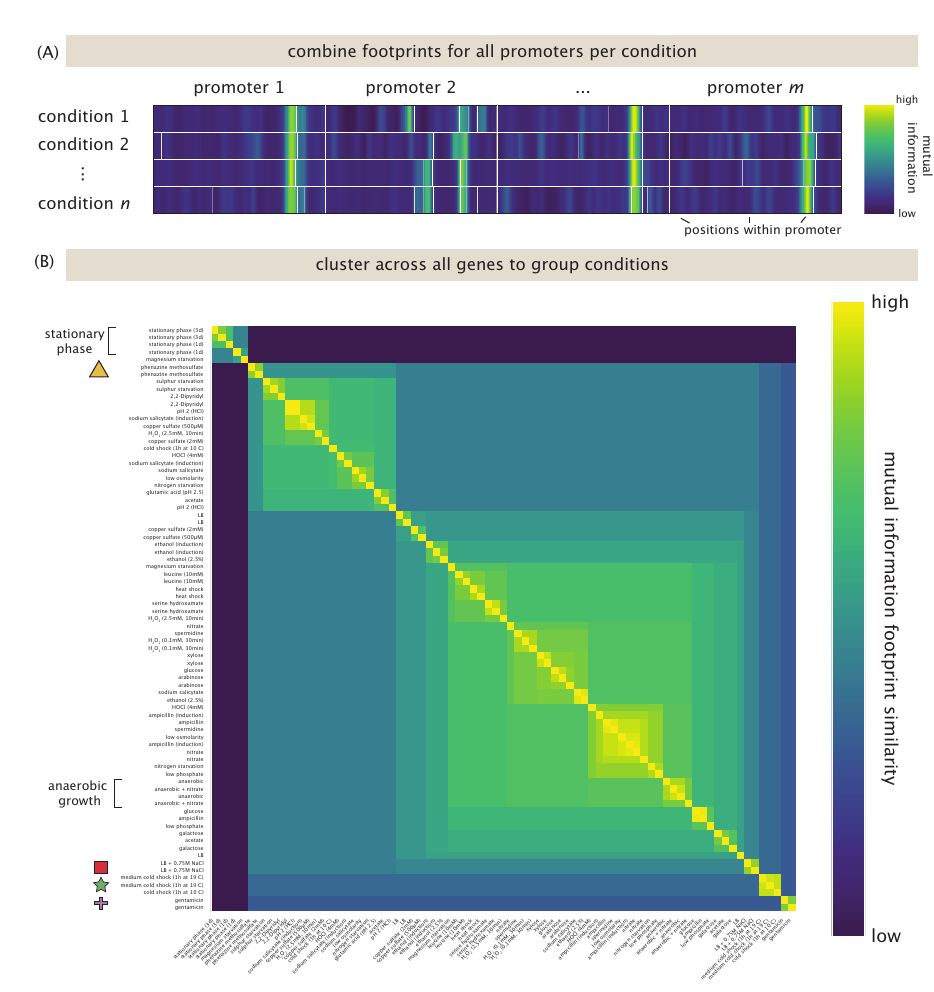}
    \caption{\textbf{Clustering of all growth conditions.} (A) Information footprints are grouped into a matrix with each row corresponding to all the information footprints for a given condition and each column a specific base within a specific promoter. Colors identify regions of high mutual information. (B) Clustering shows the similarity of footprints across all conditions. This figure reveals that the largest distinction is between stationary phase and not, with this shift implying a wholesale change in regulation across all the genes we examined.  Shock in LB with high salt concentration (red square), induction with phenazine methosulfate (yellow triangle) and induction with gentamicin (purple cross) also lead to drastic perturbations to the regulatory state of all the genes we considered. Highlighted are also growth in LB for two of the three replicates (blue circle) and both cold shock conditions (green star).}
    \label{condition_clustering_examples}
\end{figure}

Another way of  determining how gene regulation reflects environmental conditions based on Reg-Seq data is to examine all genes in all conditions, as seen in Figure~\ref{condition_clustering_examples}.
Figure~\ref{condition_clustering_examples}(A) shows
the concept of the analysis, in which all information footprints are combined into a matrix, where each row contains all footprints for one condition, and each column contains the mutual information at a certain base in a promoter. We then ask whether there is a clustering of the data such that all conditions with consistently similar regulatory profiles are grouped together. Figure~\ref{condition_clustering_examples}(B) shows the result of that  analysis. The biggest separation is between growth in stationary phase, and all other conditions. Note that one replicate for magnesium starvation groups with the stationary phase conditions, indicating that in this specific experiment, the cells entered stationary phase. On the other end of the clustering, another strong perturbation is shock in LB with 750 mM of NaCl. In this condition, we have identified multiple interesting new binding sites, as discussed below. In general, replicates for the same conditions cluster well together, indicating a general reproducibilty of the experiments. For the genes studied here, induction with phenazine methosulfate or gentamicin also lead to a significant change in regulatory patterns.

\subsection{Dissecting the Regulatory Architecture of the y-ome}

Completely understanding the physiology and evolution of an organism will require figuring out the function and regulation of all of its genes. Even in the ostensibly ``well understood'' model organism {\it E. coli}, the presence of the y-ome and the allosterome make it clear that there is  much left to discover.  We were excited to explore genes from
the large list of y-ome genes.  In Figure~\ref{yome} we show results for  \textit{yjbJ}, \textit{ygiW}, and \textit{ybaY}, which were identified from the Schmidt et al. dataset~\cite{schmidt2016quantitative}; \textit{yadE} and  \textit{yadI}, which we chose from Ecocyc; and  \textit{ybiYW} from the YgeV-iModulon and \textit{intE} and \textit{icdC} from the YmfT iModulon identified by Lamoureux et al.~\cite{lamoureux2021precise}. None of these genes had an annotated transcription start site, so we predicted transcription start sites using the model from LaFleur et al.~\cite{lafleur2022automated}. In each of these cases, the predicted transcription start site was active in at least one growth condition. For seven of these genes, we identified putative binding sites for transcription factors in at least one condition, with four genes having more than one site. For three binding sites we were able to identify the transcription factor by mass spectrometry, and for one binding site we were able to identify CRP as the transcription factor by sequence comparison to the CRP binding motif.

\begin{figure}
    \centering
    \includegraphics[width=5.4truein]{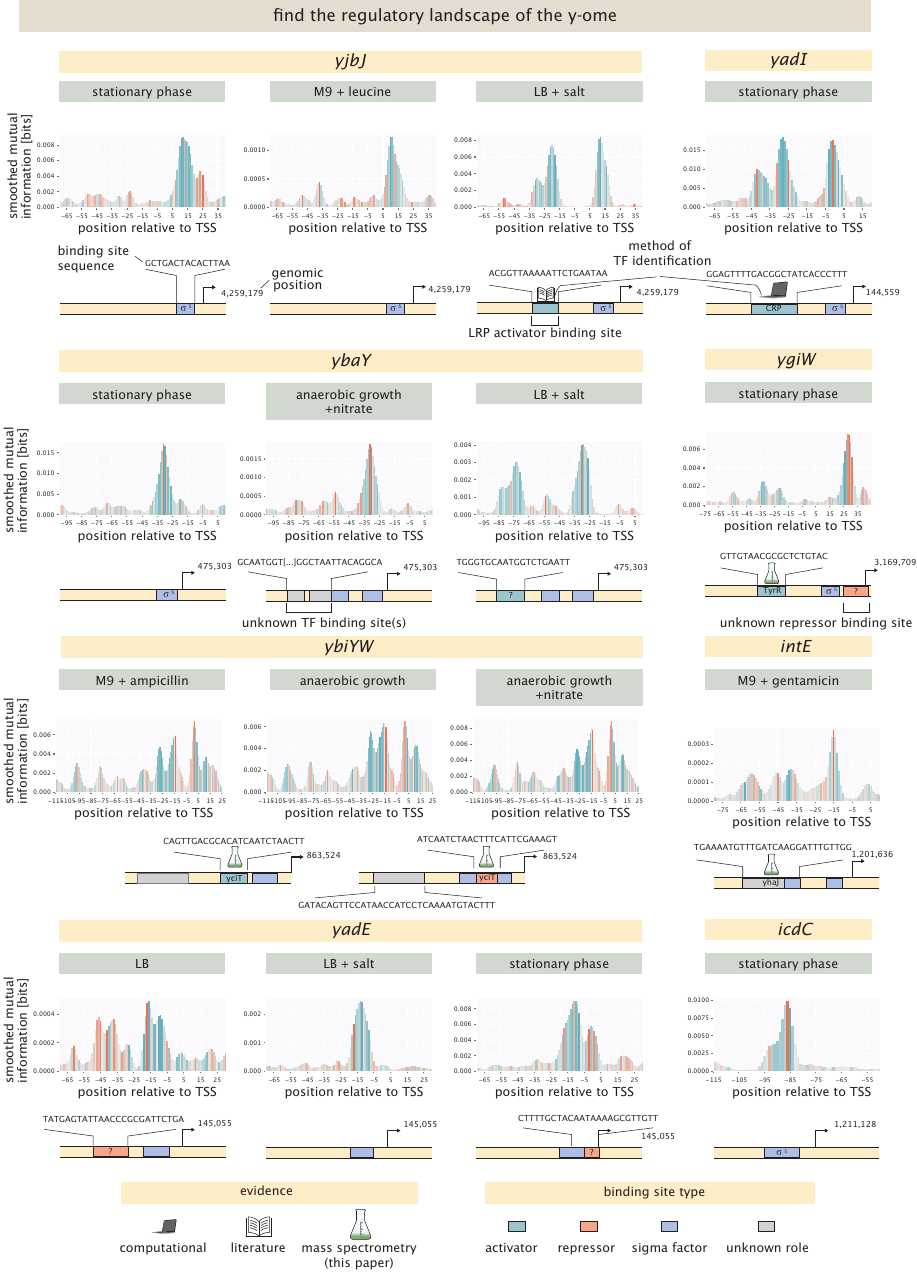}
    \caption{\textbf{Discovering the regulatory architecture of y-ome genes.} Each panel shows a different y-ome gene, the conditions under which it was queried, and the resulting binding site positions and sequences. When a transcription factor was identified, the type of evidence resulting in that identification is also shown. }
    \label{yome}
\end{figure}


The putative stress response gene \textit{yjbJ} is thought to be part of the RpoS regulon and expressed in stationary phase \cite{schmidt2016quantitative,link1997comparing}, and indeed we find an active transcription start site in stationary phase. It has also been observed to be  upregulated when cells are exposed to high salt concentrations~\cite{weber2006time}. Here, we find an activator site when cells are induced in LB media with 750 mM of NaCl, but not when cells are grown without additional salts (for more details, see Figure~\ref{fig:SI_yjbJ_BS}). This putative binding site overlaps with a reported binding site for LRP from ChIP-chip experiments \cite{cho2008genome}, so we propose that LRP activates expression of YjbJ. However, we did not detect enrichment for LRP (or any other transcription factor) in DNA chromatography and mass spectrometry experiments. (These experiments, which were performed in high-salt conditions generally yielded no enrichment.)   We do not find expression from the \textit{yjbJ} promoter when cells are grown in minimal media with glucose, unless cells are induced with L-leucine in exponential phase growth. This also indicates regulation by LRP.

The gene \textit{yadI} was identified as a y-gene in the first complete genome assembly of \textit{E. coli}~\cite{blattner1997complete}, but its function remains elusive.  It is a predicted PTS enzyme IIA \cite{riley2006escherichia}. We find an active transcription start site, and a putative activator binding site in a few conditions. We find expression in LB, as well as minimal media with acetate and galactose as shown in  Figure~\ref{fig:SI_yadI_BS}. Expression seems to be especially strong in stationary phase, as shown in Figure~\ref{yome}. Computational motif scanning predicts this site to be a binding site for CRP, as shown in Figure~\ref{yadI_tomtom}, which has not been identified before. Our mass spectrometry results for \textit{yadI}, however, did not show enrichment of CRP, despite being able to detect CRP at levels well above the median of all protein abundances. This is consistent with the general observation that we do not find significant CRP enrichment even for genes where CRP binding is expected (e.g. \textit{mglB} and \textit{araB}). 

\textit{ybaY} undergoes supercoiling-dependent transcription, which is associated with the osmotic stress response and acts through RpoS, which has been found in media with osmolarity of 0.8~Osm \cite{cheung2003microarray}. Indeed, we find an active transcription start site for \textit{ybaY} when grown in LB supplemented with 0.75M NaCl. We also find an activator-like site, which is similar to the activator site for \textit{yjbJ}. The sequences for these binding sites are very similar, with 12 of 19 bases shared and a conserved segment of TCTGAAT, suggesting that the same transcription factor is regulating these two genes.  However, the identity of this TF remains unknown. The identified transcription start site is also active in stationary phase -- indicating binding of RpoS -- as well as in many other conditions, as shown in Figure~\ref{fig:SI_ybaY_BS}.

\textit{ygiW} is thought to be upregulated in stationary phase~\cite{schmidt2016quantitative,tani2002adaptation}, and we found an active transcription start site, which is slightly shifted compared to the predicted transcription start site at $\approx+40$ and only active in stationary phase. In DNA chromatography and  mass spectrometry experiments, TyrR was enriched at a putative activator binding site around the -30 region. A potential repressor binding site can be seen at +30, but no candidate transcription factor was found by mass spectrometry or computational motif scanning. This region could alternatively be an imperfect sigma factor site, where the repressor-like mutations are actually increasing the binding affinity of the sigma factor.

Another y-ome gene that we considered is  the \textit{ybiYW} operon, which until now has had no functional annotation. Using our broad suite of environmental conditions, we found an active transcription start site when cells are grown anaerobically both with and without supplemented nitrate. We also found expression when minimal media is supplemented with glucose and sub-inhibitory concentrations of ampicillin, but only in one replicate, which could indicate that in this specific experiment cells entered anaerobic growth conditions. As shown in Figure~\ref{fig:SI_ybiW_mass-spec}, DNA chromatography and mass spectrometry with lysate from cells grown in minimal media with glucose shows high enrichment for the transcription factor YciT, suggesting YciT binds in the vicinity of the -5 site relative to the predicted transcription start site. This finding suggests that YciT and YbiY-YbiW are involved in the cell's response to anaerobic conditions, which has not been reported before. The information footprint suggests binding of an activator around the -20 region, but we found no candidate transcription factors through the mass spectrometry experiments or the computational motif scanning. It should be noted that cells grown aerobically were used for DNA chromatography, as producing the needed amount of cell lysate anaerobically was technically out of scope.

The \textit{intE-xisE-ymfH} operon is part of the e14 prophage and has no annotated function. We find an active transcription start site only in the presence of gentamicin, shown in Figure~\ref{yome}, and only in one of the two replicates tested for this condition. However, we do find a binding site for the transcription factor YhaJ in the vicinity of the -55 position relative to the predicted transcription start site, as shown by mass spectrometry results shown in Figure~\ref{fig:SI_intE_mass-spec}.

\textit{yadE} has been identified as a possible envelope assembly factor \cite{paradis2014genome}. We find an active transcription start site across many conditions, with most conditions that reach exponential phase showing a repressor binding site around the -40 region. However, when either salt or phenazine methosulfate is added to the media, the site disappears, suggesting that the repressor is deactivated. Additionally, there is expression from this promoter by RpoS under stationary phase, with an apparent repressor binding site downstream. No enrichment for any transcription factors was found using mass spectrometry.
The pseudogene \textit{icdC} was identified as part of the YmfT-iModulon by Lamoureux et al. \cite{lamoureux2021precise}. We found an active transcription start site for this gene in stationary phase, indicating that this gene is part of the RpoS regulon.

\subsection{Transcription Start Sites: Old and New}
\label{Section:TSSOldNew}

One of the surprising outcomes of our experiments was regulatory discoveries other than  novel transcription factor binding sites. As seen in Figure~\ref{emerging}, in some cases a single mutation sufficed to produce entirely new transcription start sites. While this phenomenon has been reported before~\cite{Gore2018},  we were intrigued by how often it occurred in our experiments. Figure~\ref{emerging}(A) shows the situation schematically  in which a promoter of interest has some known transcription start site.  However, as a result of a single mutation (see also the example of \textit{araBp}-xylose in Figure~\ref{figure_method}) an entirely new transcription start site emerges. The most important sequence in determining if $\sigma^{70}$, the most abundant $\sigma$-factor, binds to DNA and initiates transcription is the -10 element, which has the consensus sequence TATAAT. As is shown in the top right panel of Figure~\ref{emerging}, the first two bases and the last base are the most important, meaning, mutations in the other three bases affect binding much less. We find that in most cases, a new transcription start site is created when all three of the important positions in the -10 element match the consensus sequence after a mutation occurs.

\begin{figure}
\centering
\includegraphics[width=5.6truein]{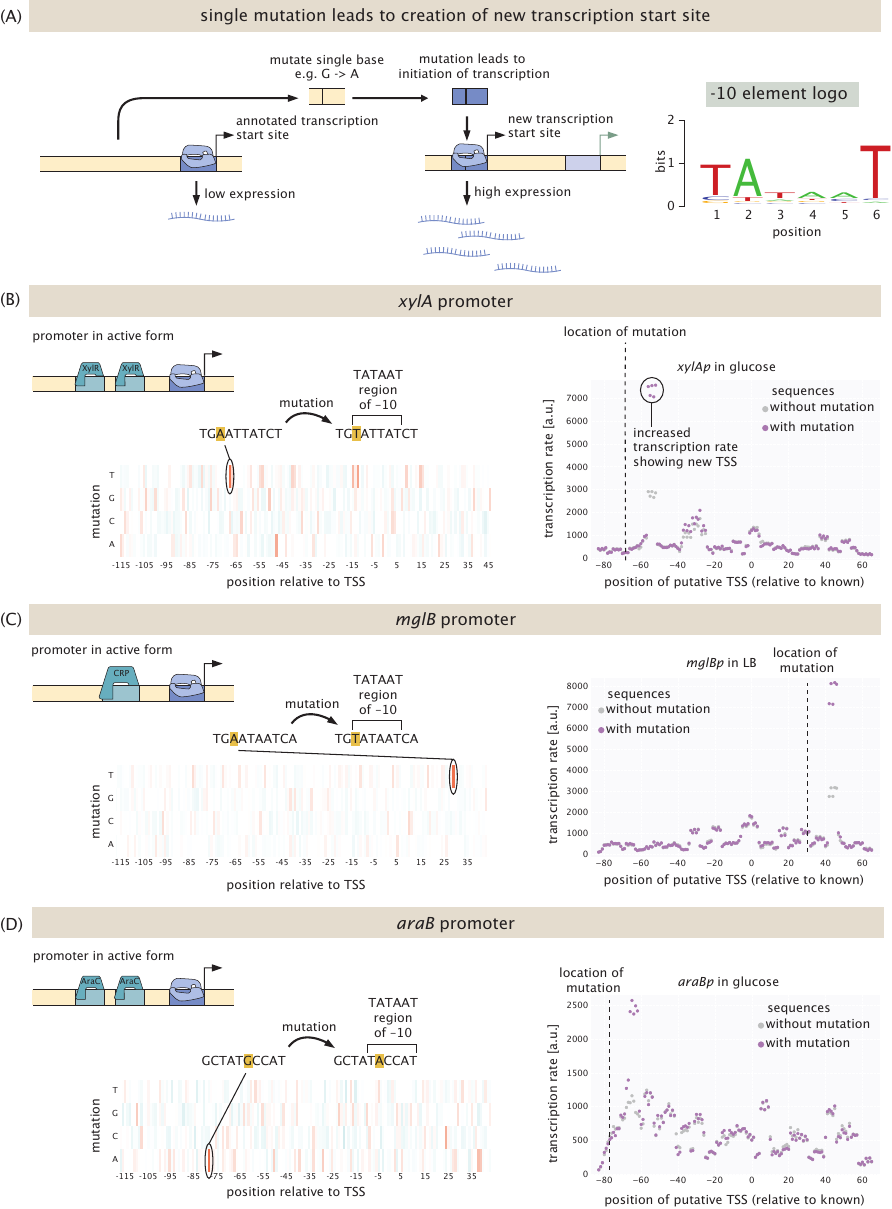}
\caption{\textbf{De novo emergence of new transcription start sites.} (A) A single mutation can lead to creation of a new transcription start site.  (B) A single mutation in the {\it xylA} promoter leads to a new transcription start site.  (C) A single mutation in the {\it mglB} promoter leads to a new transcription start site. (D) A single mutation in the {\it araB} promoter leads to a new transcription start site. In panels (B)-(D), the left side shows the expression shift matrix which highlights the mutation leading to the emergence of a transcription start site. }
\label{emerging}
\end{figure}

Figures~\ref{emerging}(B-D) give three specific examples of this result.  In the case of the {\it xylA} promoter, which is activated by XylR in the presence of xylose, we see that in the absence of xylose, changing an A to a T at position -68 in the promoter creates the
TATAAT region of a new core promoter, giving rise to a large increase in transcription rate, as shown in the right panel of Figure~\ref{emerging}(B).  A similar story plays out in the case of the {\it mglB} promoter, which is activated by CRP. Here, a mutation at the +30 position leads to a consensus -10 sequence, leading to a new transcription start site.  In the {\it araB} promoter, a G becomes an A at the -77 position, but with the same consequence which is the production of a new -10 region. In all three of these examples, a single mutation creates a new transcription start site under conditions other than those in which it is normally activated. 
These observations inspire the question of whether they are a matter of chance mutations on random sequences~\cite{Gore2018}, or instead, that it is a beneficial feature to have inoperative promoters poised a single mutation away from being functional.


Another set of intriguing new results involved several cases where earlier work had hypothesized the existence of multiple distinct transcription start sites.  For example, for the {\it ompR} promoter, as shown in Figure~\ref{ompR_promoters}(A), inspecting the 5' ends of mRNAs using primer extension assays~\cite{huang1992positive} revealed multiple distinct mRNA species, suggesting the existence of at least four transcription start sites with a window of 116 bases.  As shown in the figure, we find that only one transcription start site - the one associated with \textit{ompRp1} - is active, as the information footprints exhibit binding peaks only in the case of this transcription start site. There is no transcription initiated from any of the other start sites, as shown in Figure~\ref{fig:ompR}.

Figure~\ref{ompR_promoters}(B) shows a second intriguing example of the {\it tolC} promoter for which we identify two transcription factor start sites. Like {\it ompR},  earlier work on the 5' ends of mRNAs found multiple species, where initially two start sites were identified~\cite{eguchi2003transcriptional}, {\it tolCp1} and {\it tolCp2} and later, two additional sites were found~\cite{zhang2008transcriptional}, {\it tolCp3} and {\it tolCp4}.
One {\it tolC} site ({\it tolCp1} and {\it tolCp2}) is active in most conditions and is known to be activated by PhoP in magnesium limiting conditions. Indeed, in one of our replicates for magnesium starvation, we find an activator binding site at the annotated position, as shown in Figure~\ref{fig:SI_tolC}. However, we do not observe two different promoters here, indicating that transcription is initiated from one site only. The other two promoters, {\it tolCp3} and {\it tolCp4}, are 50 bases downstream, and have been found to be activated by MarA, SoxS and Rob through a mar-box~\cite{aono1998involvement,zhang2008transcriptional}. Here we only find activation by Rob under its activating conditions, which is induction with 2,2-dipyridyl. The other annotated activators, i.e. MarA and SoxS, are not found to activate {\it tolC} from this site even under inducing conditions.  In previous work, activation of these two transcription factors relied on overexpressing the proteins~\cite{aono1998involvement, zhang2008transcriptional}, which may have led to supra-physiological concentrations. As shown in Figure~\ref{fig:SI_acrZ}, we do find activation by these transcription factors in said conditions on other promoters, such as the promoter for {\it acrZ}. To summarize, for {\it tolC} we find two distinct transcription start sites, instead of the four annotated sites.
We find more examples for promoters with multiple annotated transcription start sites in our dataset, where only a subset of the annotated start sites are active, such as the promoter for CRP, see Figure~\ref{fig:crp1}, and the promoter or {\it galE}, see Figure~\ref{fig:galE1}.

As the examples shown in Figures~\ref{emerging} and \ref{ompR_promoters} illustrate, the question of transcription start sites demonstrates unequivocally that physiological and experimental context both matter.  In some conditions, a single mutation suffices to yield an entirely new transcription start site.  In other cases, we hypothesize that the existence of different species of mRNAs as revealed by their 5' ends may not be a valid signature of alternative start sites.  What is certain is that sorting out the complexity and nuance of how genes are regulated, even in this ``simplest'' of model organisms, is very challenging.

\begin{figure}
    \centering
    \includegraphics[width=5.7truein]{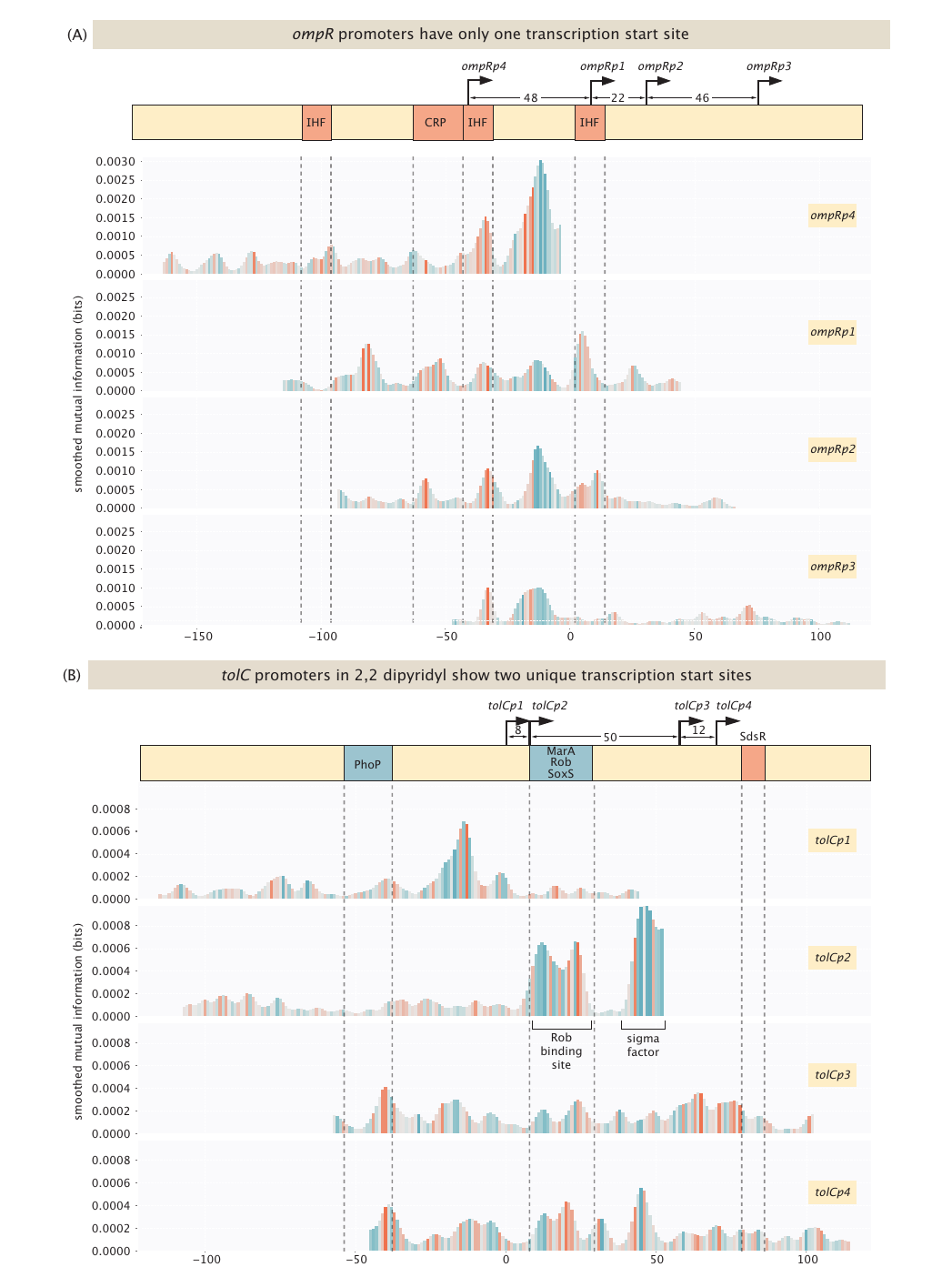}
    \caption{{\bf Transcription start sites for the
    \textbf{\textit{ompR}} and \textbf{\textit{tolC}}} promoters. (A) Footprints for the four reported {\it ompR} promoters are shown for medium cold shock (19C for 1h), where the genomic locations of the footprints are aligned. Annotated binding sites are shown for IHF and CRP. (B) Footprints for the four reported {\it tolC} promoters are shown for induction with 2,2-dipyridyl, where the genomic locations of the footprints are aligned. Annotated binding sites are shown for PhoP, MarA, Rob, SoxS and the small RNA SdsR.}
    \label{ompR_promoters}
\end{figure}

\section{Discussion}

The study of gene regulation is one of the centerpieces of modern biology.
Indeed, one of its great success stories  is our increasing mastery over the reading, writing and control of the genome.  That said, despite an impressive status quo, there still remain gaping holes in our understanding.  We are surprised that with more than $10^{17}$ nucleotides of DNA sequences on the Sequence Read Archive -- orders of magnitude more ``letters'' than are present in all the English language books of the Library of Congress or Wikipedia -- our knowledge of the regulatory part of genomes remains elusive.  Even in our best understood model organisms,  we don't know how the majority of genes are controlled in response to changes in environment.

One of the central motivations of the present work is the hope that we can make more systematic and quantitative inroads into discovering the regulatory architectures that govern the physiological and evolutionary responses of different classes of organisms.  
Genes are regulated in response to both internal and external signals.  As a result, part of the quest to understand the regulatory genome demands including this environmental dependence, which is often mediated by allosteric transcription factors.   The molecular aspect of this problem has been christened the ``allosterome,''~\cite{lindsley2006whence} referring to the fact that many of the transcription factors that control gene expression change their activity as a result of binding by effector molecules.   Greater mastery over regulatory architectures and their environmental dependence  will make a actionable contribution to efforts to understanding physiology and evolution, as well as forming the basis for more systematic approaches to synthetic biology.

In this paper, we have tackled the parallel and entangled challenges of discovering regulatory architectures and the environmental conditions that affect them.  By using the power of massively parallel reporter assays in conjunction with the tools of information theory, we are able to peer into the regulatory ``dark matter'' of the {\it E. coli} genome. Further, by carrying out our experiments over a broad array of different growth conditions,  we can get a glimpse of the rich context dependence of those regulatory architectures.  Said differently, in principle, there is a different regulatory architecture for each and every environment.
To reveal these architectures, we chose roughly 100 biologically interesting genes with which to carry out this environment-dependent regulatory dissection. 
By studying ``gold standard'' genes for which much about their regulatory response was already known, we were able to confirm and clarify previous data, and to classify environments by the commonality of their regulatory response.   Already, in the case of these gold standards we uncovered new insights, including the discovery of new binding sites, clarifications about the nature of transcription start sites, and the ability to characterize a given promoter by the categories of distinct environments that it recognizes.  Beyond the gold standard genes, we undertook a systematic analysis of a broad variety of other genes, including a representative sampling of genes from the y-ome, the set of genes within the {\it E. coli} genome which remain of unknown function.  The results of those experiments  led to the discovery of a collection of new environment-dependent regulatory architectures.

How well does Reg-Seq actually work?  To begin, we consider the 16 gold standard genes.  Current databases have 32 annotated binding sites for these gold standard genes.  We successfully re-identified 18. However, upon further review, we only expect to find 22 of these binding sites.  In particular, some of the annotated binding sites will escape detection using our Reg-Seq approach because either the interactions between the transcription factor and the DNA are very weak; they require long range interactions implying binding sites outside the region of our mutation library;  or because the binding site annotation resulted from {\it in vitro} experiments with no corresponding {\it in vivo} evidence.   To be more precise, many of the binding sites we ``missed'' were found {\it in vitro} or under non-physiological conditions, such as overexpression of the transcription factors. We interpret these findings not as a failure of previous efforts, but instead as showcasing the difficulty of harmonizing {\it in vitro} and {\it in vivo} conditions and methods. Additionally, if we did not identify a previously found binding site, it could indicate that the transcription factor does not interact directly with the polymerase or sigma factor to regulate transcription, or more generally that its function might not relate directly to its binding affinity. 

More importantly, our approach led to a number of new insights into the environment-dependent regulatory genome of {\it E. coli}.  First, our work led to  intriguing hypotheses for previously unknown transcription start sites whose emergence was condition dependent.  These emergent transcription start sites lead us to wonder and speculate about a possible evolutionary role for such mutationally ``close'' promoters as well as changes to binding site strength through small mutational changes.
Writ large, changes in
transcriptional regulation are known to be one of the main ingredients of evolutionary change.  Evolutionary adaptation itself is influenced to a large extent by fluctuations or wholesale changes in environments. A story that is far from complete are the ways in which evolutionary patterns and dynamics are altered by the regulatory genome.  We are hopeful that wide-ranging measurements such as the Reg-Seq approach described here  will provide a powerful substrate for dissecting regulatory evolution.  There have already been a variety of quantitative models that link basepair resolution binding strengths of transcription factors to evolutionary fitness~\cite{Gerland2002,Berg2004,Lynch2015} and
the basepair-by-basepair dissection of gene expression changes and corresponding energy matrices provide the data to sharpen the predictions from such models.


Although the work here reflected an ambitious attempt to systematically and quantitatively dissect promoter regulatory architectures (i.e. binding sites and their associated transcription factors), there is still much work to be done to realize the aim of completing the regulatory genome of a single organism.  One immediate step that lies within reach is to increase the size of the mutated regions in our libraries 
which is mostly limited by the availability and affordability of synthesized oligo pools.  Second, we believe that an even broader range of environmental conditions are needed to conclude what conditions a given gene ``cares'' about.  Further, the transcription factor identification part of the problem requires complementary experimental and bioinformatic approaches, and even then, is unable to find all transcription factors in all environments.  We hope that new approaches will be found for identifying transcription factors and their allied effectors.

Given this overview of our results, it raises a number of questions about the broader significance of our approach and findings.  As we expand the diversity of microbial organisms investigated in environments ranging from the human gut to the ocean floor, we ask how can we understand the physiological and evolutionary response of those organisms?  One approach that is becoming nearly routine is to   sequence the genome of some organism of interest and then use bioinformatic and AI tools to annotate genes, infer biochemical pathways and place the organism within the current known microbial phylogeny.
However, what the current approach does not give us
 is systematic insight into the regulatory landscape, a minimal version of which requires knowing which external signals induce and activate the many genes of the organism, as well as something about the apparatus that converts the environmental input into a gene expression output. Despite a number of outstanding
challenges, the work described in this paper provides a viable route to achieving a whole genome reconstruction of its regulatory landscape in a single experimental approach and corresponding analytical pipeline.  

Microbes perform a huge variety of different functions in conjunction with signals from their environments. In that sense, microbes are environmental transducers, they sense their environments and then they act upon what they sense by altering their gene expression profiles. 
The regulatory genome is both the sensory and response apparatus. It is that part of the genome that indirectly senses the concentration of effector molecules and their allied transcription factors, and together, collectively modulates the degree of gene activity. 
Carrying this analogy further, there are two questions one might ask. How many different ``senses'' does a cell have? And how many ``input-output systems'' does it have? 
How can we begin to ask such questions for organisms that lie beyond the library of model organisms?  We suspect
answering these questions is not  merely a matter of scaling up existing methods. Rather, this involves  systematic experiments, measurements, and analyses that give concrete quantitative answers to these questions. The diversity of genes and environmental conditions that the Reg-Seq approach allows us to investigate  permits  a form of statistical analysis that address both  the cellular senses and the cellular input-output repertoire by analyzing correlations across rows vs columns of one of the essential summary statistics of our study, the mutual information footprint per base pair in each condition.

There is much that has been written and said about the increasing and powerful role of AI in biology.  As demonstrated by some of the massive recent successes of AI approaches such as AlphaFold~\cite{jumper2021highly,mirdita2022colabfold} and Virtual Cell projects~\cite{Bunne2024}, we seem to be at the precipice of a time when modern computation combined with high-throughput biology data will give us answers to questions about the physiology and evolution of microbial organisms that we have been struggling to achieve for decades.   For example, synthetic biologists might 
want to design a promoter that has a bespoke gene expression profile across a cohort of environments of their own choosing. Or,  we  might want to design minimal genomes whose large scale patterns of expression are controllable by the induction of a small number of genomic designed elements~\cite{posfai2006emergent,hutchison2016design}. Or, we might wish to identify a condition or molecule that might repress the activity of a microbial pathogen. Ultimately, these are all questions about gene regulation.
However, projects such as AlphaFold are predicated upon the existence of huge quantities of high-quality data with great similarities from one protein structure to the next. This high quality training data is one of the prerequisites for these successes.  If one of the next frontiers is the data-driven modeling of the regulatory genome to drive engineering and synthetic biology goals then what is needed is systematic, quantitative examples of regulatory landscapes.   In the work presented here, we attempted to measure and systematize the kind of regulatory data that might serve as a  basis of data-driven generative models. 
The approach outlined in our paper, when taken to scale, across a diversity of organisms, promises to enhance such efforts. Once done for a sufficiently large and diverse ensemble of microbial organisms, data-driven generative models will be able to perform things that are currently out of reach.

\section{Acknowledgements}
We are grateful to N. Belliveau, H. Garcia,  E. G\"{o}kman, S. Grill, B. Ireland, F. J\"ulicher,  J. Kinney, S. Madhavi, G. Urtecho, V. Vitelli and C. Wiggins for useful discussions, and V. Garcia for help with cell lysis for mass spectrometry experiments. 
We are grateful to the NIH for support through award numbers DP1OD000217 (Director's Pioneer Award) and NIH MIRA 1R35 GM118043-01.
TR was supported by a  fellowship from Boehringer Ingelheim Fonds.
This work was supported by Igor A. Antoshechkin and by the Millard and Muriel Jacobs Genetics and Genomics Laboratory at the California Institute of Technology.

\section{Data and Code Availability}
Sequencing data will be available on the SRA. Mass-spectrometry data will be available on Caltech Data. Code written to process and analyze data, as well as to generate figures, will be made available on Github.
An interactive dashboard to explore information footprints for every gene in every condition can be found at \url{http://rpdata.caltech.edu/data/interactive_footprints.html}. A joint display of all information footprints can be found at \url{http://rpdata.caltech.edu/data/all_data.html}. The compendium with a discussion of the results for every gene, as well as an overview of the known binding sites in its promoter can be found at \url{http://rpdata.caltech.edu/data/reg-seq_compendium.pdf}.
\section{Materials and Methods}
\label{methods}
\subsection{Oligo Pool Design} \label{sec:oligo_pool}
\subsubsection{Identification of Transcription Start Sites}
\label{sec:ident_tss}
For each gene chosen for this study, shown in Table \ref{tab:genes_chosen}, we first looked for its promoter on EcoCyc \cite{karp2023ecocyc}. If the promoter was found, the annotated transcription start site (TSS) was used. If multiple promoters were identified, each promoter was included in the experiment. If no promoter was found, we looked for transcriptionally active sites in the data set from Urtecho et al.\cite{urtecho2023genome}. In their work, the genome was fragmented and every fragment was tested for transcription initiation in LB. If we could find a site that was identified as active close to the gene of interest, we chose this site as the origin for computational promoter mutagenesis. If no TSS could be identified for a gene, the model from LaFleur et al.~\cite{lafleur2022automated} was used to computationally predict a TSS in the intergenic region. The site predicted to be the most active within 500 bp upstream of the coding region was chosen as the TSS. Initially, 119 promoters were chosen, however, 7 promoters (mglBp, hdeAp2, mtnp, ybeDp, cpxRp2, galEp1, and ompFp), had an identical TSS as another promoter annotated in Ecocyc. The duplicated promoters were treated as independent when mutated variants were created, leading to twice the number of variants in the total pool.
\clearpage

\subsubsection{Computational Promoter Mutagenesis}
\label{sec:comp_prom_muta}
Once a TSS is identified, the 160 bp region from 115 bp upstream of the TSS to 45 bp downstream is taken from the genome. Most of the cis regulation has been shown to occur within this window. Based on the approach of Kinney et al.~\cite{kinney2010using}, each promoter sequence is randomly mutated at a rate of 0.1 per position. 1500 mutated sequences are created per promoter, following the approach of \cite{ireland2020deciphering}, which creates sufficient mutational coverage in the promoter region. The promoter oligonucleotides are flanked by restriction enzyme sites (SpeI at 5' and ApaI at 3') which are used in subsequent cloning steps. The restriction sites are flanked by primer sites that are used to amplify the oligo pool. The primer sequences were chosen from a list of orthogonal primer pairs, designed to be optimal for cloning procedures \cite{subramanian2018set}. Oligo pools were synthesized (TwistBioscience, San Francisco, CA, USA) and used for subsequent cloning steps.

\subsection{Library Cloning}
\subsubsection{Cloning Oligo Pool into a Plasmid Vector}
\label{sec:library_cloning}
The oligo pool was amplified using a 20 bp forward primer (SC142) and a 40 bp reverse primer (SC143), which consists of a 20 bp primer binding site and a 20 bp overhang. PCRs were run with minimal amplification until faint bands appear on an agarose gel in order to minimize amplification bias, using 10 ng of the oligo pool as template, as recommended by TWIST. The PCR was run for with 10 amplification cycles using
 a volume of 12.5 \textmu l. PCR products were cleaned and concentrated (DNA Clean \& Concentrator-5, ZymoResearch) and used for a second amplification step. The 20 bp overhang on the reverse primer from the first amplification was used as primer site for a reverse primer (SC172), which contains randomized 20 bp barcode, flanked by two restriction enzyme sites (SbfI and SalI, 5' to 3' direction)
. The forward primer is the same as in the first amplification step (SC142). PCR amplification is run again with minimal amplification to minimize amplification bias, which we found to be 8 cycles. PCR products are run on a 2\% agarose TAE gel and subsequently extracted and purified (Zymoclean Gel DNA Recovery Kit, ZymoResearch). In the next step, a restriction digest is performed on the outer restriction enzyme sites (SpeI-HF and SalI-HF). Unless noted otherwise, all restriction digests were run for 15 minutes at 37 \degree C, using 1 \textmu g of DNA as template and using 1 \textmu l of each restriction enzyme. The plasmid vector was digested with different restriction enzymes which create compatible sticky ends (XbaI and XhoI). Most restriction enzyme sites are palindromes, so by choosing different enzymes with compatible ends, we avoid having palindromes flanking the plasmid inserts after ligation.  The oligo pool is combined with the plasmid vector using T7 DNA ligase (New England Biolabs, Ipswich, MA, USA) following the supplier's protocol. Ligation products were cleaned and concentrated (DNA Clean \& Concentrator-5, ZymoResearch) and a drop dialysis (MF-Millipore VSWP02500, MilliporeSigma, Burlington, MA, USA)  was performed for 1 hour to improve sample purity. Electroporation using \textit{E. coli} pir116 electrocompetent cells (Lucigen, Middleton, WI) was performed at 1.8 kV in 1 mm electroporation cuvettes, followed by 1 hour recovery at 37 \degree C and 250 rpm in 1 ml LB media (BD Difco). The entire culture was plated on 150 mm LB + kanamycin (50 $\mu$g/ml) petri dishes and grown overnight at 37 \degree C. The following day, plates were scraped and the colonies resuspended. Freezer stocks were prepared using a 1:1 dilution of resuspended colonies and 50\% glycerol.
Cultures were inoculated with $5\times 10^8$ cells in 200 ml of LB + kanamycin (50 $\mu$g/ml) and grown at 37 \degree C until saturation. Plasmids
 were extracted (ZymoPURE II Plasmid Maxiprep Kit, ZymoResearch) and used for subsequent sequencing. For details, see \ref{sec:barcode_mapping}. The plasmid library is then used as template in a restriction digest using restriction enzymes ApaI and SbfI-HF. The resulting product was cleaned and concentrated (NEB Monarch) and the DNA concentration was measured on a NanoDrop. Similarly, the RiboJ::GFP element was PCR amplified (primers SC191 and SC192), adding restriction sites as overhangs (ApaI and PstI). For details about the restriction sites, see Table~\ref{tab:re_sites}. The PCR product was cleaned and concentrated (NEB Monarch) and digested with the respective restriction enzymes. The plasmid library is combined with the RiboJ::sfGFP element using T7 DNA ligase (New England Biolabs, Ipswich, MA, USA) following the supplier's protocol. Ligation products were cleaned and concentrated (NEB Monarch) and a drop dialysis (MF-Millipore VSWP02500, MilliporeSigma, Burlington, MA, USA) was performed for 1 hour to improve sample purity. Electroporation using \textit{E. coli} pir116 electrocompetent cells (Lucigen, Middleton, WI) was performed at 1.8 kV in 1 mm electroporation cuvettes, followed by 1 hour recovery at 37 \degree C and 250 rpm in 1 ml LB media. Entire cultures were plated on 150 mm kanamycin (50 $\mu$g/ml) + LB petri dishes aweren grown overnight. The following day, plates were scraped and the colonies resuspended. Freezer stocks were prepared using a 1:1 dilution of resuspended colonies and 50\% glycerol. Cultures were inoculated with $5\times 10^8$ cells in 200 ml of LB + kanamycin (50 $\mu$g/ml) and grown at 37 \degree C until saturation. Plasmids were extracted (ZymoPURE II Plasmid Maxiprep Kit, ZymoResearch) and used for subsequent genome integrations. 

\begin{table}[]
\centering
\begin{tabular}{|l|l|l|}
\hline
Part s            & 5' restriction site & 3' restriction site \\ \hline
Plasmid Vector   & XbaI                & XhoI                \\ \hline
RiboJ::GFP       & ApaI                & PtsI               \\ \hline
Oligo Pool       & SpeI               & ApaI                \\ \hline
Barcoding Primer & SbfI                & SalI              \\ \hline
\end{tabular}
\caption{\textbf{Restriction sites used.} All enzymes were ordered from NEB. If available, high fidelity versions of the enzymes were used.}
\label{tab:re_sites}
\end{table}
\subsection{Barcode Mapping}
\label{sec:barcode_mapping}
The plasmid library is used for barcode mapping. Purified plasmid is PCR amplified using forward primer (SC185) outside the promoter region and a reverse primer outside the 20 bp barcode (SC184). The PCR is run with minimal amplification, and the product is gel purified (NEB Monarch). The purified DNA was used as template for a second PCR using a primer (SC196), adding an Illumina P5 adapter to the promoter side, using a primer (SC199), and adding an Illumina P7 adapter to the barcode side. The PCR is again run with minimal amplification and gel purified (NEB Monarch). The product was used for sequencing on a Illumina NextSeq 2000 with a P2 flow cell with pair-end reads using primers SC185 for read 1, SC184 for read 2 and SC201 for the index read. Reads were filtered and merged using custom bash scripts, which are available in the Github repository. After processing, each promoter/barcode pair was identified in each read, and pairs with less than 3 total reads were discarded. An alignment algorithm was used to identify the identity of each sequenced promoter variant. This allowed us to include additional promoter variants that were in the initial oligo pool because of synthesis errors in the production of the oligos. The barcode mapping was used in the analysis of libraries grown in various growth conditions. The code used to perform processing of sequencing data can be found in the associated Github repository. Processing is done with the help of various software modules \cite{bushnell2014bbmap, chen2018fastp, tange_2023_7761866}. Custom Python code used for the analysis and visualization of results can be found in the associated Github repository.

\subsubsection{Genome Integration}
\label{sec:genome_int}
\paragraph{Creation of Landing Pad Strain}
We used ORBIT \cite{saunders2024orbit} to integrate reporter libraries into the \textit{E. coli} chromosome. ORBIT uses a targeting oligo containing an attB site, and an integration plasmid using an attP site. An additional helper plasmid facilitates the integration of the targeting oligo into the replication fork, followed by recombination of the attB and attP sites catalyzed by a \textit{bxb-1} gene in the helper plasmid. To increase the efficiency of genome integration, we created a landing pad strain that contains an attB site close to the \textit{glmS} gene in the \textit{E. coli} chromosome. Wild type \textit{E. coli} (K12 MG1655) is streaked on a LB plate and grown overnight at 37 \degree C. A single colony is picked and prepared to make elecotrocompetent cells as follows. The picked colony is grown in 3 ml of LB at 37 \degree C and shaken at 250 rpm overnight. The overnight culture is diluted 1:1000 into fresh LB (e.g. 200 ml) and grown at 37 \degree C and 250 rpm until exponential phase, reaching an optical density at 600nm (OD600) of $\sim$ 0.4. The cultures are then immediately put on ice and spun in a centrifuge at 5000 g for 10 min. Following the spin, the supernatant is discarded, and the cells are resuspended in the same volume as the initial culture of deionized water at 4 \degree C. The cells are spun again at 5000 g for 10 min. This wash step is repeated 3 times with 10\% glycerol. After the last wash step,the supernatant is discarded and cells are resuspended in the remaining liquid and distributed into 50 $\mu$l aliquots. Aliquots are frozen on dry ice and kept at -80 \degree C until they are used for electroporation. For electroporation, aliquots are thawed on ice and 1 mm electroporation cuvettes are pre-chilled on ice. 100 ng of helper plasmid (Addgene \#205291) is added to a 50 $\mu$l cell aliquot and mixed by slowly pipetting up and down. The aliquot is then added to the electroporation cuvette and electroporation is performed at 1.8 kV. The aliquot is recovered with 1 ml of LB media pre-warmed to 37 \degree C. The culture is recovered for 1 hour at 37 \degree C and shaken at 250 rpm. After recovery, aliquots at various dilutions are plated on LB + gentamicin (gentamicin sulfate 15 \textmu g/ml). Plates are grown overnight and a single colony is picked to prepare electrocompetent cells and frozen stocks as described above. The cells are electroporated with 2 mM of the targeting oligo (SC219) and an integration plasmid containing \textit{kanR} for selection and the \textit{sacB} gene for counterselection. After recovery, the cultures are plated on LB + kanamycin  (50 $\mu$g/ml). A colony is picked and electrocompetent cells are prepared again as mentioned above. Another electroporation is performed using only the targeting oligo (SC219). This time, cells are plated on LB + 7.5\% sucrose for selection of loss of the integrated cassette, leaving only an attB site in the locus. This results in a scarless insertion of the attB site into the chromosome. 
\paragraph{Integration of the Library.}
To perform genome integration, the host strain carrying the helper plasmid is made electrocompetent (follow growing and washing steps described above), and the plasmid library is electroporated into the host strain, using about 100 ng of plasmid per transformation. The cells are recovered in 3 ml of prewarmed LB + 1\% arabinose and shaken at 37 \degree C at 250 rpm for 1 hour. The entire volume is plated on LB + kanamycin plates and colonies are grown over night. The next day, all colonies are scraped, resuspended in LB and diluted to an OD600 of 1. The helper plasmid used for genome integration causes growth deficits, hence, the library needs to be removed of the plasmid. Therefore, the library is inoculated with 0.5 ml of culture at an OD600 of 1 in 200 ml of LB, and grown until exponential phase at 37 \degree C shaken at 250 rpm. The helper plasmid carries the \textit{sacB} gene, which is used for negative selection in the presence of sucrose. At exponential phase, the culture is plated on LB + 7.5\% sucrose agarose plates. Plates are grown overnight, scraped and made into frozen stocks at an OD600 of 1. The frozen stocks are then ready for growth experiments.

\subsection{Growth Media and Culture Growth}
\label{sec:media_and_growth}
\subsubsection{Base Media}
Lysogeny Broth (LB) was prepared from powder (BD Difco, tryptone 10 g/l, yeast extract 5 g/l, sodium chloride 10 g/l), and sterilized by autoclaving. M9 Minimal Media pre-mix without carbon source was prepared in the following way, similar to \cite{schmidt2016quantitative}: to 700 ml of ultrapure water, 200 ml of 5 × base salt solution (BD Difco, containing disodium phosphate (anhydrous) 33.9 g/l, monopotassium phosphate 15 g/l, sodium chloride 2.5 g/l, ammonium chloride 5 g/l, in H2O, autoclaved), 10 ml of 100X trace elements (5 g/l EDTA, 0.83 g/l FeCl3-6H2O, 84 mg/l ZnCl2, 19 mg/l CuSO4 - 5 H2O, 10 mg/l CoCl2 - 6H2Oin H2O, 10 mg/l H3BO3, 1.6 mg/l MnCl2 - 4H2O, prepared as described in \cite{flamholz2020functional}), 1 ml 0.1 M CaCl2 solution, in H2O, autoclaved, 1 ml 1 M MgSO4 solution, in H2O, autoclaved and 1 ml of 1000 × thiamine solution (1 mg/ml in water, filter sterilized) were added. The resulting solution was filled up to 1 l with water and filter sterilized. M9 minimal medium was complemented with carbon source by mixing appropriate amounts of carbon-source-free M9 minimal medium and carbon source stock solutions. Carbon source stock solutions were prepared as 20\% solutions and filter sterilized.
\subsubsection{Cultivation} 
\label{sec:cult}
Overnight cultures were incubated from frozen stock in 200 ml M9 minimal media with 0.5\% glucose and grown at 37 \degree C  while shaken at 250 rpm. Cultures were diluted 1:100 into the respective growth media (prewarmed to 37 \degree C, 200 ml) and grown to exponential phase (OD600 of 0.3). To ensure steady state growth, the cultures were diluted a second time 1:100 into the same growth media and grown again to an OD600 of 0.3, ensuring at least 10 cell divisions in the growth media. At this step, there are four different paths for a culture: 1. It is immediately harvested (called standard growth).; 2. A compound is added to the culture and the culture is harvested at a later specified time (called induction); 3. the culture is moved to water bath of a different temperature and then harvested at a later specified time ; or 4. the culture is spun down in four 50 ml aliquots at 3500 rpm for 7 min, washed in a different media twice, and then grown in that media for 1 hour. Unless otherwise mentioned, glucose was used as carbon source. Each condition was done in duplicate with some conditions being done in triplicate when the initial replicates did not seem to correlate well. To compare how experiments correlate, for each pair of conditions, we computed the Pearson correlation coefficient across all mutual information footprints. Figure~\ref{SI-fig:gc_corr} shows the correlation between all experiments and shows which experiments were excluded from analysis due to poor correlation to the rest of the experiments.
\subsubsection{Specific Growth Conditions}
 \label{sec:growth_cond}
\noindent\textbf{Glucose}: For standard growth, 5 ml of 20\% glucose solution added to 200 ml of M9 minimal media pre-mix for a final concentration of 0.5\%.\\ 
\textbf{Arabinose}: For standard growth, 5 ml of 20\% arabinose solution added to 200 ml of M9 minimal media pre-mix for a final concentration of 0.5\%.\\ 
\textbf{Xylose}: For standard growth, 5 ml of 20\% xylose solution added to 200 ml of M9 minimal media pre-mix for a final concentration of 0.5\%.\\ 
\textbf{Galactose}: For standard growth, 2.3 ml of 20\% galactose solution added to 200 ml of M9 minimal media pre-mix for a final concentration of 0.23\%.\\
\textbf{Acetate}: For standard growth, 5 ml of 20\% sodium acetate solution added to 200 ml of M9 minimal media pre-mix for a final concentration of 0.5\%.\\
\textbf{Sodium Salicylate}: 1 M sodium salicylate stock was prepared and filter sterilized. For standard growth, 500 \textmu l of the stock was added to 200 ml of M9-glucose media for a final concentration of 2.5 mM. For 1 hour induction, 2 ml of the stock was added to 200 ml M9-glucose media for a final concentration of 10 mM.\\
\textbf{Ethanol}: For standard growth, 5 ml of 200 proof ethanol was added to 200ml of M9-glucose media for a final concentration of 2.5\%. For 1 hour induction, 10 ml of 200 proof ethanol was added to 200 ml M9-glucose media for a final concentration of 5\%.\\
\textbf{Ampicillin}: For both standard growth and 1 hour induction, ampicillin was added to M9-glucose media for a final concentration of 2 mg/l.\\
\textbf{LB}: For standard growth, cultures were grown in LB media\\
\textbf{Stationary Phase}: Cultures were grown in M9-glucose media for an additional one day (1d) or three days (3d) after reaching an OD600 of 0.3.\\
\textbf{Leucine}: For 1 hour induction, leucine was added for a final concentration of 10 mM.\\
\textbf{Phenazine Methosulfate}: For 1 hour induction, 61 mg of 2,2 phenazine methosulfate (SigmaAldrich) was added for a final concentration of 100 \textmu M.\\
\textbf{2,2 Dipyridyl}: For 1 hour induction, 156 mg of 2,2 dipyridyl (SigmaAldrich) was added for a final concentration of 5 mM.\\
\textbf{Gentamicin}: For 1 hour induction, gentamicin was added for a final concentration of 5 mg/l.\\
\textbf{Copper Sulfate}: 1M stock of CuSO$_4$ was prepared. 1 hour inductions were performed with final concentrations of both 500 \textmu M and 2 mM.\\
\textbf{Hypochlorous Acid}: For 1 hour induction, sodium hypochlorite solution (Sigma-Aldrich \#425044) was added for a final concentration of 4 mM.\\
\textbf{Spermidine}: For a 1 hour induction, spermidine was added for a final concentration of 5 mM.\\
\textbf{Serine Hydroxamate}: For a 30 min induction, serine hydroxamate (Sigma-Aldrich) was added for a final concentration of 0.4 mg/ml.\\
\textbf{Cold Shock}: Cultures were grown in M9-glucose media to an OD600 of 0.3 and then were immersed in a 10 \degree C water bath and shaken for 1 hour.\\
\textbf{Medium Cold Shock}: Cultures were grown in M9-glucose media to an OD600 of 0.3 and then were immersed in a 19 \degree C water bath and shaken for 1 hour.\\
\textbf{Heatshock}: Cultures were grown in M9-glucose media to an OD600 of 0.3 and then were immersed in a 42 \degree C water bath and shaken for 5 min.\\
\textbf{H$_2$O$_2$}: For 30 min induction, H$_2$O$_2$ was added to M9-glucose media for a final concentration of 0.1 mM. For 10 min induction, H$_2$O$_2$ was added to M9-glucose media for a final concentration of 2.5 mM.\\
\textbf{Nitrogen Starvation}: Minimal media premix was prepared with only 10\% NH$_4$Cl. Cultures were grown in M9-glucose media to an OD600 of 0.3, then washed and grown for 1 hour in M9-glucose media with reduced NH$_4$Cl.\\
\textbf{Magnesium Starvation}: Minimal media premix was prepared where MgSO$_4$ was replaced with NaSO$_4$ at the same concentration. Cultures were grown in M9-glucose media to an OD600 of 0.3, then washed and grown for 1 hour in M9-glucose media with NaSO$_4$.\\
\textbf{Sulphur Starvation}: Minimal media premix was prepared where MgSO$_4$ was replaced with MgCl at the same concentration. Cultures were grown in M9-glucose media to an OD600 of 0.3, then washed and grown for 1 hour in M9-glucose media with MgCl.\\
\textbf{pH2}: Minimal media was prepared as usual, but the pH is adjusted to 2.0 using sodium hydroxide. Cultures were grown in M9-glucose media to an OD600 of 0.3, then washed and grown for 1 hour in M9-glucose media with pH2.\\
\textbf{Low Osmolarity}: Minimal media pre-mix was diluted by a factor of two before adding glucose. Cultures were grown in M9-glucose media to an OD600 of 0.3, then washed and grown for 1 hour in low osmolarity M9-glucose media.\\
\textbf{High Osmolarity}: LB was supplemented with 0.75 M NaCl. Cultures were grown in M9-glucose media to an OD600 of 0.3, then washed and grown for 1 hour in LB with 0.75 M NaCl.\\
\textbf{Low Phosphate}: Minimal media was prepared with only 10 \% of disodium phosphate and monopotassium phosphate. Cultures were grown in M9-glucose media to an OD600 of 0.3, then washed and grown for 1 hour in low phosphate M9-glucose media.\\
\textbf{Nitrate}: For standard growth, potassium nitrate was added to standard M9-glucose media for a final concentration of 80 mM.\\
\textbf{Anaerobic}: For anaerobic growth, M9-glucose media was kept in a glove box containing nitrogen for multiple days to equilibrate and remove oxygen from the media. Cultures were inoculated in 20 ml of this media inside the glove box in glass tubes which are sealed with rubber plugs. Tubes were grown in a shaker at 37 \degree C and shaken at 250 rpm. When the culture reached an OD600 of 0.3, a 1:100 dilution was performed inside the glove box and grown to an OD600 of 0.3 again. Cultures were then harvested. \\
\textbf{Anaerobic and Nitrate}: 80 mM potassium nitrate was added to standard M9-glucose media. Cultures were grown in anaerobic conditions as described above.

\subsection{Barcode Sequencing}
\label{sec:barcode_seq}
Once a culture is ready for harvesting, 750 \textmu l of culture are mixed with 750 \textmu l of freshly prepared 1X Monarch DNA/RNA Protection Reagent (NEB) and pelleted by spinning at 20000 g for 1 minute. The supernatant is discarded, and the pellets are frozen on dry ice. Genomic DNA is extracted from four pellets for each sample using a Monarch Spin gDNA Extraction Kit (NEB), following the manufacturer's protocol for gram-negative bacteria. RNA was extracted and reverse transcription was performed using a custom protocol. 500 ng of gDNA and 5 µl of cDNA was used as template for library preparation. First, the template is amplified by PCR using primers SC184 and SC88. 12 cycles are run for gDNA and between 20 and 25 cycles for cDNA, depending on the sample. The PCR product was run on a 2\% agarose gel and bands were gel purified (Monarch DNA Gel Extraction Kit, New England Biolabs). Then, 5 ng of amplified DNA was used for a second PCR (50 \textmu l volume), using forward primer SC80 and one of 92 reverse primers (SC354-SC445), which add an index for demultiplexing. The PCR is run for 6 cycles and the product is run in a 2\% agarose gel, followed by gel extraction. The extracted DNA is used for sequencing. Sequencing runs were perfomed on a NextSeq 2000.  A summary of all the sequencing runs used for this paper is shown in Table~\ref{tab:my-table}. Primer SC450 was used for read 1, and primer SC270 for the index read.
Sequencing data is filtered for quality and trimmed using \textit{fastp} \cite{chen2018fastp}. Barcodes are extracted and counted from sequencing files using custom Bash scripts, which are available on Github.

\subsection{Sequencing Runs}
\begin{table}[H]
\fontsize{4}{8}\selectfont
\begin{tabular}{|l|l|l|l|l|}
Date       & SRA Number & Content                                   & Run Type              & Sequencer                   \\\hline
05/14/2022 &            & Mapping of promoter variants and barcodes & Paired-End            & Next-Seq 2000 - P2 flowcell \\
03/27/2023 &  & \begin{tabular}[c]{@{}l@{}}Comparison of plasmid reporters and genome \\ integrated reporters\end{tabular} & Single-end & MiSeq v2 flowcell \\
09/07/2023 &            & Barcode counting in 9 Conditions          & Single-end, 50 cycles & MiSeq, v2 flowcell          \\
12/07/2023 &            & Barcode counting in 27 Conditions         & Single-end, 27 cycles & Next-Seq 2000, P3 flowcell  \\
06/21/2024 &            & Barcode counting in 44 Conditions         & Single-end, 27 cycles & Next-Seq 2000, P4 flowcell  \\
09/07/2024 &            & Barcode counting in 24 Conditions         & Single-end, 27 cycles & Next-Seq 2000, P2 flowcell 
\end{tabular}
\caption{\textbf{Sequencing runs}. Every sequencing run containing data used in this work. Each run has its own code for processing, which can be found in the associated Github repository.}
\label{tab:my-table}
\end{table}
\subsection{DNA Chromatography and Tandem Mass Spectrometry}
\subsubsection{Cultivation for Lysate}
\label{sec:cult_lysate}
Similar to the procedure described in \ref{sec:cult}, overnight cultures were grown in 5 ml of M9 minimal medium with 0.5\% glucose at 37 \degree C and then diluted 1:100 into the growth media listed in \ref{sec:growth_cond}.  Cultures were carried out in 2800 ml Fernbach-style flasks filled with 500 ml of media. The total volume of liquid culture for a given growth condition ranged from 1000 to 6000 ml, depending on the number of required DNA chromatography experiments. After a given growth condition duration is completed, the cells were harvested by centrifuging at 8000 g for 30 minutes at 4 \degree C. Cell pellets were stored at -80 \degree C or subsequently lysed.
\subsubsection{Lysate Preparation}
\label{sec:lysate_prep}
Cell pellets were re-suspended in lysis buffer (70 mM potassium acetate, 50 mM HEPES pH 7.5, 5 mM magnesium acetate, 2.5 mM DTT, and cOmplete Ultra EDTA-free protease inhibitor). Mechanical cell lysis was performed using a high pressure cell disruptor (Constant Systems). Afterwards, to help solubilize membrane proteins, n-dodecyl-β-D-Maltoside (DDM) detergent was added to the crude lysate for a final concentration of 1 mg/ml. Lysates were clarified of non-soluble cell debris by centrifuging at 30000 g for 1 hour at 4 \degree C. The collected supernatants were further concentrated to $\sim$100 mg/ml using centrifugal protein concentration filter (Amicon Ultra -15) with a molecular weight cut-off of 3 kDA.  Protein concentrations were determined using a fluorometer (Qubit fluorometer) and proprietary dyes that specifically label proteins (Qubit Reagent). The lysates were further cleared of non-specific DNA binding proteins by incubating with a competitor salmon sperm DNA (Invitrogen) at 0.1 mg/ml for 10 minutes at 4 \degree C. An additional 1 hour incubation at 4 \degree C is performed by adding sacrificial streptavidin-coated magnetic beads without any attached DNA oligos (Dynabeads MyOne Streptavidin T1) at $\sim$3 mg/ml in order to clear proteins that may non-specifically bind to the beads surfaces. Lysates are centrifuged one final time to pellet the sacrificial beads and any remaining insoluble component.  The resulting supernatants are stored at -80 \degree C or aliquoted into volumes of 200 \textmu l for subsequent DNA chromatography experiments.  
\subsubsection{DNA Chromatography}
\label{sec:DNA-chrom}
DNA affinity chromatography is used to isolate a transcription factor of interest from a given cell lysate. The procedure detailed below is similar to the one we have used previously~\cite{mittler2009silac,belliveau2018systematic}. In brief, DNA oligos that have putative transcription factor binding sites are attached to magnetic beads.  These beads with tethered DNA are incubated with cell lysates to "fish out" proteins that bind to the oligos. They are spatially separated from the remaining lysate by magnets, allowing for extraction of the bound proteins.  The relative enrichment of a given protein is determined by comparison to a control DNA sequence that has no specific binding sites.    
\paragraph{DNA Oligos for Magentic Beads}
\label{sec:dna-oligo}
The binding sequence of an oligo (IDT) is taken from the native \textit{E. coli} genome, where the sequence region is hand-selected to match the putative binding sites determined by RegSeq. For the control sequence, a region near the TSS associated with the promoter ymjF was used, since this sequence had no discernible binding sites, as determined by RegSeq. Each oligo has the 5' end biotinylated (to ensure attachment to streptavidin-coated magnetic beads) and starts with a cut site sequence for the PstI restriction enzyme (New England Biolabs), which allows for the bound protein to be recovered by a restiction digest. 
\paragraph{Bead Incubations and Protein Recovery}
\label{sec:bead_inc}
A batch volume of magnetic beads (Dynabeads MyOne Streptavidin T1) is measured out, according to the total number of DNA chromatography experiments being performed and assuming each individual chromatography experiment requires 160 \textmu l of stock beads per 200 \textmu l of aliquoted lysate. The total volume of beads is washed twice in TE buffer (10 mM Tris-HCl pH 8.0, 1 mM EDTA), washed twice in DW buffer (20 mM Tris-HCl pH 8.0, 2 M NaCl, 0.5 mM EDTA), and re-suspended in annealing buffer (20 mM Tris-HCl pH 8.0, 10 mM MgCl2, 100 mM KCl).  The beads are aliquoted according to the number of oligos used. DNA oligos are added to the aliquoted beads to a final concentration of 5 \textmu M and incubated for at least 3 hours at room temperature or overnight at 4 \degree C. After oligo incubation, beads are washed twice in TE buffer and then twice DW buffer. All wash buffers are supplemented 0.05\% TWEEN-20 detergent to minimize bead loss related to sticking to surfaces.  After washing, beads are incubated in a blocking buffer (20 mM HEPES pH 7.9, 300 mM KCl, 0.05 mg/ml bovine serum albumin, 0.05 mg/ml glycogen, 2.5 mM DTT, 5 mg/ml polyvinylpyrrolidone, and 0.02\% DDM) for 1 hour at room temperature to reduce nonspecific protein binding to the bead surfaces. The beads are then washed three times in lysis buffer.  The beads are added to the aliquoted lysates to a final concentration of ~5 mg/mL.  When applicable, a supplement of the reagent defining a given growth condition (carbon source, antibiotic, chemical stress,...) is added to the lysate to approximate the internal cell environment.  For example, for the M9-glucose growth condition, glucose is added to the lysate to a final concentration of 0.5\%. See Table~\ref{tab:lysate_supplement} for details of all lysate supplements used. The beads and lysate are incubated overnight at 4 \degree C on a rotating rack. The next day, the beads are washed three times in lysis buffer and once in the reaction buffer (NEB buffer r3.1) for the restriction enzyme PstI.  1000 units of PstI is added to each bead reaction. The beads are incubated for 90 min at room temperature.  The supernatant containing the DNA and bound proteins is collected for solution-based protein digestion. 
\subsubsection{Protein Digestion, Labeling and Desalting for Proteomic Analysis}
\label{sec:digest_label}
The samples were subjected to an isobaric-labeled filter-aided sample preparation (iFASP) protocol (PMID: 23692318) with minor changes. Briefly, supernatant from each sample was loaded onto a 10 kDa Amicon filter (Pierce), and washed with 8 M urea in 100 mM HEPES (urea buffer) 3 times. Each washing step includes adding 200 \textmu l of the corresponding solution followed by 14000 g centrifugation for 15 min. After 3 washes with urea buffer, 200 \textmu l of urea buffer containing 5 mM tris(2-carboxyethyl)phosphine was added into each filter to break disulfide bonds. The reaction was incubated for 1 hour at room temperature, and 200 \textmu l of urea buffer containing 20 mM of chloroacetamide was added into each filter to alkylate free thiols. The alkylation reaction was incubated for 15 min at room temperature, and the filters were centrifuged for 14000 g for 15 min. The filters were further washed 3 times with 150 \textmu l of 100 mM of triethylamine bicarbonate (TEAB) in water. After the TEAB washes, 120 \textmu l of 100 mM TEAB containing 1 \textmu g of Trypsin (Pierce) was added into each filter. The enzyme to substrate ratio is estimated be from 1:5 to 1:10. The trypsinization step occurred for 16 hours at 37 \degree C. After trypsinization, 5 \textmu l of DMSO containing 0.05 mg of TMTpro reagent (Thermo) was added into each filter, and the labeling reaction incubated for 1 hour. 1 \textmu l of 5\% hydroxylamine was added into each filter to quench the TMT labeling. The samples were then eluted from the filters by 14000 g centrifugation for 15 min. The filters were further washed 3 times with 50 \textmu l of 0.5 M NaCl in water, and all elutions were pooled together. The pooled sample was dried using a CentriVap concentrator (LabConco), and was desalted with a monospin C18 column (GL Science) according to manufacturer’s instructions. The desalted sample was dried again using a CentriVap concentrator and was stored at -80 \degree C.
\subsubsection{Liquid Chromatography and Tandem Mass Spectrometry}
\label{sec:liq_chrom}
 Samples were reconstituted in 20 \textmu l of 2\% acetonitrile and 0.2\% formic acid in water. The peptide concentration was determined using the Pierce Colorimetric Quantitative Peptide Assay. An aliquot of 500 \textmu g of the peptide was loaded onto a Thermo Vanquish Neo liquid chromatography (LC) system, where the peptides were separated on an Aurora UHPLC Column (25 cm × 75 \textmu m, 1.6 \textmu m C18, AUR2-25075C18A, Ion Opticks). The LC gradient was increased from 2\% to 98\%  of mobile phase B over 130 min. See Table~\ref{tab:LC_table} for the gradient settings. The LC processed peptides were analyzed on a Thermo Eclipse Tribrid mass spectrometer using a data-dependent acquisition (DDA) method, where the mass spectrometer selects the most intense peptide precursor ions from the first scan of tandem mass spectrometry (MS1) and then fragments and analyzes the precursors in a second scan (MS2).  MS1 scans were acquired with a range of 375–1600 m/z in the Orbitrap at a resolution of 120000. The maximum injection time was 50 ms, and the AGC target was 250. MS2 scans were acquired using the quadrupole isolation mode and the higher-energy collisional dissociation (HCD) activation type in the Iontrap. For these scans, the resolution was 50000, the isolation window was 0.7 m/z and the collision energy was 35\%.  See Table~\ref{tab:MS_table} for the detailed parameters used for the mass spectrometry scans. 
 \subsubsection{Mass Spectrometry Data Analysis}
\label{sec:mass_spec_data}
The raw data generated by the mass spectrometer is analyzed using Proteome Discoverer 2.5 based on the Sequest HT algorithm~\cite{tabb2015sequest}. The data is compared against the \textit{E. coli} proteome from UniprotKB for protein identification. The fragment mass tolerance was set to 10 ppm. The maximum false peptide discovery rate was specified as 0.01 using the Percolator Node validated by a q-value. See Table~\ref{tab:PD_table} for the complete list of parameters used for Protein Discoverer 2.5. The resulting data was exported and analyzed with custom Python-based code, which is available on Github.   
			}{}
			\printbibliography[segment=\therefsegment]
		\end{refsegment}

	\clearpage

	\title{\textbf{Supplemental Information for: The Environment-Dependent Regulatory Landscape of the {\it E. coli} Genome }}
	\setboolean{sitext}{true}
	\ifthenelse{\boolean{sitext}}{
	\maketitle


	\addtocontents{toc}{\protect\setcounter{tocdepth}{2}}

		\begin{refsegment}
			\beginsupplement
			\tableofcontents
\section{Existing methods for dissection of gene regulation in bacteria}
\label{sec:SI_literature}
A huge effort has been expended in uncovering how genes in \textit{E. coli} are regulated~\cite{Echols2001,Ptashne2004,Oppenheim2005,Oehler1990, Oehler1994, Muller-Hill1996,Muller1996,Kuhlman2007,Lewis2002,Lee1989,Lee1992,Schleif2000}. For much of the history of modern molecular biology, genes were usually studied on a one-by-one basis due to the lack of high throughput methods.  This led to major success stories including insights into the lysis-lysogeny decision in bacteriophage lambda, discoveries on how bacteria use different carbon sources including lactose, galactose and arabinose, insights into the role of DNA looping and a myriad of other examples.  Over the past few decades, a variety of high-throughput methods have re-enlivened the subject by  enabling the identification of many binding sites for a single transcription factor in one experiment, or identifying binding sites for all transcription factors without knowing the identity of the transcription factors specifically binding those sites. Here we provide an overview of some examples of previous work to give a sense of where our own efforts fit into this enormous subject, highlighting both
 the successes and open questions resulting from previous work.
 We begin by discussing in vitro approaches in which DNA and proteins are interrogated outside of their natural cellular environments.  Much has been learned from these approaches.  Then, we turn to the analysis of in vivo methods which attempt to capture DNA-protein interactions in the context of living cells.

\begin{itemize}
    \item SELEX: Systematic evolution of ligands by exponential enrichment (SELEX) was developed in 1990~\cite{ellington1990vitro} with the purpose of identifying \textit{in vitro} which DNA sequence or ligand a protein binds to. The method uses a library of synthesized DNA that is incubated with purified proteins. Unbound DNA is removed and bound DNA is eluted from the protein and subsequently amplified. This process is repeated over multiple rounds to find DNA with high binding affinity to the protein. The level of specificity of the DNA obtained in these experiments can be tuned by choosing a different number of cycles. Once a DNA sequence with high affinity for the protein is identified, the genome of interest can be scanned for potential transcription factor binding sites by looking for sequence similarity between the genomic DNA and the SELEX DNA. In the context of \textit{E. coli}, binding sites for hundreds of transcription factors have been identified genome wide~\cite{ishihama2016transcription}.
    \item PBM: Protein-binding microarrays (PBM) use a large array of synthesized DNA oligonucleotides that are fused to a surface. Binding of purified protein to the DNA oligonucleotides is measured by fluorescence microscopy of tagged proteins. PBMs are able to better detect less specific binding sites for TFs than SELEX~\cite{berger2009universal}. PBMs have been used to identify the motifs for more than 1000 transcription factors~\cite{weirauch2014determination}.
    \item DAP-Seq: One modification to \textit{in vitro} binding assays that has been useful is to choose genomic DNA as template instead of synthesized DNA.  This approach is the basis of DNA affinity purification sequencing (DAP-Seq) \cite{o2016cistrome, bartlett2017mapping}. One of the advantages of using genomic DNA is that such DNA maintains chemical modifications to the DNA such as methylation and reveals that such methylation  can be important for DNA - TF interactions.
    \item ChIP-Seq: Chromatin immunoprecipitation with sequencing (ChIP-Seq) is one of the most commonly used methods to identify binding sites for TFs in the \textit{in vivo} setting. Identifying binding sites \textit{in vivo} is beneficial as potential co-factors and enzymes modifying the conformation of the TF are present if the correct growth condition is chosen. In the first iterations of the method the resolution of identified binding sites was low, however, the use of endonucleases in ChIPexo-Seq \cite{rhee2012chip} lead to higher accuracy. To pull down TFs that are crosslinked to DNA, the TF needs to be modified with a tag, e.g., His-tag \cite{peano2015characterization}, or antibodies against the TF need to be available, which can limit the throughput and often means that only one or a few TFs can be studied at the same time. However, binding sites across the entire genome can be found in a single experiment, giving high throughput on this axis. There have been drastic differences in the number of binding sites identified for certain TFs between ChIP-Seq and SELEX \cite{peano2015characterization}. ChIP-Seq has been used to discover a variety of different DNA-Protein interactions, such as the RpoS regulon in \textit{E. coli}~\cite{peano2015characterization}, the PhoB regulon~\cite{fitzgerald2023genome}, nucleoid organization by H-NS and MukBEF~\cite{lioy2018multiscale}, genome wide binding of CRP (using DNA microarrays)~\cite{grainger2005studies} and binding of Sigma70~\cite{chung2013dpeak}. In a recent study, ChIp-Seq was performed on 139 TFs for cells grown in minimal media with glycerol~\cite{lally2024predictive}.
    \item DNAse footprinting: In contrast to ChIP-Seq,  DNAse footprinting does not require a pull down on a specific TF. This allows for the discovery of binding for all DNA binding proteins across the entire genome at the same time, but also comes with the loss of the identity of the protein binding to each site. This method has been combined with RNAP occupancy studies to verify the function of identified binding sites \cite{freddolino2021dynamic, trouillon2023genomic}.
     \item MPRA: Massively-parallel reporter assays are one of the signature recent achievements in high-throughput approaches for dissecting promoter function.  The method is based upon creating large libraries of genetic variants and measuring their function in parallel. Urtecho et al. use genome-integrated massively parallel reporter assays to catalog and characterize promoters throughout the \textit{E. coli} genome \cite{urtecho2023genome}.
    This approach has been used impressively in {\it E. coli} to explore not only binding sites, but also ribosomal binding site sequences \cite{kosuri2013composability}, etc.
    \item Sort-Seq: In Sort-Seq \cite{kinney2010using, belliveau2018systematic} binding of transcription factors is identified by mutating bases in the vicinity of a transcription start site and measuring expression of a downstream reporter gene using fluorescence activated cell sorting (FACS), followed by DNA sequencing of the mutated promoter variants. By identifying bases where a mutation leads to a large change in expression putative binding sites are discovered. The identity of the transcription factor binding to these sites is then identified by DNA chromotography and mass spectrometry. This approach not only identifies binding of transcription factors, but also shows that the binding is functional, i.e., binding of the transcription factor effects expression of a gene. It has been found that binding sites that have been identified \text{in vitro} do not necessarily imply that the binding has regulatory function \cite{fitzgerald2023genome}.
\end{itemize}

All of these methods have been used to gain insights into the regulatory landscape of \textit{E. coli} and other organisms. Databases such as EcoCyc \cite{karp2023ecocyc} and RegulonDB \cite{salgado2024regulondb} contain a large number of annotated binding sites. Our goal is to use data from these methods as well as approaches in our paper to provide a systematic, rigerous and complete description of all promoters in \textit{E. coli} with standardized annotations in databases that can be used for e.g. phylogenetic modeling and building blocks for synthetic biology.
In particular, as shown in several of the figures in the paper, in those cases where we are able to find putative binding sites
and identify the TFs that bind to those sites, the aim is to be able to go from a promoter of unknown regulatory architecture all
the way to an environment-dependent regulatory architecture including energy matrices describing transcription factor binding and statistical mechanical models of the input-output response of
the promoter of interest as a function of key regulatory knobs such as DNA-TF binding site strength, TF copy number and effector concentration.  Beyond that, the aim is for all of these disparate sources to come together to make excellent databases such as EcoCyc \cite{karp2023ecocyc} and RegulonDB \cite{salgado2024regulondb} a more complete and internally consistent source as a basis for rigorous understanding of the physiology and evolution of {\it E. coli} and that will serve as a template for how to structure such databases for other organisms.

			\clearpage
\section{Theory of the experiment}
\label{sec:theory}
For each promoter-growth condition pair we produce a unique dataset consisting of counts of both DNA and RNA  for each barcode. An example dataset is shown in Table~\ref{tab:fakedata}.
\begin{table}
\centering
    \begin{tabular}{|c r r|}
    \hline
    Sequence & DNA counts & RNA counts \\
    \hline
    ACTA & 5 & 23\\
    \hline
    ATTA & 5 & 3\\
    \hline
    CCTG & 11 & 11\\
    \hline
    TAGA & 12 & 3\\
    \hline
    GCGC & 2 & 0\\
    \hline
    ACCA & 8 & 7\\
    \hline
    AGTA & 7 & 3\\ 
    \hline
    \end{tabular}
    \caption{\textbf{Synthetic dataset used to explain the logic of creating summary statistics.} Hypothetical dataset where for each sequence there is an associated count from DNA sequencing and a count from RNA sequencing. For this hypothetical case, the tiny wild-type ``sequence'' is ACTA.} 
    \label{tab:fakedata}
\end{table}
To identify potential binding sites in the sequence, we are looking for positions in the sequence of interest that lead to a high change in expression of the reporter when the base is mutated.  We hypothesize that this coupling between sequence and expression indicates that the binding of a regulatory factor has been modified. There are multiple ways of displaying the connection between the identity of each base and the level of   associated with such sequence changes. Ultimately, we have opted to use ``summary statistics'' that take as input the sequencing reads and result in output of some basepair by basepair picture of the importance of a given base to the level of expression. 
 In particular, here we use two such summary statistics, namely,  expression shifts and mutual information as a way to generate hypotheses for binding site positions and sequences. A detailed discussion of both quantities can be found in~\cite{Pan2024}.  Here, we give a brief reminder of these two summary statistics to make our approach self-contained, though the reader can see that reference for more details.

\subsection{Expression Shifts}
The expression shift summary statistic is an average way of determining how much the gene expression will be changed if the base at site $i$ is changed from its wild type value to one of the three other alternatives.
Expression shifts can be calculated directly from the data, and measure directly how much the expression is changed given a mutation to a specifc base at each position. The result is a $4\times L$ matrix, where $L$ is the length of the promoter sequence. 
Specifically, for a dataset where each sequence $i$ is associated with a measure for expression $c_i$ the value in the expression shift matrix at position $l$ corresponding to base $b$ is given by
\begin{equation}
\Delta s_{b,l} = \left\{
\begin{array}{ll}
\frac{1}{n} \sum_{i=1}^{n} \xi_{i,l} \left( \frac{c_i}{\langle c \rangle} - 1 \right), & \text{where} \quad \langle c \rangle = \frac{1}{n} \sum_{i=1}^{n} c_i, \quad \text{if \( b \) is mutated} \\
0, & \text{if \( b \) is wild type}
\end{array}
\right.
\end{equation}
where $ \xi_{i,l} = 1 $ if the base at position $ l $ in the $ i $-th promoter variant corresponds to base identity $ b $ and $ \xi_{i,l} = 0 $ otherwise. For our experiments, the measure of expression $c_i$ is the ratio of RNA to DNA barcode counts. The same measure can be applied to Sort-Seq datasets, where the measure of expression is the expression bin the sequence was assigned to. This matrix form of the expression shift allows us to capture how mutations in specific bases affect the promoter change in expression.

\subsection{Mutual Information}
As noted above, a second useful summary statistic is
the information footprint. Details for how to go from
sequence reads to this summary statistic can be found in~\cite{Pan2024}.
This way of displaying how mutations change expression of the reporter gene is achieved by computing the mutual information between the identity of a given base and the level of gene expression.   High mutual information indicates that the identity of the base is significant in governing the level of
expression and leads to the hypothesis that 
that base is part of a binding site, for example.  However, as we showed in Figure~\ref{figure_method}, not only does
the information footprint reveal binding sites that had not been seen before, but also sometimes hides the existence of mutations that created new transcription start sites.  Ultimately, the information footprint is one of many possible summary statistics for trying to tame the enormous datasets, and needs to be used judiciously.

From the data we can compute the ratio of reads that contain a mutation at each position for both DNA and RNA reads, $p(m, 
\mu)$, where $m$ indicates the base is wild-type ($m=0$) or mutated ($m=1$), and $\mu$ indicates if the read belongs to DNA ($\mu=0$) or RNA ($\mu=1$). Then, we compute mutual information at position $i$ as
\begin{equation}
    I_i =  \sum_{m=0}^1  \sum_{\mu=0}^1 p(m,\mu)\log_2\left(\frac{p(m,\mu)}{p(m)p(\mu)}\right),
    \label{eqn:SI_mutual_info}
\end{equation}
\noindent where $p(\mu)$ and $p(m$) are the marginal distributions of $p(m, \mu)$. In the example dataset introduced in Table~\ref{tab:fakedata} for the purposes of explaining notation and the algorithm, there are a total of 50 DNA counts and 50 RNA counts. In the first position,  the wild-type base is A, and 25 DNA reads contain the wild-type base, hence we compute $p(0, 0) = 0.25$, while 36 RNA reads contain the wild-type base, thus $p(0, 1) = 0.36$. 25 DNA reads contain a mutation in the first base, which gives $p(1, 0) = 0.25$, and 14 RNA reads contain a mutation in the first base, therefore, $p(1, 1) = 0.14$.
Using these values, using equation~\ref{eqn:SI_mutual_info} we can compute mutual information for the first base  to be $I_1=0.037$ bits. At this point in time, the meaning of an absolute value is still unclear, and we can only compare the value of mutual information at one position to the values at other positions in the same promoter. 

\subsection{Identification of Binding Sites}
\label{sec:ident}
Once an information footprint is computed, one has to identify where binding sites are. For a handful of footprints this can 
be done by hand, however, for the scale of this experiment when on the order of 10000 unique footprints are obtained, an unbiased and
automated method to detect binding sites from the data is needed. We have explored multiple different methods which are 
discussed below.

\subsubsection{Triaging}
\label{sec:theory_tria}
At first, we put each footprint into one of three categories: 1. Footprints containing putative binding sites for
transcription factors and/or sigma factors 2. One single mutation in the promoter sequence leads to high expression
from a transcription start site that is different from the annotated transcription start site, a phenomena we call
\textit{de novo promoters} 3. The footprint is not distinguishable from noise. 
The mutual information computed from equation~\ref{eqn:SI_mutual_info} is at base-pair resolution. To smoothen the 
footprint and make it easier to identify peaks, we use a sliding gaussian kernel across each footprint.
An important quantity for our classification of footprints is the coefficient of variation (CV) of mutual information across
the promoter. A footprint with distinct peaks compared to a noise background has a large CV compared to a footprint
with lots of fluctuations, as shown in Figure~\ref{SI-fig:triaging}(A). To estimate a noise floor for our dataset, we use 
a shuffling approach. We take our datasets (measurements of promoters in growth conditions), and assign sequencing counts to 
random mutated promoter variant that is different from the 
sequence the counts were identified initially. This removes the measured correlation between base identity and 
effect of the mutation. We perform this shuffle 100 times per promoter-growth condition pair, and for each iteration the footprints
are smoothed using the gaussian kernel and the CV is computed. All resulting CV values are pooled, and we compute a 95\% confidence 
interval for the distribution, where
we choose the upper limit of the interval as the noise threshold, which turns out to be 0.75, as shown in 
Figure~\ref{SI-fig:triaging}(B). This allows us to distinguish measurements that are dominated by noise from footprints
that contain a distinguishable signal. To identify if a footprint contains joint regions of high mutual information, a sign
of a binding site, or a single mutation that leads to activity from the promoter, we look at how the smoothing of the footprint
changes the CV across the positions. We simulate footprints with varying number of neighboring positions that have high
mutual information compared to a noisy background, and compute the CV before and after smoothing of these footprints. We find
that there is a distinct group of datasets that follow the trend for a single position with high mutual information, as can be seen
by the blue line in Figure~\ref{SI-fig:triaging}(C). This allows us to separate the datasets between footprints that have putative 
binding sites, and footprints with single positions of high mutual information. Figure~\ref{SI-fig:triaging}(D) shows the thresholds
and example footprints for the \textit{araB} promoter. We then use a Hidden Markov model to identify binding sites for promoters
in the first class and the computational model by LaFleur et al. \cite{lafleur2022automated} to look for putative new transcription
start sites in promoters with single mutations of high mutual information, both cases are discussed in detail below.

\begin{figure}
    \centering
    \includegraphics[width=\linewidth]{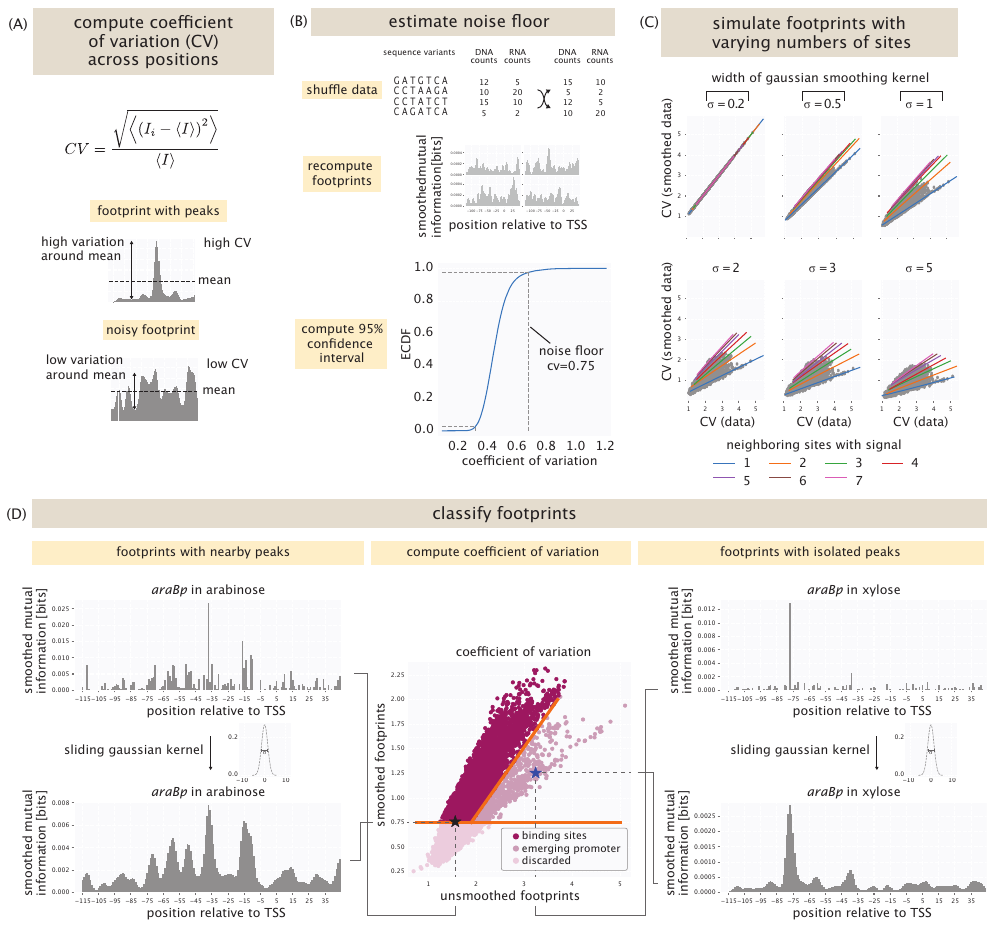}
    \caption{\textbf{Triaging of the data into different classes.} (A) The coeffiecent of varation of mutual information across positions
    in the footprint is a proxy for noise. (B) The noise floor of measurements can be estimated by shuffling datasets and recomputing
    footprints and coeffiecents of variation. (C) Datasets follow different trends under smoothing depending on the number of positions
    with high mutual information. Footprints with single positions of high mutual information separate from the rest of the data.
    (D) Based on the noise floor and the distinct behavior of footprints with single positions of importance, thresholds are drawn in the
    plane of coefficient of variation before and after smoothing of the data. Shown are examples for \textit{araB}, which has activator binding
    sites when cells are grown in arabinose, but not when grown in xylose. In that case, a single mutation upstream of the annotated TSS leads
    to a new active TSS emerging.}
    \label{SI-fig:triaging}
\end{figure}

\subsubsection{Hidden Markov Model}
\label{sec:theory_hmm}
Once a sequence is identified as containing binding sites, we use a two state Hidden Markov Model to distinguish binding sites from positions that do not contain binding sites. Hidden Markov models are often used for sequence analysis~\cite{yoon2009hidden}, and here we present an application specific to the nature of our experiment and datasets.
The two hidden states in the model are binding sites and background, where we assume that binding sites contain significantly higher mutual information that noisy background. Positions that do not contain binding sites have non-zero mutual information due to experimental noise but also due to the nature of our mutated library, where each sequence contains on average 16 mutations (160 bases at a 0.1 mutation rate). Hence, mutations in and outside binding sites occur at the same time. Since we are using about 1500 mutated sequences per promoter, we are using only a tiny fraction of all possible mutated sequences. The observable in the model is the mutual information in the footprints, which are treated like time-series, in the sense that each position can be interpreted as a subsequent "time point". The transition probability matrix in the model captures the fact that binding sites are not at single base pairs, but instead usually made out of groups of 10-20 bases at a time. 
We use gaussian distributions to parameterize the emission probabilities, and use the "hmmlearn" Python package to fit and evaluate the models. For each information footprint we fit
a individual model, repeating the process 10 times to exlude possible diverging runs. From the 10 separately fit models, we choose the one that produces the highest log-likelihood when
the sequence is evaluated using the model. The information footprint itself is unsigned and hence, does not contain information about binding sites being repressor-like or activator-like. When a region of the promoter is identified as binding site, we look at the expression shift to identify the sign of change in expression when the site is occupied. In Figure~\ref{SI-fig:HMM} we show the result of the Hidden Markov Model for the \textit{araB} promoter for growth in arabinose, and the identified binding sites, which overlap with the annotated activator sites for AraC. In the expression shift we can see that within the binding sites, nearly every mutation leads to a decrease in expression, showing how mutations
in the binding site weaken binding of the activator and decrease expression.   

As quality control, fit the model both for the forward sequence, but also the reverse sequence, to ensure consistency. Examples are shown for the \textit{araB} promoter for growth with arabinose and the \textit{tisB} promoter for growth in glucose. Another important parameter is the width of the gaussian kernel used for smoothing. If the smoothing kernel is too narrow, binding sites are not identified as continuous units and many single positions outside actual binding sites are identified as binding sites by the model. If the kernel is too wide, binding sites merge and we lose the resolution to distinguish neighboring binding sites, as shown in Figure~\ref{SI-fig:HMM}. Thus, we choose a kernel width of $\sigma=2$ for our analysis.
\begin{figure}
    \centering
    \includegraphics[width=\linewidth]{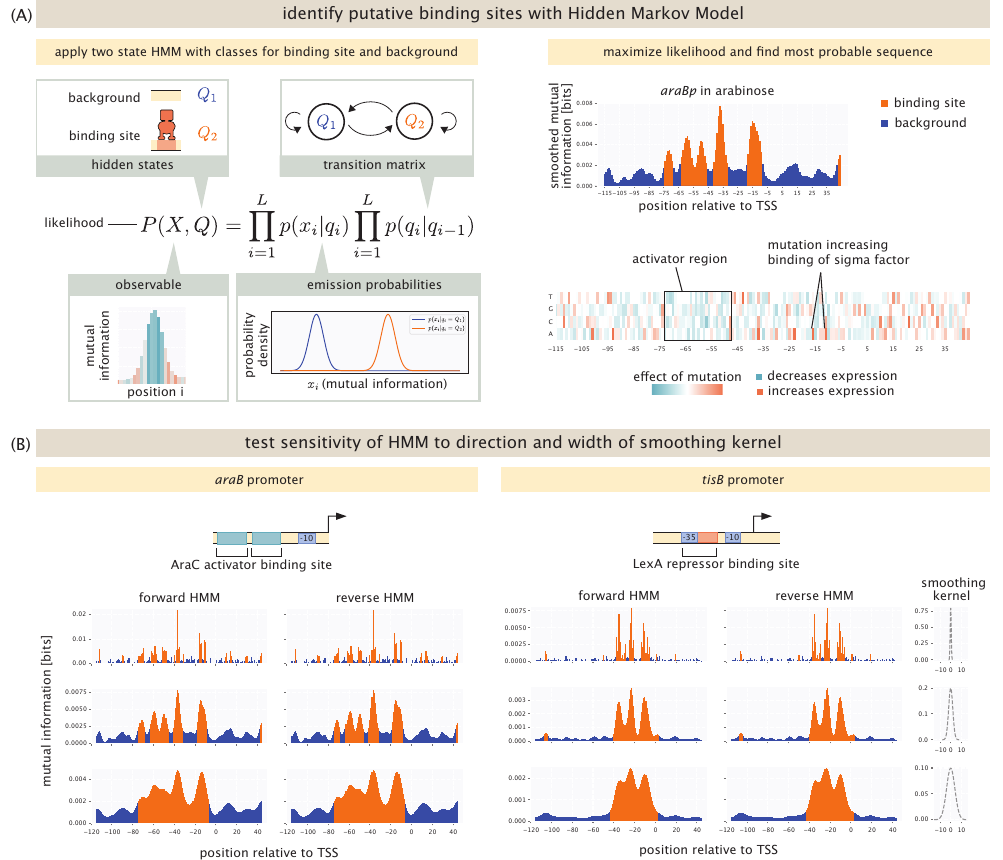}
    \caption{\textbf{Using Hidden Markov Models to find binding sites in information footprints.} (A) There are two hidden states associated with each basepair.  We identify a given basepair either as being part of a binding site or not (these sites are labeled as background). The observable used is the mutual information per position. To maximize the likelihood, a transition matrix is fit which gives the prior probability that the $i+1^{th}$ position is a certain state given the state of the $i^{th}$ position. Emission probabilities describe how likely it is to find a value for mutual information in a binding site or in background. Once the model is fit, the sequence of states which returns the maximum likelihood is returned, and binding sites are identified. The expression shift matrix is used to find if the binding site is activator-like or repressor-like. (B) Test of the Sensitivity of the Hidden Markov model by examining the consistency of the analysis when going from right to left rather than left to right and as a function of the width of the smoothing kernel.}
    \label{SI-fig:HMM}
\end{figure}

\subsubsection{{\it De novo} promoters}

As seen in the main text, one of the key and surprising outcomes of our experiments was the emergence of new transcription start sites based only upon a key driver mutation. This phenomenon was discovered in earlier experiments by the Gore Lab~\cite{Gore2018}. 
The presence of such results in our experiments led us to develop a systematic approach for identifying these new transcriptional start sites as shown in Figure~\ref{SI-fig:emerging}.  

The most useful way of visualizing the emergence of new promoters in our data is using the summary statistic we call the expression shift matrix.  In the left panel of 
Figure~\ref{SI-fig:emerging} we show an example of such an expression shift matrix in the context of the {\it araBp} promoter when grown in xylose.   As seen in the figure, there is a mutation just to the left of the -75 position in which a G is replaced by an A and for which there is a very large resulting shift in the expression.  
To investigate if this mutation indeed led to formation of a new transcription start site, we split the dataset into two groups, one that harbors the mutation at this site, and a second group that has the wild type base at that site.  At this point in the procedure, we invoke bioinformatics in the form of a model
from the Salis lab for predicting transcription rates from putative promoter sequences~\cite{lafleur2022automated}. 
As seen in the middle panel of the figure, our approach is to slide along the promoter region in single base pair increments, where for each position, we extract a block of sequence and feed into into the Salis Lab transcription rate calculator.  The result is a predicted transcription rate for every possible transcription start site for each sequence in both groups. To summarize the outcomes, we compute the average transcription rate per transcription start site across all sequences within each group.
The average predicted transcription rate is then plotted as a function of the position of our query block of sequence as shown in the right panel of the figure.   As is evident in the figure, there is significant enhancement in the predicted transcription rate that corresponds precisely with the presence of the mutation of interest.  As noted in the main text, such mutation are precisely in the -10 region of a new promoter.

\begin{figure}
    \centering
    \includegraphics[width=\linewidth]{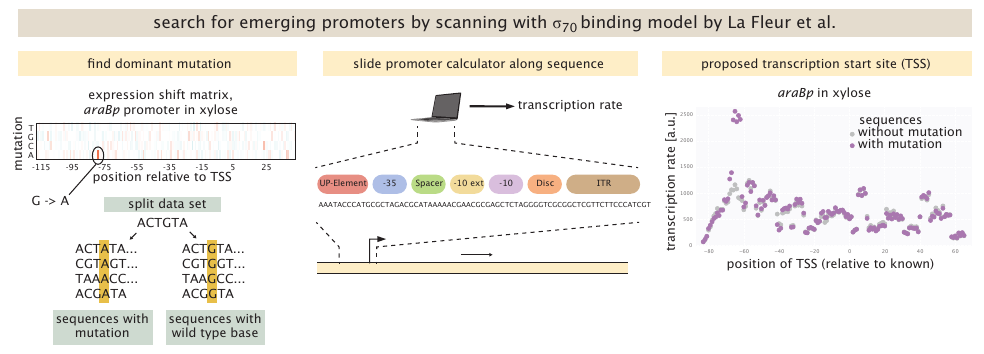}
    \caption{\textbf{Using a computational model to identify putative transcription start sites.} When a promoter is identified to contain a putative new transcription start site, the mutation with the largest expression shift is identified. The sequences for this promoter are split into a group that contain that mutation and all sequences that do not carry that mutation. The model of LaFleur et al.~\cite{lafleur2022automated} is used to predict a transcription rate for every possible transcription start site in the sequence. The model takes different elements of the $\sigma^{70}$ factor into account. We then visualize the predicted transcription rate for every possible transcription start site and look for a {\it de novo} promoter.}
    \label{SI-fig:emerging}
\end{figure}

\subsection{Computational identification of transcription factor binding partner using Tomtom}
\label{sec:theory_comp}
After generating hypotheses about putative binding sites in each promoter region, our next task is to identify which transcription factor binds to each of these binding sites. To do this, one approach is by comparing our putative binding sites with datasets of transcription factor binding sites that have been previously reported in the literature. If a query binding site shares large similarities with a known binding site, we can then hypothesize that the query binding site likely has the same binding partner as that known binding site.

To perform this motif comparison, we use the algorithm Tomtom, which identifies which known binding motifs have statistically significant similarities with a given query binding site sequence. In particular, Tomtom searched for hits for both the forward sequences of the known binding motifs as well as the reverse complement of the known binding motifs. For each query-target pair, Tomtom produces an optimal alignment and scores the overlapping region using a chosen distance metric. In our case, we choose to use Euclidean distance. Tomtom then computes a p-value under the null hypothesis that the two motifs are drawn independently from the same probability distribution, and reports a corresponding q-value that estimates the false discovery rate. We retain only motif matches that exceed a minimum q-value of 0.5 and a minimum overlap length of 15.

In this analysis, the known binding sites are downloaded from RegulonDB and Ecocyc. RegulonDB contains 4897 binding sites and EcoCyc contains 2781 binding sites. There are many duplicated or overlapping binding sites within each database and between the two databases. We first removed the duplicated entries and combined regions with more than 90\% overlap within RegulonDB and EcoCyc databases separately. This resulted in 3319 binding sites in RegulonDB and 2763 binding sites in EcoCyc. Finally, we consolidated the two databases by merging sequences with over 90\% overlap, and we have a final list of 4426 binding sites across the two databases.

After a transcription factor binding partner is found at a particular binding site, we use the sequence of that particular binding site as the query sequence for Tomtom and search it against all target sequences with a minimum q-value of 1. This allows us to generate plots to plot the distribution of transcription factor binding p-values and q-values at that particular site and visualize the p-value and q-value of the transcription factor that is determined as a hit relative the p-values and q-values of other transcription factors.

			\clearpage
\section{Compendium of All Promoters in Our Study}
\label{sec:all_promoters}
In this section we give a description of our compendium of every promoter we studied in our experiments. 
In some ways, the compendium is an organized but informal  notebook
of all the genes we considered as an aid to those who are interested in a summary of what was found for each promoter under different conditions.  In another sense, the compendium is a visual summary of the huge dataset that emerged from our study.  We also refer interested readers to the online resources
\url{http://rpdata.caltech.edu/data/interactive_footprints.html} and  \url{http://rpdata.caltech.edu/data/all_data.pdf}.

For each promoter, we show a cartoon figure of annotated binding sites and the conditions they were identified in. To set both notation and to define the visual icons used in the paper and in this section, Figure~\ref{fig:SI_legend}(A) presents an example which describes our color scheme for labeling activator binding sites, repressor binding sites, binding sites of dual function and binding sites for sigma factors.   Each of the figures in our compendium also shows the information footprints that reveal our results for these promoters in relevant growth conditions. Figure~\ref{fig:SI_legend}(B) gives an example of the way we will show our information footprints throughout the compendium.  Figure~\ref{fig:SI_legend}(C) shows one of our most useful ways of visualizing our data with a compact figure that shows for each growth condition of interest the regions of the promoter in which something interesting was found.  In this figure, we show both the emergence of {\it de novo} transcription start sites as well as putative binding sites.   

\begin{figure}
    \centering
    \includegraphics[width=0.6\linewidth]{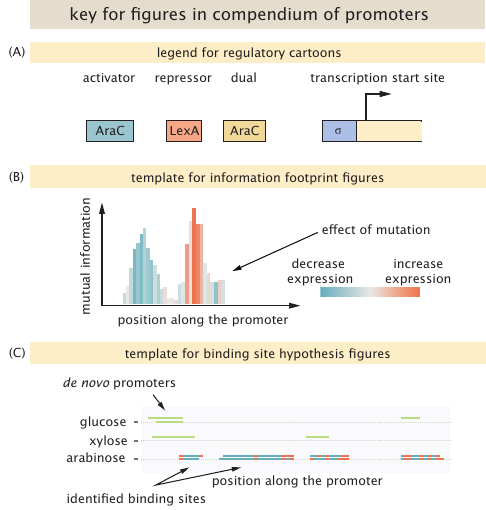}
    \caption{\textbf{Key for regulatory cartoons and information footprints.} (A) Binding sites for activators are shown using teal boxes. Repressor binding sites are shown using orange boxes. -10 and -35 regions of $\sigma^{70}$ are shown in blue boxes. Transcription start sites are indicated by arrows. (B) Information footprints are displayed as mutual information per position. For easier interpretation, each position is averaged with its two neighbors on each side. Each position is colored by the average expression shift at that position, indicating if a mutation decreases expression on average (shown in teal, indicative of a binding site for an activator) or increases expression on average (shown in orange, indicative of a repressor). (C) For each promoter where regions in information footprints were identified as putative binding sites, we show a figure that highlights such regions. Each base is colored by being either repressor-like (increasing expression when mutated), in red, or activator-like (decreasing expression when mutated), in blue. If a footprint was classified to contain a {\it de novo} promoter, the bases are highlighted in green instead.}
    \label{fig:SI_legend}
\end{figure}

In this long compendium, for each promoter, we briefly review the literature to identify binding sites that have been previously annotated  and judge if the binding sites were identified in our experiments.   Previously annotated binding sites from the literature can escape detection in our experiments for a variety of reasons, e.g., the growth conditions in which the original discovery of the binding site was made are not included in the set of conditions that we used.  As a result,  the transcription factor of interest might not bind at all in our experiments.   It is also possible that binding sites were identified in previous work using strains or conditions in which there was over-expression of a transcription factor. Since our \textit{E. coli} strain is not over-expressing any transcription factors, the effect of lower transcription factor copy numbers might 
result in circumstances in which binding is not measurable.  Our own earlier theory-experiment dialogue has used transcription factor copy number and plasmid copy number as a key tuning variable that can lead to 1000-fold changes in gene expression, so it is clear that copy number effects are crucial and can make the difference between a measurable effect and not~\cite{garcia2011quantitative, brewster2014transcription}.   Transcription factors that regulate their promoter by interacting with other binding sites through action at a distance, e.g. by DNA looping, are also hard to identify, since the additional binding sites required for regulation might not be included in our reporter construct, as we only study a 160 base pair region around the transcription start site.

We are cognizant that this is an extremely long compendium.  Our reason for including it is driven by a philosophical conundrum of this era of big data in biology.  How do we find ways to talk about our data other than in giant spreadsheets?  We decided here to make the specific details of every promoter available for those who might have interest in a given promoter.
The full compendium can be found under this link: \url{http://rpdata.caltech.edu/data/reg-seq_compendium.pdf}.
            \clearpage
\section{SI Tables and Figures}


\begin{table}[ht!]
\centering
    \begin{tabular}{|c c|}
    \hline
    \textbf{Growth Condition of Lysate} & \textbf{Final Concentration of Supplement(s) Added}\\
    \hline\hline
    M9-Glucose & 0.5\% Glucose, 1 mM cAMP\\
    \hline
    M9-Arabinose & 0.5\% Arabinose, 1 mM cAMP\\
    \hline
    M9-Xylose & 0.5\% Xylose \\
    \hline
    LB & None Added\\
    \hline
    Stationary Phase (1d) & 1 mM cAMP\\
    \hline
    Leucine & 10 mM Leucine \\
    \hline
    Phenazine Methosulfate & 100  \textmu M Phenazine Methosulfate\\
    \hline
    2,2 Dipyridyl & 5 mM 2,2 Dipyridyl \\
    \hline
    Gentamicin & 5 mg/l Gentamicin\\
    \hline
    Copper Sulfate & 2 mM Copper Sulfate\\
    \hline
    Heat Shock & None added\\
    \hline
    H$_2$O$_2$ & 2.5 mM H$_2$O$_2$\\
    \hline
    High Osmolarity (LB + 750 mM NaCl) & 200 mM NaCl, 150 mM KC1, 1 mM cAMP\\ 
    \hline
    \end{tabular}
    \caption{\textbf{Growth Conditions for lysates used for mass spectrometry and associated supplements added.} Notes: A supplement of 1 mM cAMP, a known co-factor for CRP binding, was used for our initial mass spectrometry runs of M9-glucose, M9-arabinose, stationary phase, and high osmolarity growth conditions, where we expected there to be annotated (e.g. \textit{mglB} and \textit{araB}) or predicted (e.g. \textit{yadI}) binding sites of CRP. Since no significant enrichment for CRP was found for any of these runs, cAMP was not added to lysates for the other growth conditions. For high osmolarity, we based the concentration of supplemented salts on Figure 3 of Shabala et.al \cite{shabala2009ion}, which show measurements of intracellular salt concentrations for growth in external media of varying NaCl concentrations.} 
    \label{tab:lysate_supplement}
\end{table}

\begin{table}[ht!]
\centering
    \begin{tabular}{|l|l|l|l|}
    \hline
    \textbf{Time} & \textbf{Duration} & \textbf{Flow (nl/min)} & \textbf{\%B}\\
    \hline\hline
    0:00 & 0:00 & 300 & 2\\
    \hline
    7:30 & 7:30 & 300 & 6\\
    \hline
    90:00 & 72:00 & 300 & 25\\
    \hline
    120:00 & 30:00 & 300 & 40\\
    \hline
    121:00 & 1:00 & 300 & 98\\
    \hline
    130:00 & 9:00 & 300 & 98\\
    \hline
    \end{tabular}
    \caption{\textbf{Liquid chromatography gradient parameters for mass spectrometry.} Mobile Phase A contains 0.2\% formic acid, 2\% acetonitrile, and 97.8\% water. Mobile Phase B contains 0.2\% formic acid, 80\% acetonitrile, and 19.8\% water.} 
    \label{tab:LC_table}
\end{table}

\begin{table}
\centering
    \begin{tabular}{|l l|}
    \hline
    \textbf{Global Settings} &  \\
    \hline\hline
    Ion source type & NSI\\
    \hline
    Spray voltage & 1500 V\\
    \hline
    Ion transfer tube temperature & 275 C\\
    \hline
    Polarity & Positive\\
    \hline\hline
    \textbf{MS1 Scan Settings} &  \\
    \hline
    Resolution & 120000\\
    \hline
    Normalized AGC target & 250\\
    \hline
    Maximum IT & 50 msec\\
    \hline
    Scan range & 375-1600 m/z\\
    \hline\hline
    \textbf{MS2 Scan Settings} & \\
    \hline
    Resolution & 50000\\
    \hline
    Normalized AGC target & Standard\\
    \hline
    Maximum IT & Dynamic\\
    \hline
    Loop time & 3 sec\\ 
    \hline
    Isolation window & 0.7 m/z\\
    \hline
    NCE & 35\\
    \hline
    Spectrum data type & Centroid\\
    \hline
    Fixed first mass & 110 Z\\
    \hline
    \end{tabular}
    \caption{\textbf{Mass spectrometry scan settings for TMT samples} } 
    \label{tab:MS_table}
\end{table}

\begin{table}
\centering
    \begin{tabular}{|l l|}
    \hline
    \textbf{\textbf{Sequest HT Settings}} &  \\
    \hline\hline
    Enzyme name & Trypsin (Full)\\
    \hline
    Max. missed cleavage & 2\\
    \hline
    Min. peptide length & 6\\
    \hline
    Max. peptide length & 144\\
    \hline
    Precursor mass tolerance & 10 ppm  \\
    \hline
    Fragment mass tolerance & 0.02 Da\\
    \hline
    Max. equal modification & 3\\
    \hline
    Dynamic modification & Oxidation/ +15.995 Da (M)\\
    \hline
    Dynamic modification (peptide terminus) & Acetyl/ + 42.011 Da (N-Terminal)\\
    \hline
    Dynamic modification (peptide terminus) & Met-loss/ - 131.040 Da (M)\\
    \hline
    Dynamic modification (peptide terminus) & Met-loss+Acetyl/ - 89.030 Da (M)\\
    \hline
    Static modification (peptide terminus) & TMTpro/ + 304.027 Da (N-Terminal)\\
    \hline
    Static modification & TMTpro/ + 304.027 Da (N-Terminal)\\
    \hline
    Static modification & Carbamidomethyl/ + 57.021 Da (C)\\ 
    \hline\hline
    \textbf{Percolator} & \\
    \hline
    Target/Decoy selection & Concatenated\\
    \hline
    Validation based on & q-Value\\
    \hline
    Target FDR (Strict) & 0.01\\
    \hline
    Target FDR (Relaxed) & 0.05\\
    \hline
        \end{tabular}
    \caption{\textbf{Search parameters for Protein Discoverer 2.5} } 
    \label{tab:PD_table}
\end{table}

\begin{figure}
    \centering
    \includegraphics[width=\linewidth]{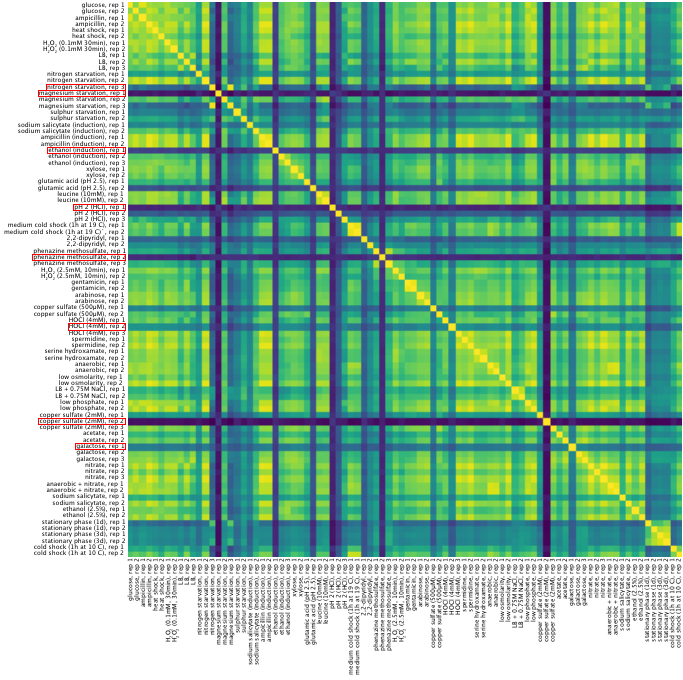}
    \caption{\textbf{Correlation between experiments.} The pearson correlation coefficient computed between all pairs of 
    experiments. Experiments removed from further analysis are boxed in red.}
    \label{SI-fig:gc_corr}
\end{figure}

\begin{figure}
    \centering
    \includegraphics[width=\linewidth]{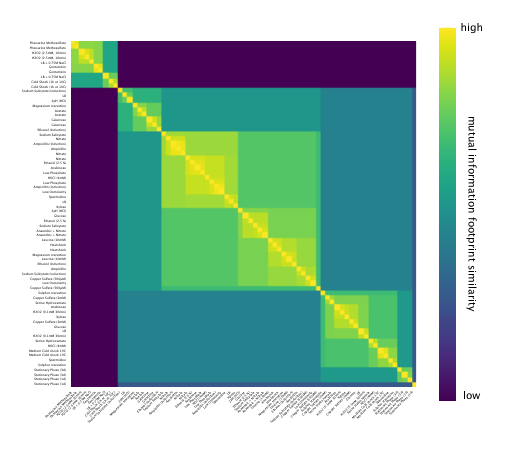}
    \caption{\textbf{The fully annotated clustering figure of Figure~\ref{condition_clustering_examples1}}. Conditions with multiple annotations are different replicates for the same condition.}
    \label{SI-fig:full_tsB_cluster}
\end{figure}
\begin{figure}
    \centering
    \includegraphics[width=\linewidth]{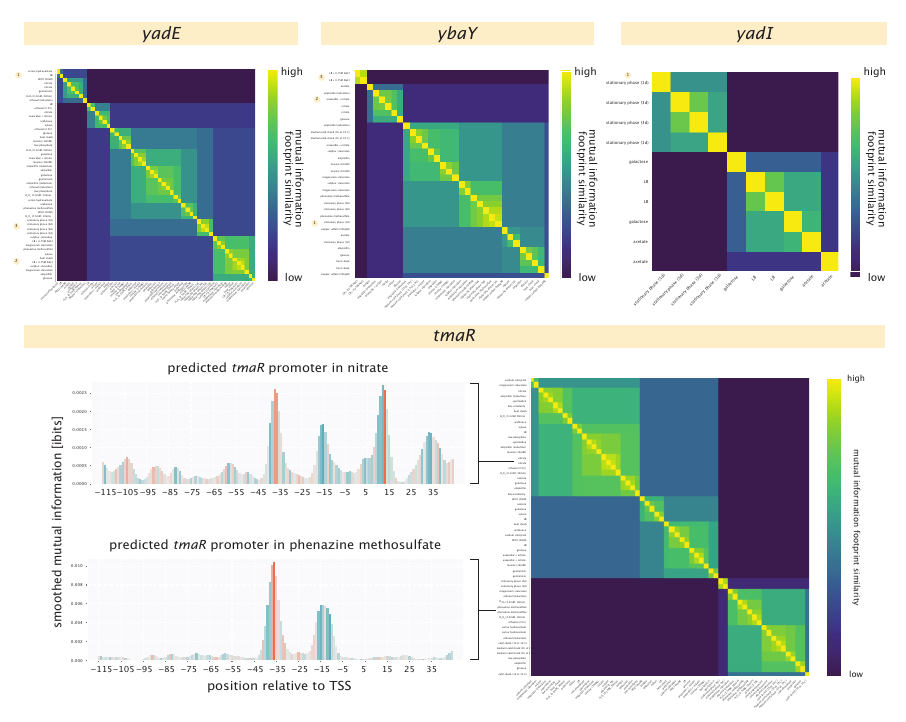}
    \caption{\textbf{Clustering of conditions for {\it yadE}, {\it ybaY}, {\it yadI} and {\it tmaR}.} Results of hierarchical clustering are displayed. Footprints are shown for the genes in the upper panel in Figure~\ref{yome}. Numbers indicate which that there is a footprint shown for this specific condition.}
    \label{SI-fig:clustering_examples}
\end{figure}

\begin{figure}
    \centering
    \includegraphics[width=0.49\linewidth]{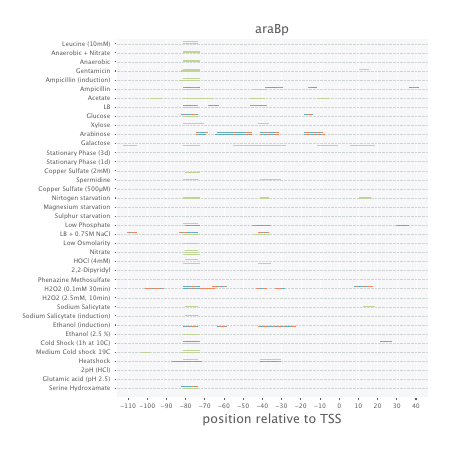}
    \caption{\textbf{Inferred putative binding sites for the {\it araB} promoter.}}
    \label{fig:SI_araB_BS}
\end{figure}

 \begin{figure}
     \centering
     \includegraphics[width=0.75\linewidth]{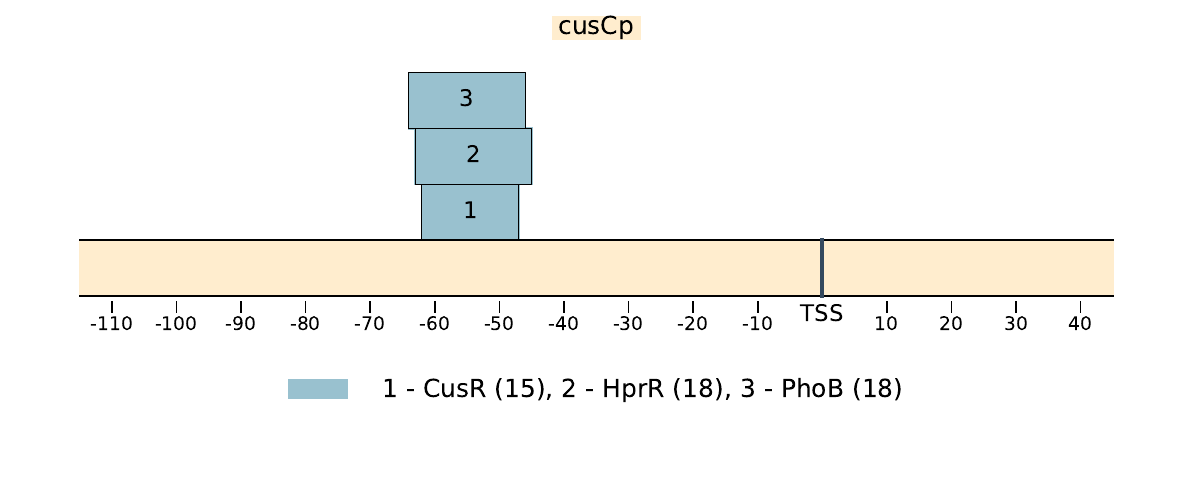}
     \includegraphics[width=0.49\linewidth]{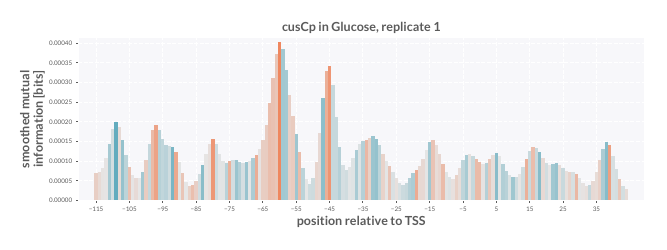}
     \includegraphics[width=0.49\linewidth]{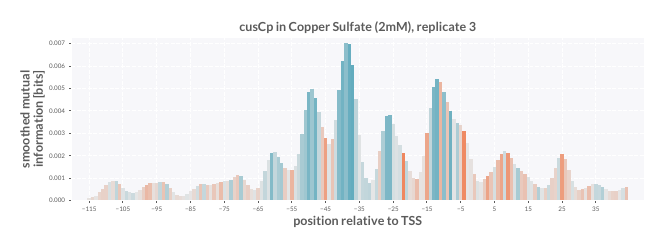}
     \includegraphics[width=0.49\linewidth]{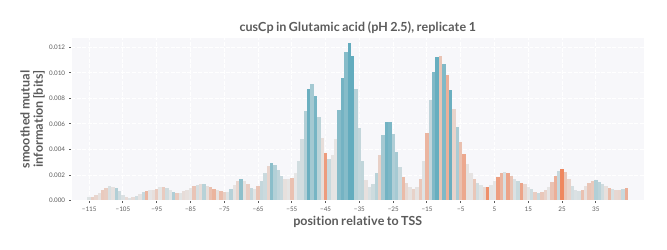}
     \caption{\textbf{Annotated regulation and information footprints for the \textit{cusCFBA} promoter}. Footprints are shown for growth in minimal media with glucose, when induced with coppers sulfate and for shock in minimal media with glucose at 2.5 pH and with 1 mM of glutamic acid.}
     \label{fig:SI_cusC}
 \end{figure}

 \begin{figure}
    \centering
    \includegraphics[]{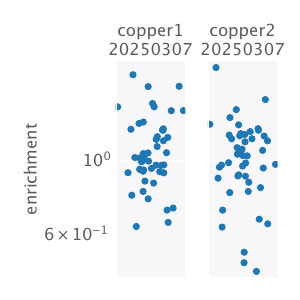}
    \caption{\textbf{Mass-spec results for {\it cusC}.} Enrichment for transcription factors is shown for mass-spec experiments where lysates from cells were used that were induced with 2 mM of copper sulfate for 1h after reaching exponential phase.}
    \label{fig:SI_cusC_mass-spec}
\end{figure}

\begin{figure}
    \centering
    \includegraphics[]{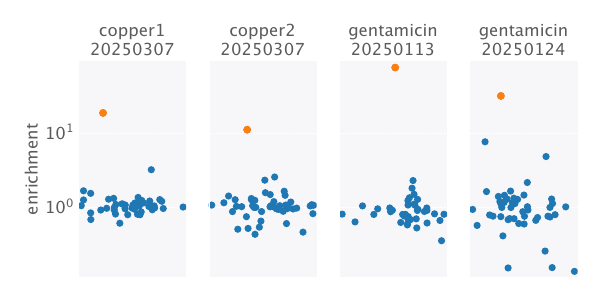}
    \caption{\textbf{Mass-spec results for {\it cpxR}.} Cells were grown and induced either with gentamicin or copper sulfate. Highlighted is CpxR in orange.}
    \label{fig:SI_cpxR_mass-spec}
\end{figure}

\begin{figure}
    \centering
    \includegraphics[width=0.5\linewidth]{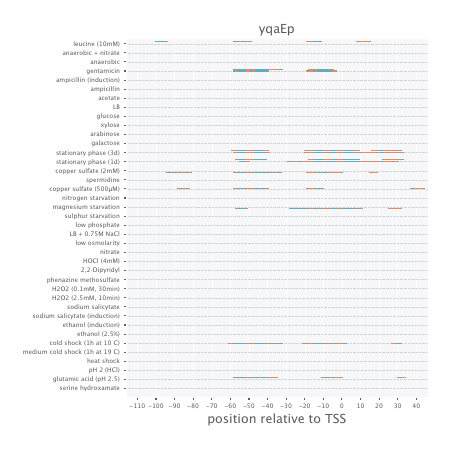}
    \caption{\textbf{Inferred putative binding sites for the {\it yqaE} promoter.}}
    \label{fig:SI_yqaE_BS}
\end{figure}

\begin{figure}
    \centering
    \includegraphics[width=0.75\linewidth]{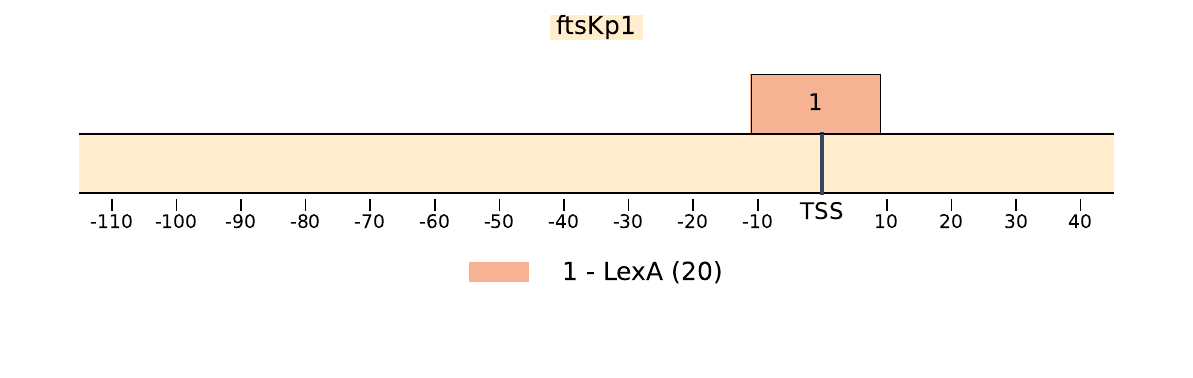}
    \includegraphics[width=0.49\linewidth]{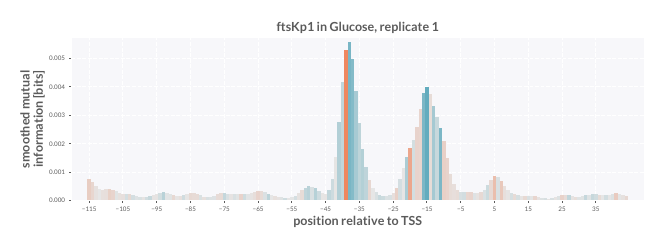}
    \includegraphics[width=0.49\linewidth]{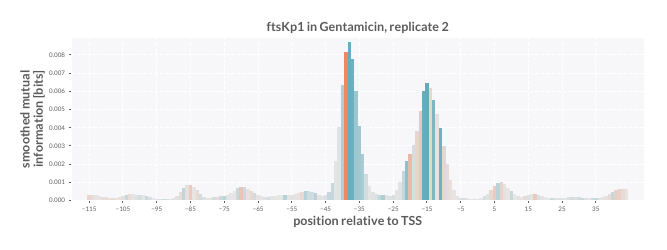}
    \includegraphics[width=0.49\linewidth]{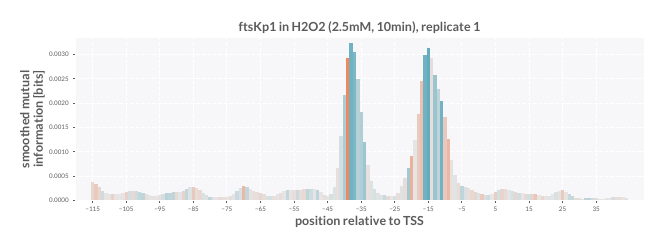}
    \caption{\textbf{Annotated regulation and information footprints for \textit{ftsK} promoter \textit{ftsKp1}}. Footprints are shown for growth in minimal media with glucose, and induction with gentamicin or hydrogen peroxide.}
    \label{fig:SI_ftsK}
\end{figure}

\begin{figure}
    \centering
    \includegraphics[width=0.75\linewidth]{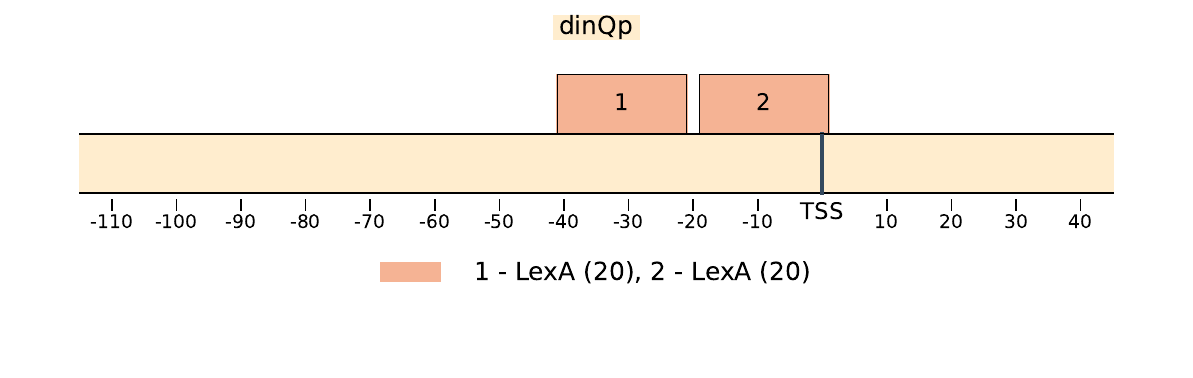}
    \includegraphics[width=0.49\linewidth]{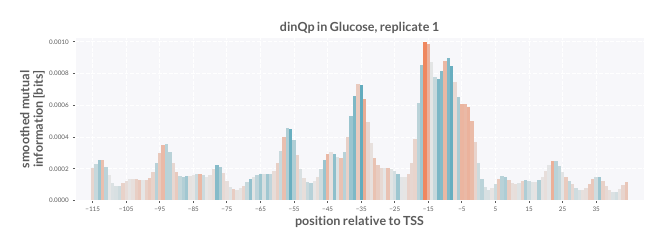}
    \includegraphics[width=0.49\linewidth]{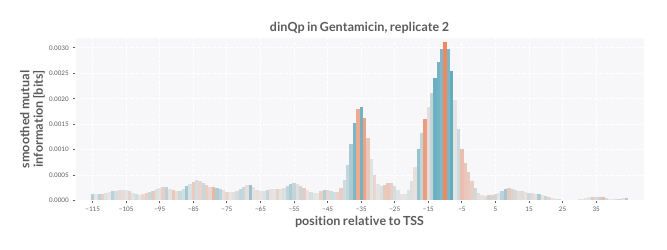}
    \includegraphics[width=0.49\linewidth]{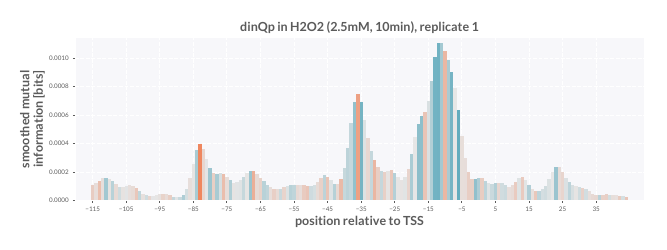}
    \caption{\textbf{Information footprints for the \textit{dinQ} promoter.} Footprints are shown for growth in minimal media with glucose, induction with gentamicin and induction with hydrogen peroxide.}
    \label{fig:SI_dinQ}
\end{figure}

\begin{figure}
    \centering
    \includegraphics[width=0.75\linewidth]{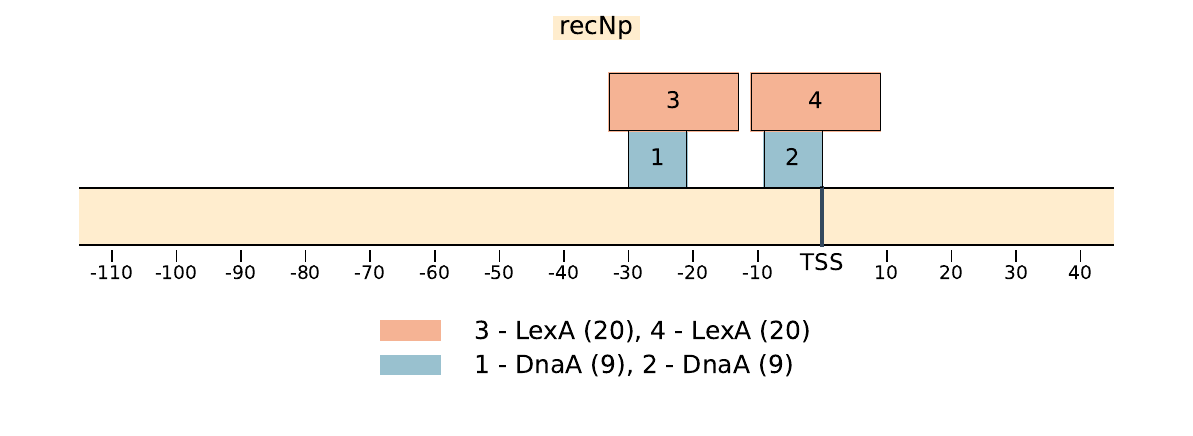}
    \includegraphics[width=0.49\linewidth]{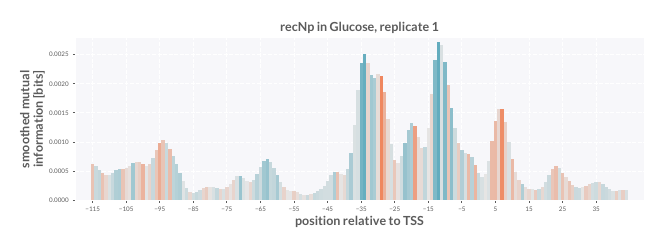}
    \includegraphics[width=0.49\linewidth]{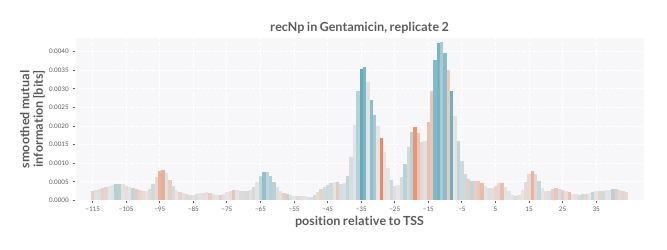}
    \includegraphics[width=0.49\linewidth]{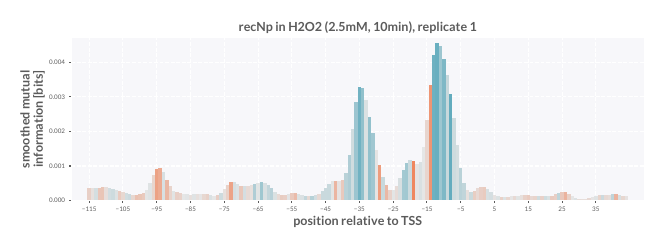}
    \caption{\textbf{Annotated regulation and information footprints for the {\it recN} promoter.} Footprints are shown for growth in minimal media with glucose, induction with gentamicin, and induction with hydrogen peroxide.}
    \label{fig:SI_recN}
\end{figure}

\begin{figure}
     \centering
     \includegraphics[width=0.75\linewidth]{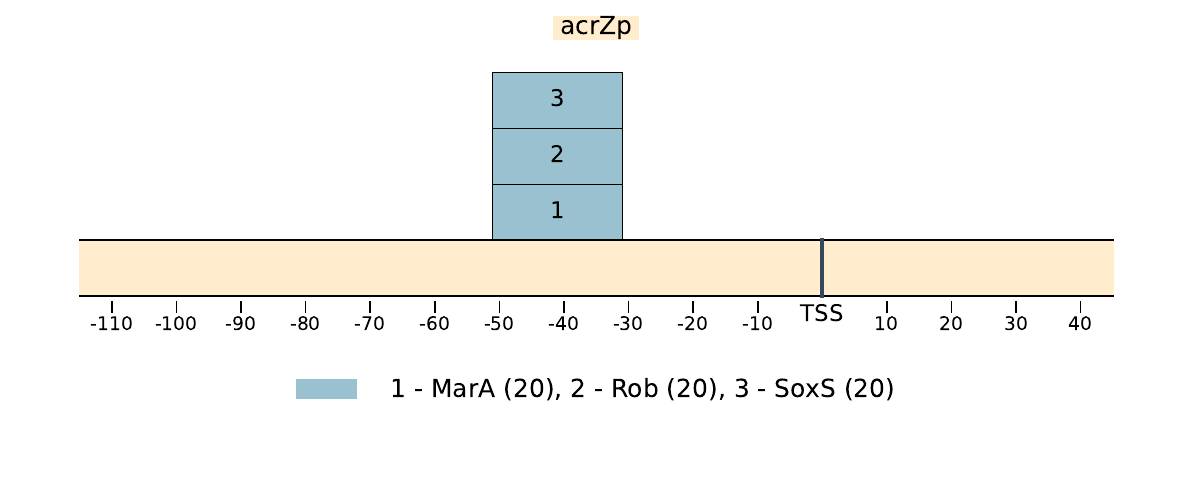}
     \includegraphics[width=0.49\linewidth]{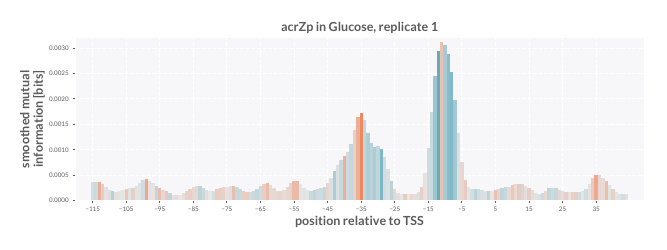}
     \includegraphics[width=0.49\linewidth]{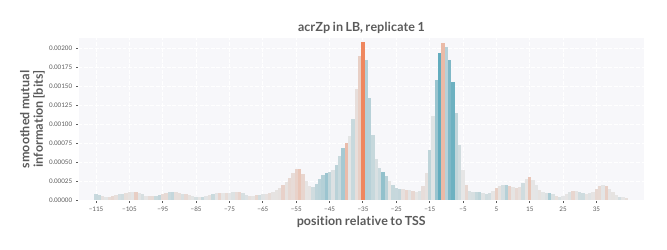}
     \includegraphics[width=0.49\linewidth]{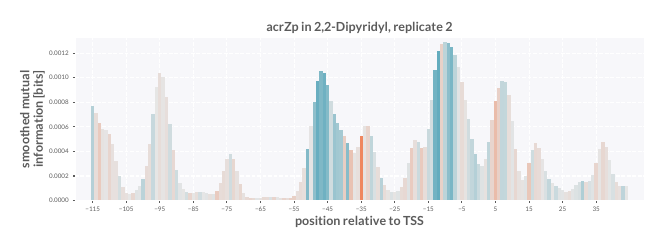}
     \includegraphics[width=0.49\linewidth]{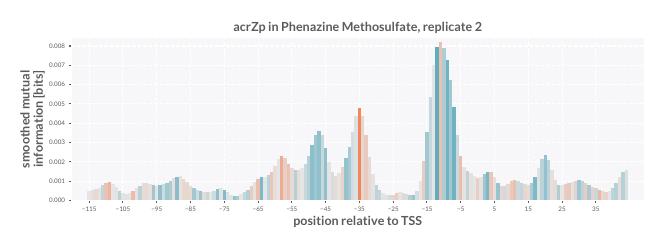}
     \includegraphics[width=0.49\linewidth]{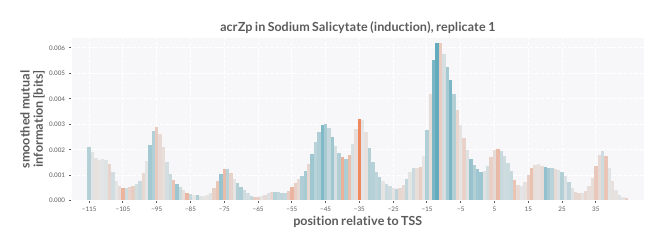}
     \caption{\textbf{Information footprints for the \textit{acrZ} promoter.} Footprints are shown for minimal media with glucose, LB, induction with 2,2-dipyridyl, induction with phenazine methosulfate, and induction with sodium salicylate.}
     \label{fig:SI_acrZ}
 \end{figure}

 \begin{figure}
    \centering
    \includegraphics[width=0.5\linewidth]{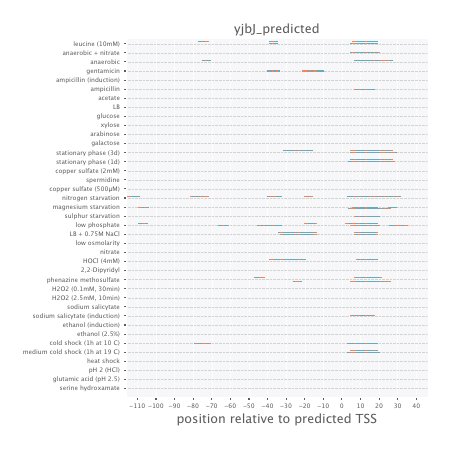}
    \caption{\textbf{Inferred putative binding sites for the predicted promoter for {\it yjbJ}.}}
    \label{fig:SI_yjbJ_BS}
\end{figure}

\begin{figure}
    \centering
    \includegraphics[width=0.5\linewidth]{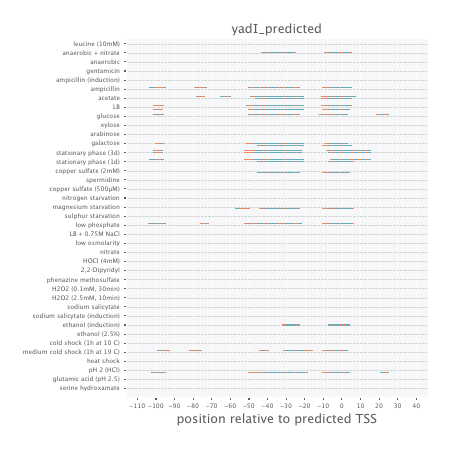}
    \caption{\textbf{Inferred putative binding sites for the predicted {\it yadI} promoter.}}
    \label{fig:SI_yadI_BS}
\end{figure}

\begin{figure}
    \centering
    \includegraphics[]
    {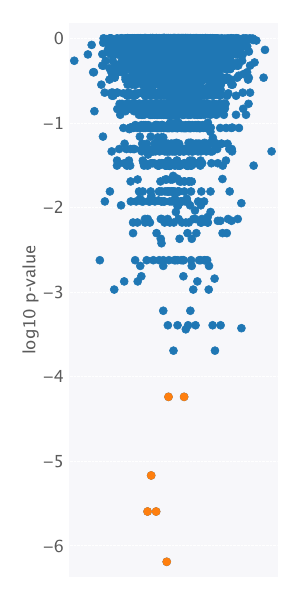}
    \caption{\textbf{P-values for known binding sites matching with the putative activator site of {\it yadI}.}  The top six hits for CRP are highlighted in orange.}
    \label{yadI_tomtom}
\end{figure}

\begin{figure}
     \centering
     \includegraphics[width=0.49\linewidth]{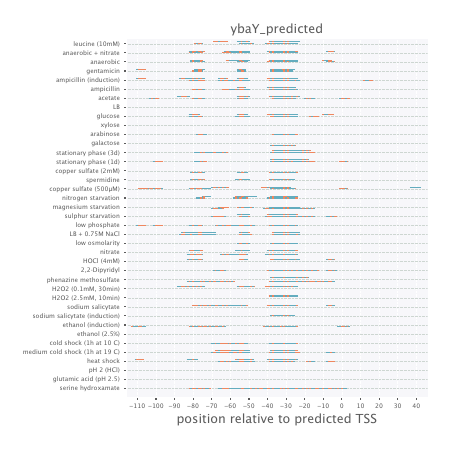}
     \caption{\textbf{Inferred putative binding sites for {\it ybaY} promoters.}}
     \label{fig:SI_ybaY_BS}
\end{figure}

\begin{figure}
    \centering
    \includegraphics[]{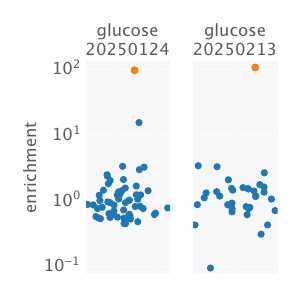}
    \includegraphics[]{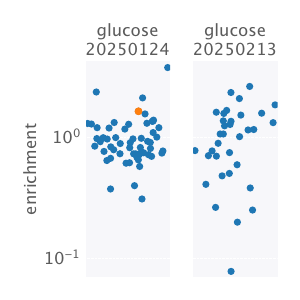}
    \includegraphics[]{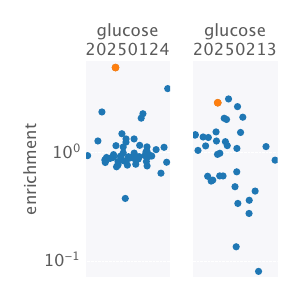}
    \caption{\textbf{Mass-spec results for {\it ybiY-ybiW.}} Left to right: {\it ybiY1} (covering -50 to +30), {\it ybiY2} (-50 to -5), {\it ybiY3} (-5 to 30). Highlighted in orange is the transcription factor YciT.}
    \label{fig:SI_ybiW_mass-spec}
\end{figure}

\begin{figure}
    \centering
    \includegraphics[]{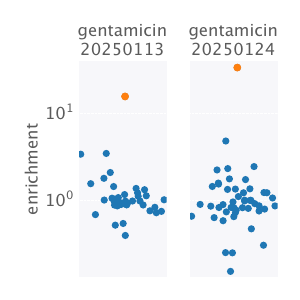}
    \caption{\textbf{Mass-spec results for {\it intE-xisE-ymfH}.} Highlighted in orange is the transcription factor YhaJ.}
    \label{fig:SI_intE_mass-spec}
\end{figure}

\begin{figure}
    \centering
    \includegraphics[width=0.75\linewidth]{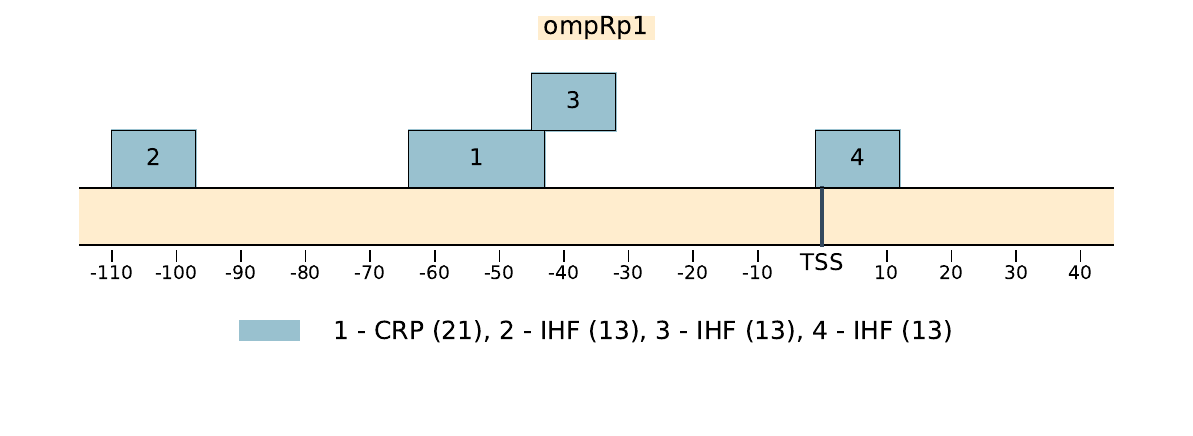}
    \includegraphics[width=0.75\linewidth]{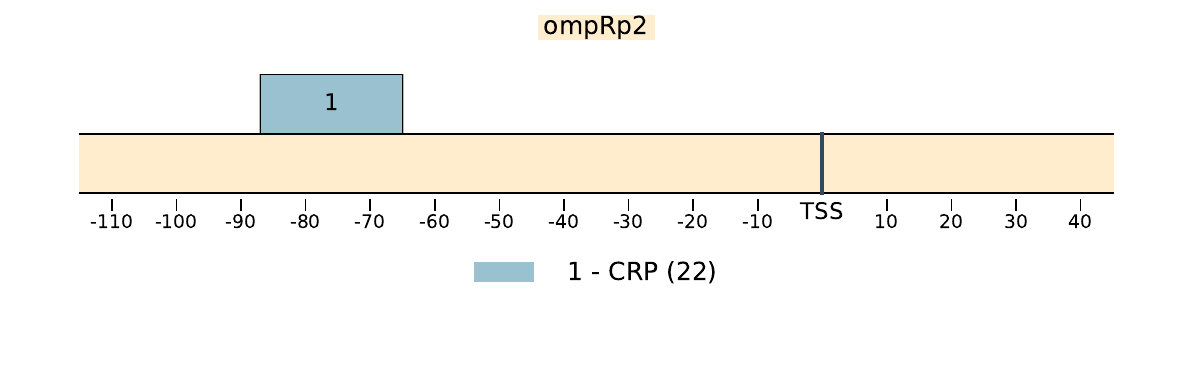}
    \includegraphics[width=0.49\linewidth]{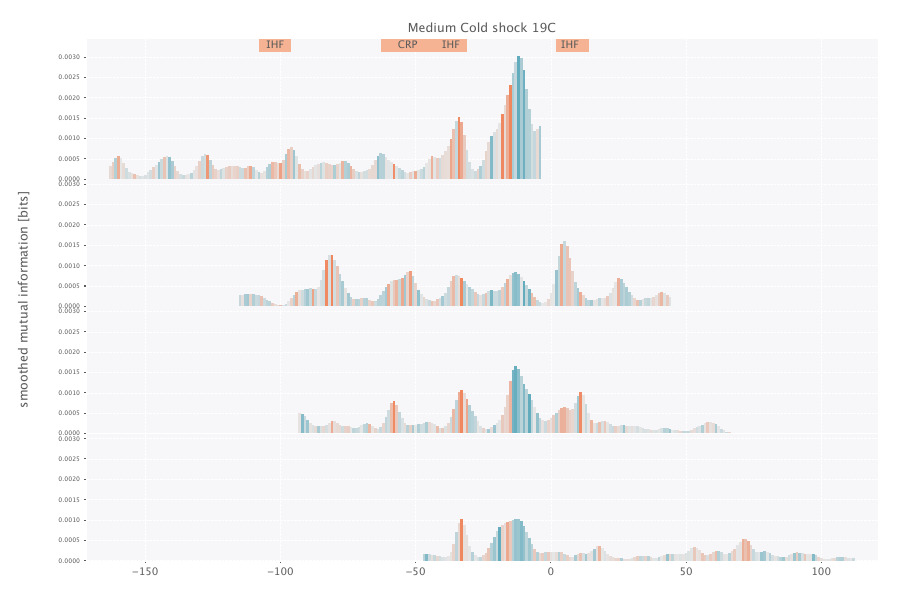}
    \includegraphics[width=0.49\linewidth]{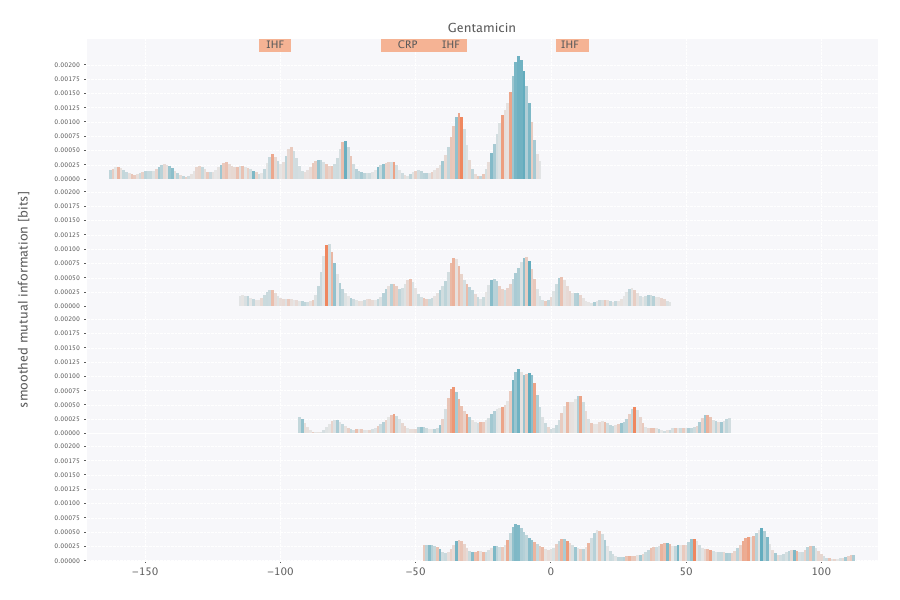}
    \includegraphics[width=0.49\linewidth]{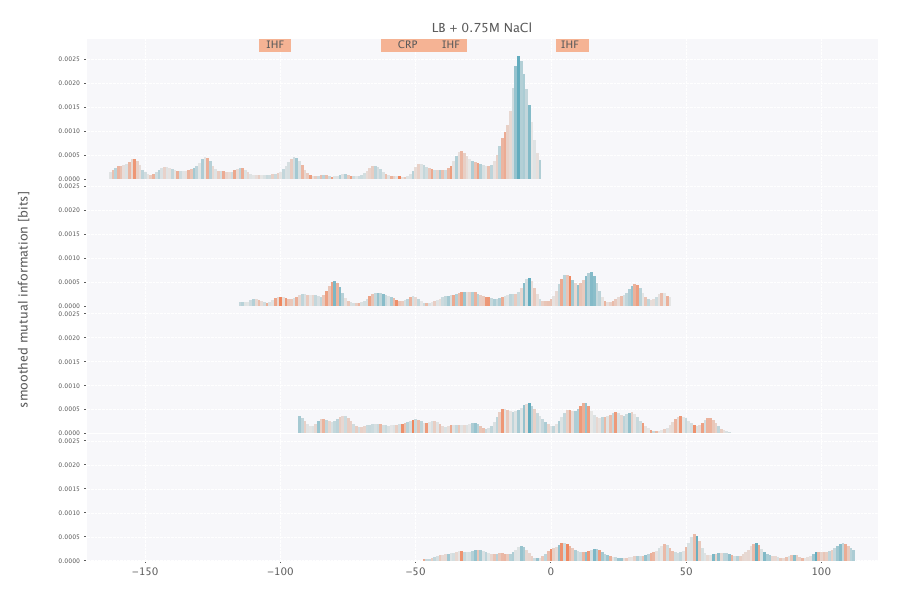}
    \includegraphics[width=0.49\linewidth]{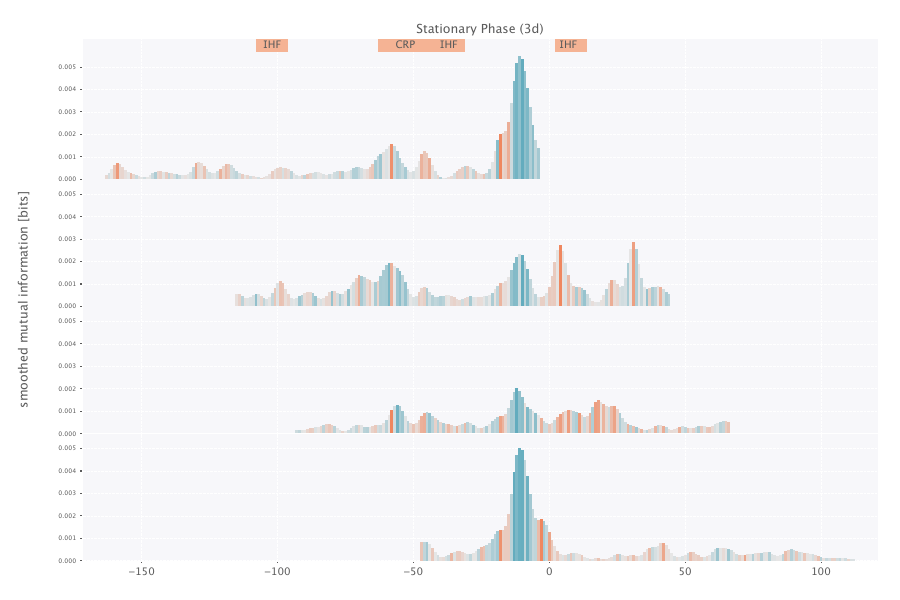}
    \caption{\textbf{Annotated regulation and information footprints for the \textit{ompR} promoters.} Footprints are shown for all four promoters, which are aligned such that 0 is the location of the transcription start site off {\it ompRp1}. Conditions shown are cold shock at 19C, induction with gentamicin, shock in LB with high salt concentration and stationary phase after 72h.}
    \label{fig:ompR}
\end{figure}

\begin{figure}
    \centering
    \includegraphics[width=0.75\linewidth]{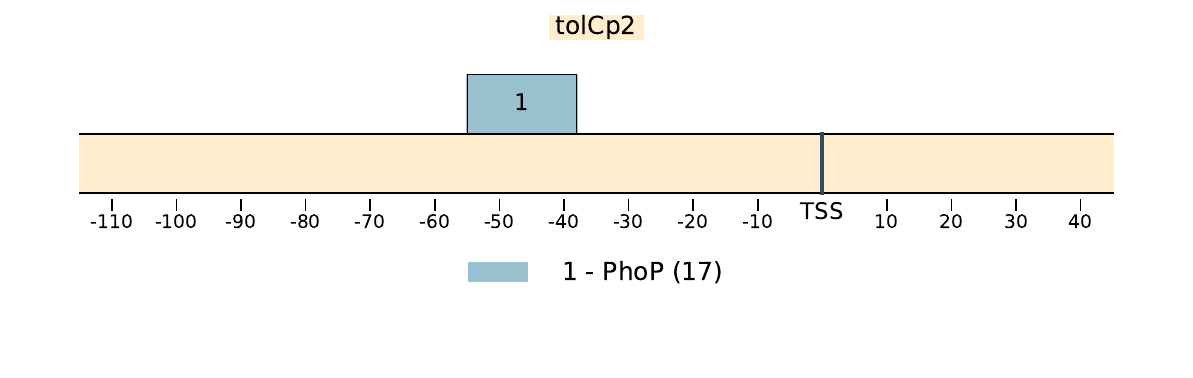}
    \includegraphics[width=0.75\linewidth]{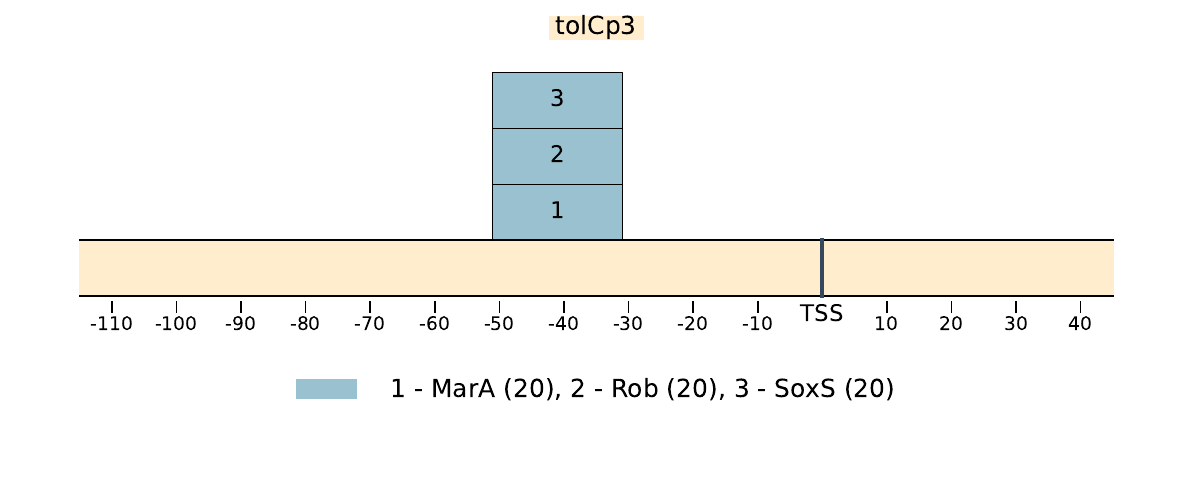}
    \includegraphics[width=0.49\linewidth]{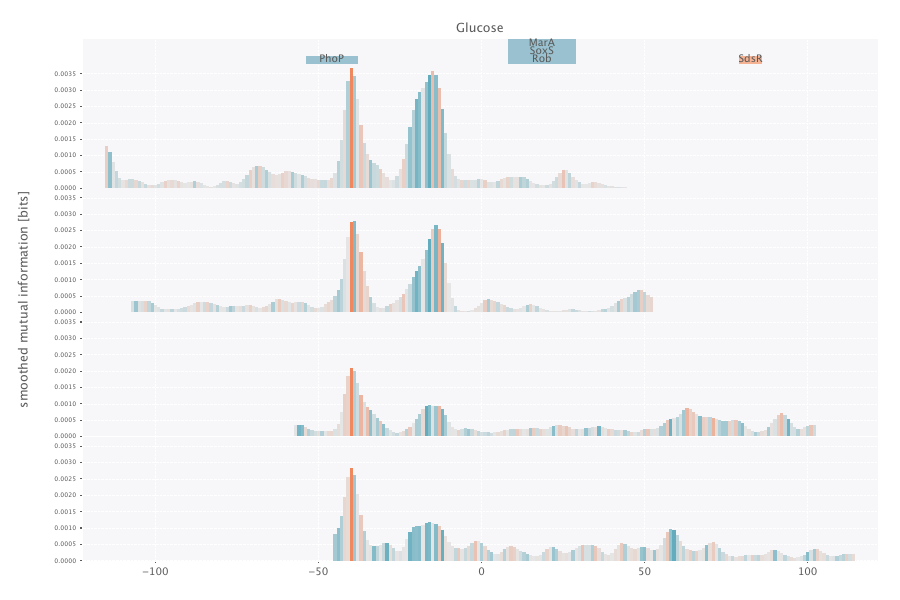}
    \includegraphics[width=0.49\linewidth]{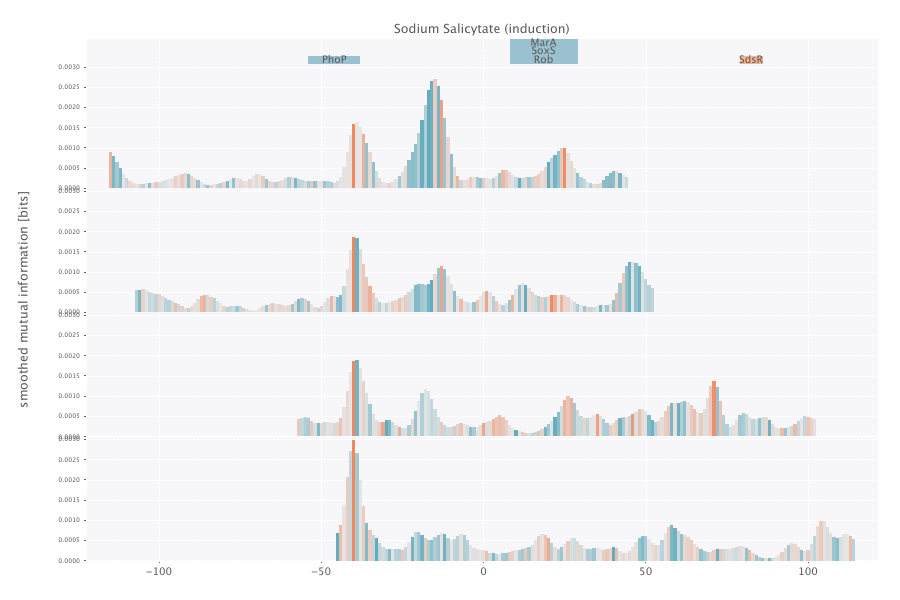}
    \includegraphics[width=0.49\linewidth]{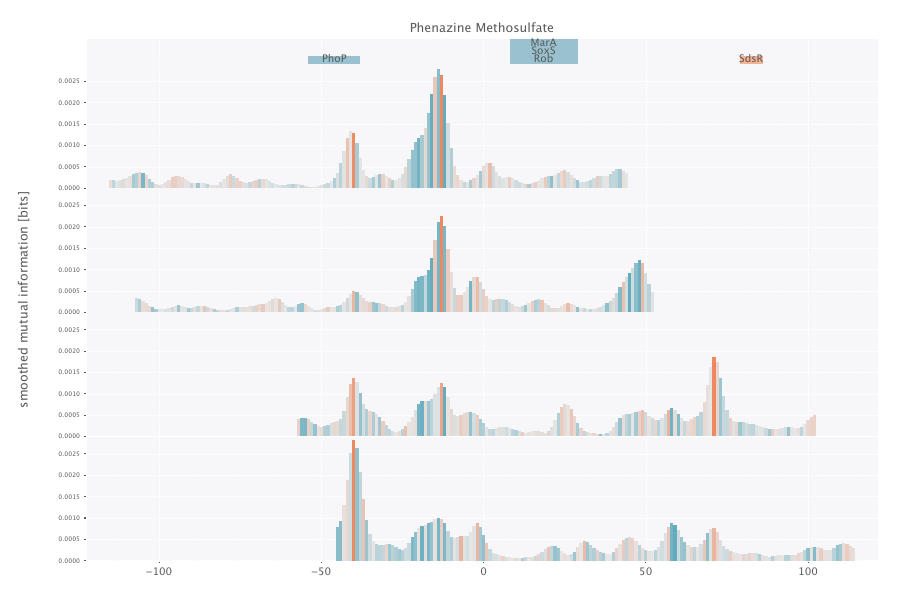}
    \includegraphics[width=0.49\linewidth]{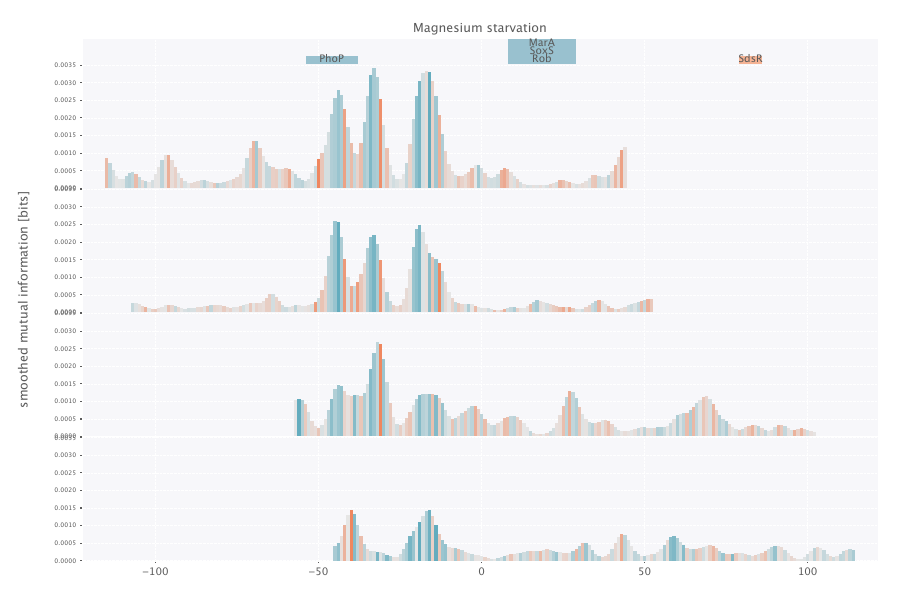}
    
    \caption{\textbf{Annotated regulation and information footprints for the \textit{tolC} promoters.} Footprints are shown for all four promoters, which are aligned such that 0 is the location of the transcription start site off {\it tolCp1}. Conditions shown are growth in minimal media with glucose, induction with sodium salicylate or phenazine methosulfate and for magensium starvation.}
    \label{fig:SI_tolC}
\end{figure}

\begin{figure}
    \centering
    \includegraphics[width=0.75\linewidth]{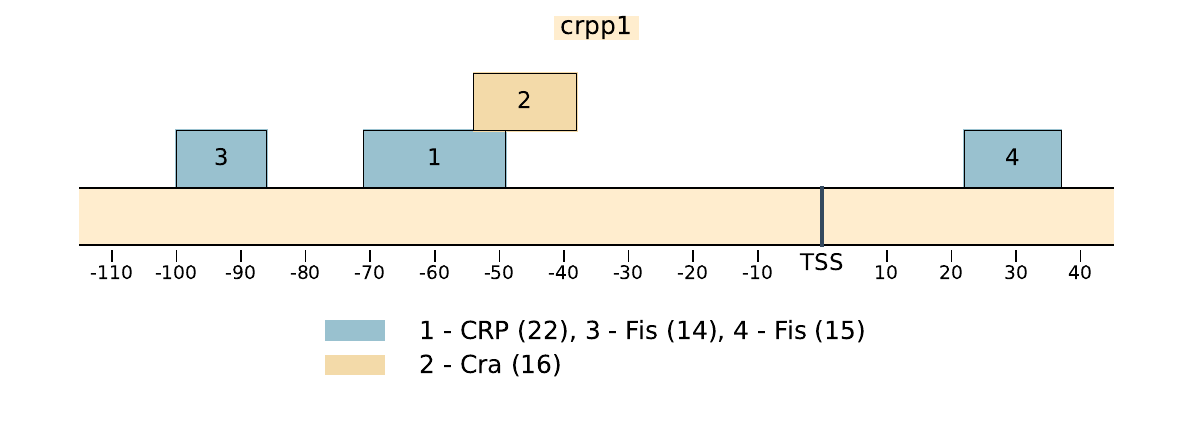}
    \includegraphics[width=0.49\linewidth]{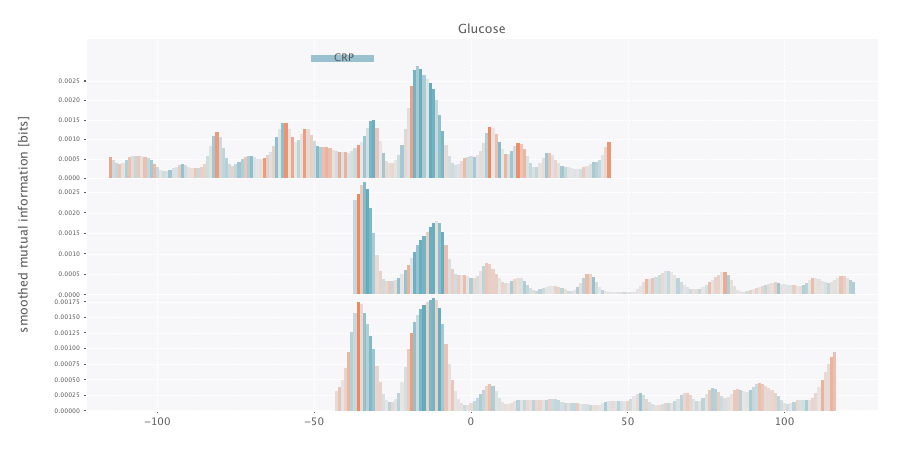}
    \includegraphics[width=0.49\linewidth]{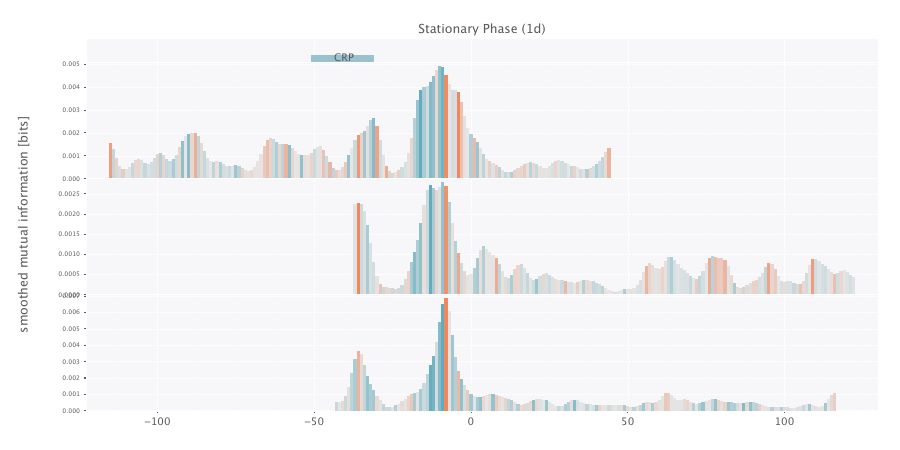}
    \includegraphics[width=0.49\linewidth]{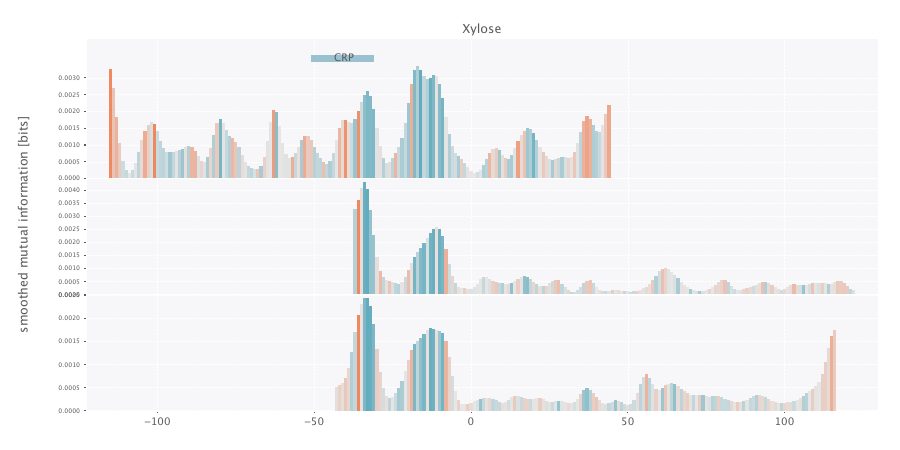}
    \includegraphics[width=0.49\linewidth]{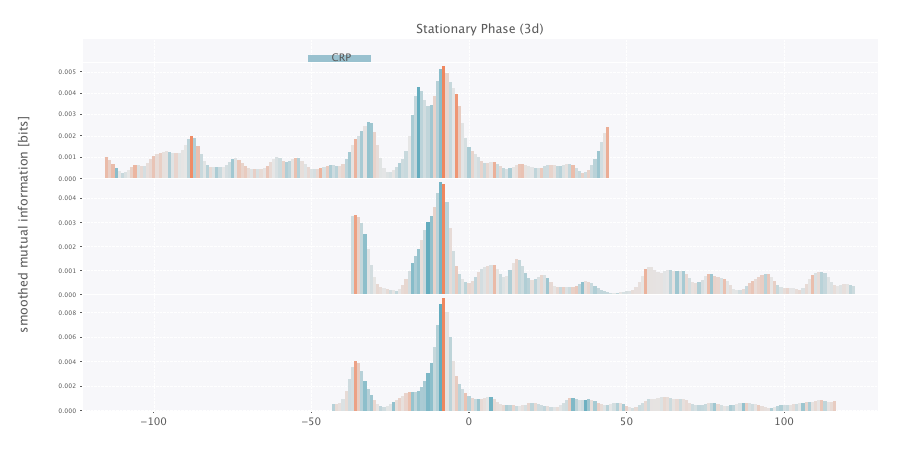}

    \caption{\textbf{Annotated regulation and information footprints for the \textit{crp} promoters.} Footprints are shown for all three annotated {\it crp} promoters (top {\it crpp1}, middle {\it crpp2}, bottom {\it crpp3}) in minimal media with glucose or xylose as carbon sourses, and for stationary phase after 24h or 72h.}
    \label{fig:crp1}
\end{figure}

\begin{figure}
    \centering
    \includegraphics[width=0.75\linewidth]{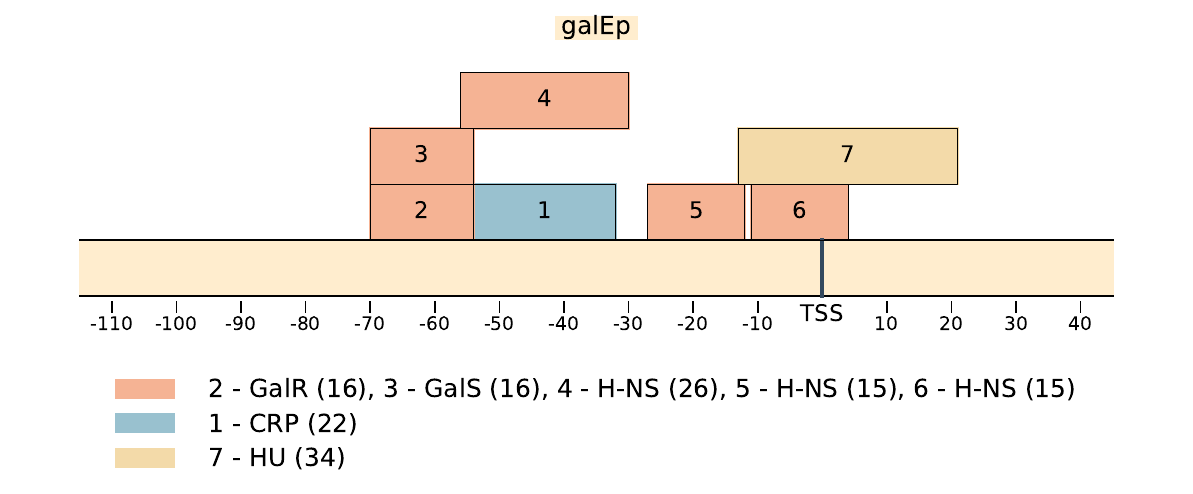}
    \includegraphics[width=0.49\linewidth]{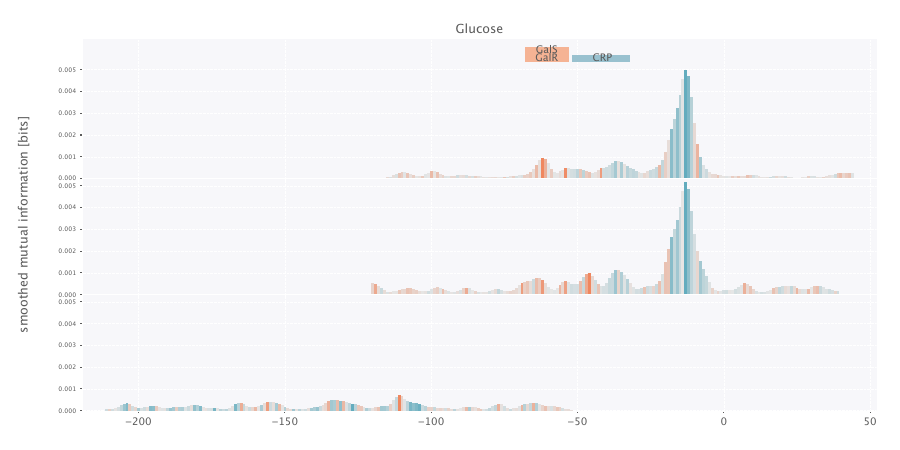}
    \includegraphics[width=0.49\linewidth]{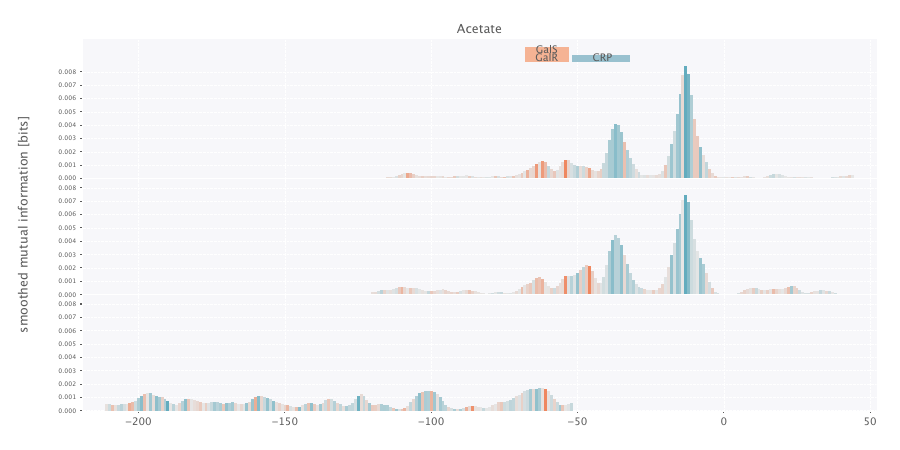}
    \includegraphics[width=0.49\linewidth]{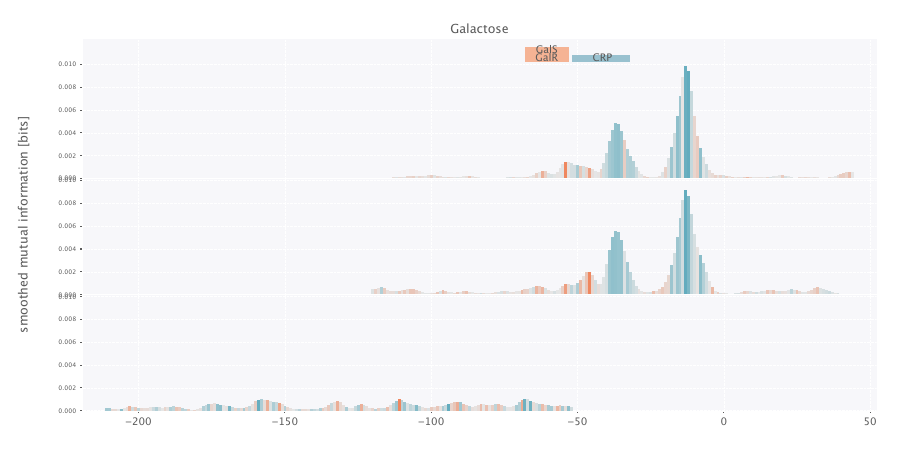}
    \includegraphics[width=0.49\linewidth]{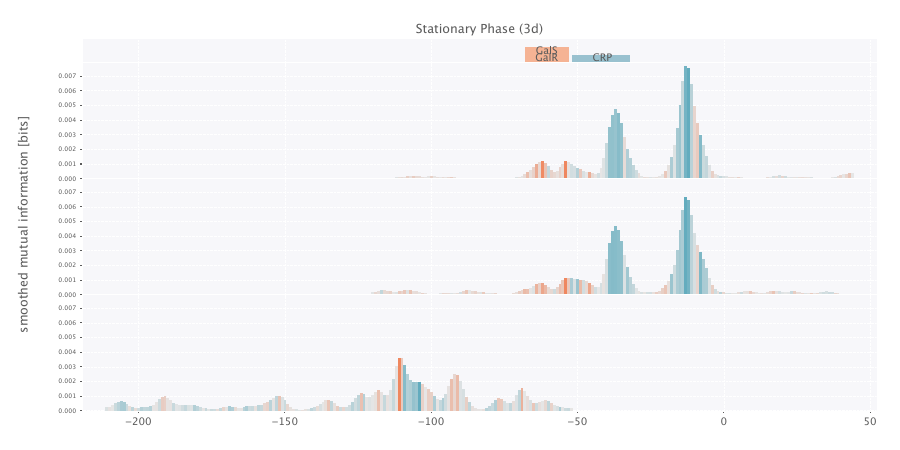}
    \caption{\textbf{Annotated regulation and information footprints for the \textit{galEp} promoters.} Footprints are shown for all three promoters, which are aligned such that 0 is the location of the transcription start site off {\it galEp1}. Conditions shown are growth in minimal media with glucose, acetate or galactose as carbon source and for stationary phase after 72h.}
    \label{fig:galE1}
\end{figure}

			\clearpage
			\clearpage
			\printbibliography[title={Supplemental References},
			segment=\therefsegment, filter=notother]
		\end{refsegment}

	}{} 
	\end{document}